\documentclass[twocolumn,aps,prd,amssymb,eqsecnum,nofootinbib,superscriptaddress,showpacs,floatfix]{revtex4}
\usepackage{ifpdf}
\ifpdf
\usepackage{hyperref}
\else
\usepackage[dvipdfm]{hyperref}
\fi

\usepackage[section] {placeins}
\usepackage{mathrsfs}
\usepackage{graphicx}
\usepackage{dcolumn}
\usepackage{bm}
\usepackage{amssymb,amsmath}
\usepackage{latexsym}
\usepackage[normalem]{ulem}
\usepackage{cancel}
\usepackage[usenames,dvipsnames]{color}
\usepackage{graphicx}
\usepackage{epsfig}
\usepackage{psfrag}
\allowdisplaybreaks
\newcommand{\Caltech}{\affiliation{Theoretical Astrophysics 350-17,
    California Institute of Technology, Pasadena, CA 91125}}
\newcommand{\Cornell}{\affiliation{Center for Radiophysics and Space
    Research, Cornell University, Ithaca, New York, 14853}}
\newcommand{\NITHEP}{\affiliation{National Institute of Theoretical Physics, Private
Bag X1 Matieland, Stellenbosch, South Africa, 7602}}

\newcommand{\thickhline}{\noalign{\hrule height 0.8pt}}

\newcommand{\beq}{\begin{equation}}
\newcommand{\eeq}{\end{equation}}
\newcommand{\bea}{\begin{eqnarray}}
\newcommand{\eea}{\end{eqnarray}}
\newcommand{\ba}{\begin{align}}
\newcommand{\ea}{\end{align}}
\newcommand{\ph}{\phantom}
\newcommand{\bma}{\begin{pmatrix}}
\newcommand{\ema}{\end{pmatrix}}

\begin{document}

\title{A geometrically motivated coordinate system for exploring 
spacetime
dynamics in numerical-relativity simulations using a quasi-Kinnersley tetrad}
\author{Fan Zhang } \Caltech
\author{Jeandrew Brink } \NITHEP
\author{B\'{e}la Szil\'{a}gyi} \Caltech
\author{Geoffrey Lovelace} \Cornell

\begin{abstract}
{ We investigate the suitability and properties of a quasi-Kinnersley tetrad and a geometrically motivated coordinate 
system as tools for quantifying both strong-field and wave-zone effects in numerical relativity (NR) simulations. We fix two of the coordinate degrees of freedom of the metric, namely the radial and latitudinal coordinates, using the Coulomb potential associated with the quasi-Kinnersley transverse frame. These coordinates are invariants of the spacetime 
and can be used to unambiguously fix the outstanding spin-boost freedom associated with the quasi-Kinnersley frame (and thus can be used to choose a preferred quasi-Kinnersley tetrad). 
In the limit of small
perturbations about a Kerr spacetime, these geometrically motivated coordinates
and quasi-Kinnersley tetrad reduce to Boyer-Lindquist coordinates and the
Kinnersley tetrad, irrespective of the simulation gauge choice.
We explore the properties of this construction both analytically and numerically, and we gain insights regarding the propagation of radiation described by a super-Poynting vector, further motivating the use of this construction in NR simulations. 
We also quantify in detail the peeling properties of the chosen tetrad and gauge.
We argue that these choices are particularly well suited for a rapidly converging wave-extraction algorithm as 
the extraction location approaches infinity, and we explore numerically the extent to which this property remains applicable on the interior of a computational domain. Using a number of additional tests, we verify numerically that the prescription behaves as required in the appropriate limits regardless of simulation gauge; these tests could also serve to benchmark other wave extraction methods.    
We explore the behavior of the geometrically motivated coordinate system in dynamical binary-black-hole NR mergers; while we obtain no unexpected results, we do find that these coordinates turn out to be useful for visualizing NR simulations (for example, for vividly illustrating effects such as the initial burst of 
spurious ``junk'' radiation passing through the computational domain).
Finally, we carefully scrutinize the head-on collision of two black holes and, for example, the way in which the extracted waveform changes as it moves through the computational domain. 
}
\end{abstract}

\date{\today}
\maketitle

\section{Introduction}\label{sec:intro}
Numerical relativity (NR) has made great strides in recent years and is now able to
explore binary black hole, 
black hole - neutron star, and neutron star - neutron star mergers in a wide 
variety of configurations (see~\cite{Centrella:2010,McWilliams:2010iq} for recent reviews).
Numerical simulations are crucial tools for calibrating and validating the large template bank of
analytic, phenomenological waveforms that will be used to search for gravitational waves in data from detectors such as the Laser Interferometer
Gravitational-Wave Observatory (LIGO)~\cite{Barish:1999,Sigg:2008}, Virgo~\cite{Acernese-etal:2006} and the Large-scale Cryogenic Gravitational-wave Telescope (LCGT)~\cite{Kuroda:2010}. Numerical simulations also make it possible, for the first time, to explore fully dynamical
spacetimes in the strong field region, such as the spacetime of a compact-binary merger. 

An important
attribute of any analysis performed on numerical simulations is the ability to extract
information in a manner independent of the gauge in which one chooses to perform the simulation.  
In this paper, we suggest one such
strategy: using a quasi-Kinnersley tetrad adapted to a choice of coordinates that are
computed using the curvature invariants of the spacetime. Most calculations
presented in this paper are local, allowing our tetrad and choice of 
geometrically motivated coordinates (and all quantities derived from them) to be computed 
in real time
during a numerical simulation (i.e. without post-processing). The
proposed scheme is also applicable throughout the spacetime, allowing 
us to study phenomena in both the strongly curved and asymptotic flat 
regions with
the same tools. 

In order to extract the $6$ physical degrees of freedom of a  
 general Lorentzian metric in four dimensions expressed in terms of
a tetrad formulation, $10$ degrees of freedom have to be specified [see e.g. \cite{hamiltonnotes}].
Of these $10$ degrees of freedom, $6$ are associated with the tetrad at a particular point on the spacetime manifold and $4$ originate 
from the freedom to label that point (the choice of gauge). 
A common choice of tetrad and the one adopted here is the
Newman-Penrose (NP) null tetrad, which consists of two real null vectors denoted $\bm{l}$
and $\bm{n}$ as well as a complex conjugate pair of null vectors $\bm{m}$ and $\bm{\overline{m}}$.
As we demonstrate explicitly in Sec.~\ref{sec:mathprelim}, where we consider the mathematical
details in greater depth, the tetrad choice is not unique. The
freedom to orient and scale the tetrad is expressed by 6 parameters associated with
a general Lorentz transformation between two different null tetrads at a fixed point
in the spacetime.

In order to extract physically meaningful quantities and to compare results from
different simulations and numerical codes, an explicit prescription for the tetrad
choice must be made.  Two  geometrically motivated prescriptions for orientating
the tetrad immediately suggest themselves: choosing i) the principle null frame
(PNF) or ii) the transverse frame (TF). (By ``frame'', we mean a set of tetrads 
related by a Type III transformation [Sec.~\ref{sec:LorentzTrans}].) 
The relationship between these two frames and
their properties are discussed in greater detail in Secs.~\ref{sec:mathprelim} and \ref{sec:Ana}; either
one of these two choices immediately removes 4 of the 6 possible tetrad degrees
of freedom. The remaining 2 degrees of freedom in the tetrad choice 
are more subtle [see the discussion in Sec.~\ref{sec:spinboost}]. 
 
The procedure we adopt in this paper is to choose a special transverse tetrad that
becomes the Kinnersley tetrad \cite{Kinnersley:1969zza} in Type-D spacetimes. 
The properties of these tetrads (known as quasi-Kinnersley
tetrads, or QKTs) and their importance for NR have previously been explored in 
detail~\cite{Beetle2005,Nerozzi2005,Burko2006,Nerozzi:2005hz,Burko:2007ps,Campanelli2006}. Part of the motivation for choosing a QKT is implicit in Chandrasekhar's
work on the gravitational perturbations of the Kerr black hole~\cite{ChandrasekharBook}: 
in this work, he
 showed that for a given perturbation of the Kerr metric (expressed in terms of the Weyl scalar $\delta \Psi_4$) it is always possible, working to linear order, to find a transverse tetrad and a gauge 
constructed from the Coulomb potential associated with this tetrad such that the Coulomb potential of the perturbed and background metrics are the same. 

We revisit these ideas in Sec.~\ref{sec:Ana}, where we investigate the properties of the quasi-Kinnersley tetrad choice. We concentrate 
on the implications of the intrinsic geometrical properties 
of the tetrad, rather than (as previous works have done) focusing on the tetrad's properties in a perturbative limit. For example, we explore the directions of energy flow
using the super-Poynting vector, showing that the choice of a QKT naturally
aligns the tetrad with the wave-fronts of passing radiation. This observation suggests that, \emph{even in the strong field regime}, the QKT is a natural, geometrically motivated tetrad choice for observing the flow of radiation and other spacetime dynamics.

After specializing to the transverse frame there exist two remaining degrees of
tetrad freedom: the freedom of the spin-boost transformations. 
We fix this remaining tetrad freedom 
by relying on a straightforward extension
of Chandrasekhar's work~\cite{ChandrasekharBook} to the strong field regime. We present the mathematical details in Secs.~\ref{sec:COORDS} and \ref{sec:SpinBoostFixingByCoords}: specifically, we use the Coulomb potential $\hat{\Psi}_2$
on the QKT to introduce a pair of geometrically motivated radial and latitudinal coordinates.
Note that $\hat{\Psi}_2$ on the transverse frame is spin-boost independent, that the resulting 
coordinates can be constructed from the curvature invariants $I$ and $J$ only, and that 
these coordinates reduce to the Boyer-Lindquist radial and latitudinal coordinates for Kerr spacetimes.
We then use these geometrically motivated coordinates to fix the spin-boost freedom by ensuring that the projection of
the tetrad base vectors onto the gradients of the new coordinates obey the relations found for the Kinnersley tetrad in the Kerr limit. 
 
One application of our chosen QKT and geometrically motivated coordinates is 
gravitational-wave extraction. For isolated, gravitating systems, gravitational radiation is only strictly defined at future null infinity; this is a consequence of  
the so-called
``peeling property'' that governs the behavior of the Weyl curvature scalars as measured on an affinely parametrized out-going geodesic. With the goal of using our tetrad and gauge prescription as a possible wave extraction method, we work out the implications that this peeling property has for the Weyl curvature scalar expressed on the QKT in Sec.~\ref{sec:Peeling}. We highlight not only the falloff behavior of the QKT Newman-Penrose quantities but also the behavior of the geometrically constructed radial coordinate. 
We explore graphically some of the implications of the peeling property for the bunching of principle null directions and argue that the directions associated with QKT are the optimal out-going directions for ensuring rapid convergence of the computed radiation quantities to the correct asymptotic results.
   
We implement our geometrically motivated coordinates and QKT numerically within the context of a pseudospectral NR code in Sec.~\ref{sec:Cons}, and we present a number of numerical simulations demonstrating the
behavior of our coordinates and QKT in Sec.~\ref{sec:Numtests}.
First, we carry out, for both non-radiative and radiative spacetimes, a few checks to verify that our scheme works correctly regardless 
of the choice of gauge in the simulation [Secs.~\ref{sec:Num} and \ref{sec:Num2}, respectively]. We confirm that we obtain numerically the correct perturbation-theory results, and we suggest that these tests could be used to benchmark other wave-extraction algorithms.

Finally, we examine the application of the QKT scheme to NR simulations of binary-black-hole collisions in Sec.~\ref{sec:BinaryBoth}, considering both the wave zone and the strong field regions. We consider a 16 orbit, equal-mass binary-black-hole in-spiral and a head-on plunge, merger, and ringdown, explicitly illustrating many of the ideas in the theoretical discussions of the previous sections. We then briefly conclude with a discussion of our
results and of prospects for the further development of our proposed scheme  
in Sec.~\ref{sec:Conclusion}.

\section{Mathematical Preliminaries} \label{sec:mathprelim}
In this section, we briefly summarize some important properties of Newman-Penrose and 
orthonormal tetrads [Sec.~\ref{sec:tetradsss}], the Weyl curvature tensor [Sec.~\ref{sec:WeylRep}], the 
Lorentz transformations of the Newman-Penrose tetrad [Sec.~\ref{sec:LorentzTrans}], and the Kinnersley
tetrad in Kerr spacetime [Sec.~\ref{sec:KerrKinnersley}]. 
Note that here and throughout this paper, letters from the front part of the Latin alphabet are used for 
four dimensional coordinate bases, those from the middle part of the Latin alphabet denotes quantities in 
three dimensional coordinate bases,  
while Greek indices are used for tetrad bases. Bold-face fonts denote vectors and tensors. 

\subsection{Newman-Penrose and orthonormal tetrads} \label{sec:tetradsss}
Two types of tetrad basis are particularly useful for the 
exploration of generic
spacetimes, such as the spacetimes of numerical-relativity simulations of compact-binary 
mergers: i) the Newman-Penrose (NP) tetrad basis
$\{e_a^{\alpha}\}= \{l_a,n_a,m_a,\overline{m}_a\}$, and ii) an orthonormal tetrad 
$\{E_a^{\alpha}\}= \{T_a,E_a^2,E_a^3,N_a \}$, 
which is closely related to the NP tetrad as follows. 
The quantities $\bm{E^2}$ and $\bm{E^3}$ are
generally associated with angular variables on a closed 
2-surface and are related
to the complex null vector $\bm{m}$ by $E_a^2 = \sqrt{2}\Re(m_a)$, $E_a^3=\sqrt{2}\Im(m_a)$.
Here $\Re(\bm{m})$ and
$\Im(\bm{m}) $ denote  the real and imaginary parts of $\bm{m}$, respectively.
The timelike vector $\bm{T}$ and spacelike vector $\bm{N}$ are 
related to the null vectors $\bm{l}$ and $\bm{n}$ by
the transformations  
\begin{align} \label{eq:OrthonormalVsNullTetrad}
l^{a} &= \frac{1}{\sqrt{2}}(T^{a}+N^{a}),& \quad n^{a} &=
\frac{1}{\sqrt{2}}(T^{a}-N^{a}).
\end{align}

The metric expressed on the orthonormal basis is the Minkowski metric,
$\gamma^{\alpha \beta}=\mbox{diag} \{-1,1,1,1\}$, while on the NP tetrad basis 
the metric
is 
 \begin{equation} \label{eq:NullNormalization}
\eta^{\alpha \beta} = \bma
0 & -1 & 0 & 0 \\
-1 & 0 & 0 & 0 \\
0 & 0 & 0 & 1 \\
0 & 0 & 1 & 0 \\
\ema .
\end{equation}
On the coordinate basis, the components of the metric are given by 
\begin{align}
g^{ab} =
\eta^{\alpha\beta}e^{a}_{\alpha} e^{b}_{\beta} = -2 n^{(a } l^{b)} + 2
m^{(a}\overline{m}^{b)} \label{metexp}.
\end{align} 

\subsection{Representations of Weyl curvature tensor}\label{sec:WeylRep}
One aim of this paper is to uniquely fix the NP tetrad basis 
to obtain a set of NP scalars from which an unambiguous measure of
the Weyl curvature (equal to the Riemann curvature in vacuum) 
can be read off. 

On the NP tetrad, the curvature content of the 
Weyl tensor can be expressed in terms
of five complex scalar functions 
\begin{eqnarray} 
\Psi_0 &=& -C_{abcd} l^a m^b l^c m^d \label{eq:Psi0} \\
\Psi_1 &=& -C_{abcd} l^a n^b l^c m^d \label{eq:Psi1} \\
\Psi_2 &=& -C_{abcd} l^a m^b \overline{m}^c n^d \label{eq:Psi2} \\
\Psi_3 &=& -C_{abcd} l^a n^b \overline{m}^c n^d \label{eq:Psi3} \\
\Psi_4 &=& -C_{abcd} n^a \overline{m}^b n^c \overline{m}^d \label{eq:Psi4}.
\end{eqnarray} 

An equivalent description of the Weyl curvature can be found on the orthonormal frame with associated timelike vector
$\bm{T}$. 
This is done by defining
 gravitoelectric $\bm{\mathcal{E}}$  and gravitomagnetic  $\bm{\mathcal{B}}$ 
tensors by, respectively, 
twice contracting $\bm{T}$ with the Weyl tensor and with its Hodge dual:
\begin{align} 
\mathcal{E}_{ij} &= h_i{}^{a}h_j^{c} C_{abcd} T^b T^d \,, \label{eq:GravE} \\  
\mathcal{B}_{ij} &= -\frac{1}{2}h_i{}^a h_j{}^c \epsilon_{abef} C^{ef}{}_{cd}T^{b}T^d \,. \label{eq:GravM}
\end{align}
Here $\bm{h}$ denotes the projection operator onto the local spatial hyper-surface orthogonal to $\bm{T}$. 
The normalization for the Levi-Civita tensors is such that $\epsilon_{0123} =1$ and $\epsilon_{123} = 1$ in right-handed orthonormal tetrads and spatial triads respectively [see~\cite{Nichols:2011pu} for a discussion of different conventions in literature]. 
These two tensors can be combined to obtain a complex tensor 
\begin{equation}\label{eq:QDef}
\mathcal{Q}_{ij} \equiv \mathcal{E}_{ij} + i \mathcal{B}_{ij}. 
\end{equation}
The curvature information contained in $\bm{\mathcal{Q}}$ is exactly the same as that contained in the five NP scalars. Recasting this information in terms of  $\bm{\mathcal{Q}}$ allows us to make use of the fact that the
 $\bm{\mathcal{E}}$ and $\bm{\mathcal{B}}$ tensors  describe the tidal acceleration and differential frame-dragging  to 
visualize the curvature of spacetime [see, e.g., Refs.~\cite{OwenEtAl:2011,Nichols:2011pu,Dennison2012,Estabrook1964,Schmid2009}]. 

To make the equivalence between $\Psi_0,\Psi_1,\Psi_2,\Psi_3,\Psi_4$ 
and $\bm{\mathcal{Q}}$ explicit, we note that the components of the complex
gravitoelectromagnetic tensor expressed on the spatial triad 
$\{\bm{E^2},\bm{E^3},\bm{N}\}$ are 
\begin{equation} \label{eq:PsiMatrix}
\bm{\mathcal{Q}}=\left[ {\begin{array}{ccc}
\Psi_2 - \frac{\Psi_0+\Psi_4}{2} & i \frac{\Psi_0-\Psi_4}{2} &
\Psi_1-\Psi_3 \\
i \frac{\Psi_0-\Psi_4}{2} & \Psi_2 + \frac{\Psi_0+\Psi_4}{2} &
-i(\Psi_1+\Psi_3)\\
\Psi_1-\Psi_3 &  -i(\Psi_1+\Psi_3) & -2\Psi_2
\end{array}} \right].
\end{equation}
Furthermore $\bm{\mathcal{Q}}$ is symmetric and trace free ($\mathcal{Q}^i_{\ i}=0$). 
These results follow from direct substitution of the definition of the orthogonal
basis vectors in terms of the NP basis vectors 
[Eq.~(\ref{eq:OrthonormalVsNullTetrad})] and the definition of the NP scalars
[Eqs.~(\ref{eq:Psi0})--(\ref{eq:Psi4})] into the definition of $\bm{\mathcal{Q}}$ 
[Eqs.~(\ref{eq:GravE}), (\ref{eq:GravM}) and (\ref{eq:QDef})].

Finally, note that for any spacetime in general relativity, 
there are a set of 16 scalar functions or Carminati-McLenaghan  
curvature invariants~\cite{Carminati:1991} that can be
constructed from polynomial contractions of the Riemann tensor. 
In vacuum, four of these scalars are non-vanishing and comprise 
a complete set of invariants. 
These four scalars can be combined into two complex functions $I$ and
$J$ and  are independent of tetrad choice.  In terms of the quantities already
computed in this section, these curvature invariants can be expressed as  
\begin{equation} \label{eq:IJ}
\begin{split}
  I =& \frac{1}{2} ( \mathcal{E}_{\ph{k}i}^k \mathcal{E}^i_{\ph{k}k}-B_{\ph{k}i}^k
B^i_{\ph{k}k} )+i( \mathcal{E}_{\ph{k}i}^k B_{\ph{k}k}^i )\\
=& \Psi_4 \Psi_0 - 4 \Psi_1 \Psi_3 + 3 \Psi_2^2 \\
       J =&   \frac{1}{6}( \mathcal{E}_{\ph{k}i}^k \mathcal{E}_{\ph{k}l}^i
\mathcal{E}_{\ph{k}k}^l - 3\mathcal{E}_{\ph{k}i}^k B_{\ph{k}l}^i
B_{\ph{k}k}^l ) \\ 
         &  -\frac{i}{6}( B_{\ph{k}i}^k B_{\ph{k}l}^i B_{\ph{k}k}^l -
3\mathcal{E}_{\ph{k}i}^k \mathcal{E}_{\ph{k}l}^i B_{\ph{k}k}^l ) \\ 
 =& 
\left| \begin{array}{ccc}
\Psi_4 & \Psi_3 & \Psi_2 \\
\Psi_3 & \Psi_2 & \Psi_1 \\
\Psi_2 & \Psi_1 & \Psi_0 \end{array} \right|
\end{split}
\end{equation}
The invariants $I$ and $J$ 
play a key role in constructing our geometrically
motivated coordinate system [Sec.~\ref{sec:Ana}].

\subsection{Lorentz transformations}\label{sec:LorentzTrans}
There are six transformations of the NP basis vectors $e_a^\alpha$ that retain the 
form of the metric given in Eq.~\eqref{eq:NullNormalization}. These are the six Lorentz
transformations, which parametrize the six degrees of tetrad freedom
\cite{ChandrasekharBook}. The Lorentz transformations can be decomposed into three types
depending on which null vector a particular transformation leaves unchanged:
\begin{itemize}
\item Type I: ($\bm{l}$ unchanged)
\begin{equation} \label{eq:Type1}
\begin{split}
\bm{l} &\rightarrow \bm{l}, \quad \bm{n} \rightarrow \bm{n} + \overline{a} \bm{m} + a \bm{\overline{m}} + a\overline{a} \bm{l} \\
\bm{m} &\rightarrow \bm{m} + a\bm{l}, \quad \bm{\overline{m}}  \rightarrow \bm{\overline{m}} + \overline{a} \bm{l} \\
\end{split}
\end{equation}
\item Type II: ($\bm{n}$ unchanged)
\begin{equation} \label{eq:Type2}
\begin{split}
\bm{l} &\rightarrow \bm{l} + \overline{b} \bm{m} + b \bm{\overline{m}} + b\overline{b} \bm{n}, \quad
\bm{n} \rightarrow \bm{n} \\
\bm{m} &\rightarrow \bm{m} + b\bm{n}, \quad
\bm{\overline{m}} \rightarrow \bm{\overline{m}} + \overline{b} \bm{n}
\end{split}
\end{equation}
\item Type III: (both $\bm{l}$ and $\bm{n}$ unchanged)
\begin{equation} \label{eq:Type3}
\begin{split}
\bm{l} &\rightarrow A^{-1} \bm{l},  \quad \quad \bm{n} \rightarrow A \bm{n}       \\
\bm{m} &\rightarrow e^{i\Theta} \bm{m}, \quad \quad \bm{\overline{m}} \rightarrow e^{-i\Theta} \bm{\overline{m}}
\end{split}
\end{equation}
\end{itemize}
Here the scalars $a$ and $b$ are complex,
while $A$ and $\Theta$ are real 
and can be combined into a single complex number $\mathcal{A}=A^{-1} \exp(i\Theta)$.
The rescaling of $\bm{l}$ and $\bm{n}$ in Eq.~(\ref{eq:Type3}) is called boost
freedom, and the phase change of $\bm{m}$ is called the spin freedom. 
We will follow the
convention of Ref.~\cite{Nerozzi2005} and call a set of tetrads related by Type III
transformations a \emph{frame}.

Under the
different Lorentz transformations, the NP scalars transform as follows:
\begin{itemize}
\item Type I: 
\begin{eqnarray}
\Psi_0 &\rightarrow& \Psi_0 \nonumber \\
\Psi_1 &\rightarrow& \Psi_1 + \overline{a}\Psi_0 \nonumber \\
\Psi_2 &\rightarrow& \Psi_2 + 2\overline{a}\Psi_1 + \overline{a}^2 \Psi_0 \label{eq:TypeIOnNPScalar} \\
\Psi_3 &\rightarrow& \Psi_3 + 3\overline{a}\Psi_2 + 3\overline{a}^2\Psi_1 + \overline{a}^3 \Psi_0 \nonumber \\
\Psi_4 &\rightarrow& \Psi_4 + 4\overline{a}\Psi_3 + 6\overline{a}^2\Psi_2 + 4\overline{a}^3\Psi_1 + \overline{a}^4 \Psi_0 \nonumber
\end{eqnarray}
\item Type II: 
\begin{eqnarray}
\Psi_0 &\rightarrow& \Psi_0 + 4b\Psi_1 + 6 b^2\Psi_2 + 4 b^3\Psi_3 + b^4 \Psi_4 \nonumber\\
\Psi_1 &\rightarrow& \Psi_1 + 3b\Psi_2 + 3b^2\Psi_3 + b^3 \Psi_4  \nonumber \\
\Psi_2 &\rightarrow& \Psi_2 + 2b\Psi_3 + b^2 \Psi_4 \label{eq:TypeIIOnNPScalar} \\
\Psi_3 &\rightarrow& \Psi_3 + b\Psi_4  \nonumber \\
\Psi_4 &\rightarrow& \Psi_4 \nonumber
\end{eqnarray}
\item Type III: 
\begin{align}
\Psi_0 &\rightarrow A^{-2}e^{2i\Theta}\Psi_0, &
\Psi_1 &\rightarrow A^{-1}e^{i\Theta}\Psi_1, &
\Psi_2 &\rightarrow \Psi_2  \notag\\
\Psi_4 &\rightarrow A^{2}e^{-2i\Theta}\Psi_4,&\Psi_3 &\rightarrow Ae^{-i\Theta}\Psi_3 
  \label{eq:TypeIIIOnNPScalar}
\end{align}
\end{itemize}

For any algebraically general spacetime, two special frame choices exist: the principle null frame
(PNF) and the transverse frame (TF). The PNF is characterized by the property that
$\Psi_4=0=\Psi_0$; 
starting from a generic tetrad a PNF can be constructed by
appropriate Type I and Type II Lorentz transformations. 
The TF is characterized by
the property that $\Psi_3=0=\Psi_1$; starting from a PNF, a TF can be
constructed by additional Type I and Type II Lorentz transformations.

There are three TFs, but 
only one contains the Kinnersley tetrad in the Kerr limit~\cite{Nerozzi2005}. 
In keeping with earlier literature~\cite{Beetle2005,Nerozzi2005}, we will call 
this frame the quasi-Kinnersley frame (QKF) and the particular tetrad we 
pick out
of this frame the quasi-Kinnersley tetrad (QKT). 

\subsection{The Kerr metric and the Kinnersley tetrad}
\label{sec:KerrKinnersley}
The no-hair theorems~\cite{Chrusciel:1994sn,Heusler:1998ua} lead us to expect all binary-black-hole collisions to ring down to the Kerr spacetime after enough time has elapsed. The limiting Kerr metric in Boyer-Lindquist
coordinates $(t,\ r,\ \theta,\ \phi)_{BL}$ can be expressed as:
\begin{align}\label{eq:KerrMetric}
ds^2 = -\left( 1-\frac{2Mr}{\Sigma} \right)dt^2 - \frac{4 M a r \sin^2\theta}{\Sigma} dtd\phi + \frac{\Sigma}{\Delta}dr^2
\notag \\
+ \Sigma d\theta^2 + \frac{\sin^2 \theta}{\Sigma}\left[\left(r^2+a^2\right)^2 -a^2 \Delta \sin^2 \theta \right] d\phi^2,
\end{align}
where $M$ and $a$ are the mass and spin of the black hole, respectively, and the
functions entering the metric are defined by
\begin{align} \label{eq:KerrParaDefs}
\Sigma &=\rho \overline{\rho},& \rho &= r - ia\cos\theta,& 
\Delta &= r^2 - 2Mr +a^2.
\end{align}
For the  Kerr spacetime, one tetrad introduced by Kinnersley is particularly
conducive for calculation. Among other things, on this tetrad the perturbation
equations  in the NP formalism decouple~\cite{Teukolsky:1972my,ChandrasekharBook}; this
feature  allows the perturbation problem to be reduced to the study of a single
complex scalar ($\delta\Psi_4$) that governs the radiation content of the perturbed
spacetime. The Kinnersley tetrad expressed on a Boyer-Lindquist coordinate basis is
given by
\begin{eqnarray} 
l^{a}    &=& \frac{1}{\Delta}       \left[r^2 + a^2,     \Delta,  0, a            \right]\label{eq:KinnL} \\
n^{a}    &=& \frac{1}{2\Sigma}      \left[r^2 + a^2,    -\Delta,  0, a            \right]
\label{eq:KinnN} \\ 
m^{a}    &=& \frac{1}{\overline{\rho} \sqrt{2}}\left[ i\ a\sin(\theta), 0,        1, i\ \csc(\theta)
\right] \label{eq:KinnM}
\end{eqnarray}
On the Kinnersley tetrad, the only non-vanishing NP curvature scalar is
\begin{align}\Psi_2=\frac{M}{\rho^3}
. \label{eq:p2k}\end{align}

In the next section, we explore the behavior of the tetrad and curvature quantities
defined in this section in cases where the physical metric is well understood.
So doing, we build up some physical intuition that motivates our QKT choice, which we then
apply to more
complicated spacetimes, such as those found in numerical simulations.

\section{Physical considerations for choosing a tetrad  \label{sec:Ana}}
In this section, we introduce several ideas that motivate the choice of 
tetrad and gauge; we will use these ideas 
to explore spacetimes produced 
by numerical-relativity simulations. 

For our purposes, we wish to adopt a tetrad and gauge with the following properties (not in order of importance):
\begin{enumerate}
\item{The tetrad (gauge) reduces to the Kinnersley tetrad (Boyer-Lindquist coordinates) 
when the spacetime is 
a weakly perturbed black hole. 
\label{crit:Kinnersley}}

\item{The choice of tetrad and gauge should be independent of the coordinate
system, including the slicing specified by 
the time coordinate, 
used in the NR simulation. 
\label{crit:gauge}}

\item{To facilitate their real-time computation during a 
NR simulation, all calculation should be local as far as possible.
\label{crit:local}}

\item{The
prescribed use for all computed quantities should be valid
in strong field regions as well
as in asymptotic regions of the spacetime. 
\label{crit:strong}}

\item{The choice of tetrad directions should as far as possible be tailored to
the physical content of the spacetime. For example, in asymptotic regions, one important direction is that of wave propagation; we seek a tetrad that
asymptotically is oriented along this direction. 
\label{crit:prop}}

\item{To facilitate gravitational-wave extrapolation 
(from the location on the NR simulation's computational domain where the waves are 
extracted to future null infinity $\mathscr{I}^+$), the falloff with radius 
of what we identify as the
radiation field should match that of an isolated, radiating system; 
i.e., it should satisfy the expected 
``peeling properties''. 
\label{crit:falloff}}

\end{enumerate}
We now consider in detail how we may achieve these criteria 
in the course of constructing our QKT.

This section roughly breaks into three parts:
\begin{enumerate}
\item We start [in Sec.~\ref{sec:prop}] by motivating the use of QKF with a new insight
regarding the relationship between its $\mathbf{l}$ basis vector and the super-Poynting vector, which 
allows it to satisfy criterion~\ref{crit:prop}. 
We then review the construction of the QKF 
in Sec.~\ref{sec:QKF}. 
\item Next, we concentrate on fixing
the spin-boost freedom to select the QKT out of the QKF. 
First of all, in Sec.~\ref{sec:spinboost} 
we discuss several methods for fixing 
this freedom that have appeared in literature. Then we present our proposal to achieve 
a global and gauge independent fixing [in Sec.~\ref{sec:SpinBoostFixingByCoords}] 
using a pair of geometrically motivated coordinates defined in Sec.~\ref{sec:COORDS}. 
We conclude this part with a brief discussion of issues related to the proposed scheme 
in Secs.~\ref{sec:Params} and ~\ref{sec:RemainingCoords}.
\item Finally, we discuss   
[in Sec.~\ref{sec:Peeling}] 
the conformity of the final QKT to criterion~\ref{crit:falloff} 
and further motivate its use. 
\end{enumerate}

\subsection{The TF and  wave-propagation direction \label{sec:Direction}}\label{sec:prop}
The Kinnersley tetrad [Eqs.~\eqref{eq:KinnL}--\eqref{eq:KinnM}]
is both a PNF and a TF~\cite{Penrose1986V2} [cf.~Sec.~\ref{sec:LorentzTrans}]; this
implies that the Kerr spacetime is Petrov Type D.
Generic non-Type-D spacetimes do not have this property: for
them no tetrad that is both a PNF and a TF exists, so one must decide which if
either of these properties to preserve. Here, we do not want $\Psi_4$, 
which plays an important role in the perturbation problem, to vanish; therefore, we
choose a tetrad that is a TF~\cite{Beetle2005,Nerozzi2005,Burko2006,Nerozzi:2005hz,Burko:2007ps}. In fact, one particular advantage of selecting the TF is its
ability to identify the direction of wave propagation in the asymptotic region [cf. criterion~\ref{crit:prop}].

In electromagnetism, a local wave vector that points in the normal direction to the
surfaces of constant phase (wavefronts) can be defined. If the medium
through which the wave is travelling is isotropic, this direction corresponds to the
direction of the waves' energy flow, or the ``wave-propagation direction'',
which is determined by the direction of the Poynting vector,
\begin{equation}
\mathcal{P}_i = \epsilon_{ijk} E^j B^k
\end{equation} where the vectors $E^j$ and $B^k$ are the electric and magnetic
field vectors. In this subsection, we summarize the relationship between the QKT and the
gravitational waves' counterpart to Poynting vector.

One approach 
for constructing a geometrically motivated tetrad follows a suggestion
by Szekeres~\cite{Szekeres1965}, which is to create a gravitational compass out of a number of
springs. Such a device is sensitive to the spacetime curvature and can be oriented
so that the longitudinal gravitational wave components vanish;
mathematically, this amounts to
reorienting the observer's tetrad so that it is a TF, which can be done
using Type I and Type II transformations to set
$\Psi_1=0=\Psi_3$.
We note that 
Chandrasekhar~\cite{ChandrasekharBook} employed the use of a TF for his
program of metric reconstruction from a small perturbation in curvature $\delta
\Psi_4$ on a background Kerr metric.

Choosing a TF turns out to orient the tetrad along the direction
of energy flow, i.e., along the super-Poynting vector
\cite{Maartens1998,ZakharovBook}
\begin{equation} \label{eq:BelRobinson} 
\mathcal{P}_{i}  
=\epsilon_{ijk}\mathcal{E}^{j}{}_{l}\mathcal{B}^{kl},  
\end{equation} 
which defines a spatial direction associated with the wave-propagation direction~\cite{Anninos:1994gp}.
The super-Poynting vector's components in the orthonormal triad $\{E_i^2,E_i^3,N_i \}$,
using the explicit form of gravitoelectromagnetic tensor in Eq.~(\ref{eq:PsiMatrix}), are
\begin{align}
\mathcal{P}_{E^2} &= -P_0(0,1)-3P_0(1,2)-3P_0(2,3)-P_0(3,4)\notag \\
\mathcal{P}_{E^3} &= P_1(0,1)+3P_1(1,2)+3P_1(2,3)+P_1(3,4)\notag \\
\mathcal{P}_{N}                  &=
\frac{1}{2}\left(-|\Psi_0|^2-2|\Psi_1|^2+2|\Psi_3|^2+|\Psi_4|^2
\right),\label{eq:superpontingcomp}
\end{align}
where the functions $P_0$ and $P_1$ are defined to be
\bea
P_0(p,q)&\equiv&\Re(\Psi_p)\Re(\Psi_q)+\Im(\Psi_p)\Im(\Psi_q) \\
P_1(p,q)&\equiv&\Re(\Psi_p)\Im(\Psi_q)-\Re(\Psi_q)\Im(\Psi_p).
\eea
By transforming to a TF, where $\Psi_1=0=\Psi_3$, Eq.~\eqref{eq:superpontingcomp} simplifies significantly, becoming 
\begin{equation} \label{eq:EnFlux}
\bm{\mathcal{P}} = \frac{1}{2}\left(|\Psi_4|^2-|\Psi_0|^2 \right) \bm{N},
\end{equation}
where its direction corresponds to spatial normal direction $\bm{N}$
fixed by our choice of TF and Eq.~(\ref{eq:OrthonormalVsNullTetrad}),
which relates $\bm{N}$ to the NP tetrad vectors $\bm{l}$ and
$\bm{n}$. By selecting the TF, we have oriented the tetrad
according to the flow of energy within the spacetime, achieving
criterion~\ref{crit:prop}. We believe this is one of the strongest
motivating factors for making the TF choice.

\subsection{Computing the quasi-Kinnersley frame on a given spacelike hyper-surface} \label{sec:QKF}
In this subsection, we review the procedure for constructing the TF
that contains the Kinnersley tetrad in the Kerr limit. This, as stated
before, is named the quasi-Kinnersley frame, or QKF. We
follow mostly the derivation of Ref.~\cite{Beetle2005}. 

\subsubsection{A spatial eigenvector problem for the QKF}
Numerical relativity simulations typically split 
the 4-dimensional spacetime to be computed into a set of
3-dimensional spatial slices. In the usual 3+1 split, the spacetime metric
$g_{ab}$ is split into a spatial metric $h_{ij}$, lapse $\alpha$, and shift $\beta^i$ according to
\begin{equation}
g_{ab} dx^a dx^b = -\alpha^2 dt^2 + h_{ij}(dx^i 
+ \beta^i dt)(dx^j + \beta^j dt),
\end{equation} while the Einstein equations in vacuum split into 
evolution equations (for advancing from one slice to the next)
\begin{equation}
R_{ij} - \frac{1}{2} g_{ij} R = 0
\end{equation} and constraint equations (satisfied on all slices)
\begin{eqnarray}
R_{TT} - \frac{1}{2} g_{TT} R = 0,\\
R_{Tj} - \frac{1}{2} g_{Tj} R = 0,
\end{eqnarray} 
where $R_{ab}$ and $R$ are the Ricci tensor and Ricci scalar of the 
spacetime, respectively, the component $T$ is in the direction normal 
to the spatial slice, and the components $i$ and $j$ 
lie within the spatial slice.

As mentioned in Sec.~\ref{sec:mathprelim}, for a given spatial slice with future directed unit normal $\bm{T}$, the curvature can be expressed in terms of the
 gravitoelectric tensor $\bm{\mathcal{E}}$ and the gravitomagnetic tensor $\bm{\mathcal{B}}$ defined in Eqs.~\eqref{eq:GravE} and \eqref{eq:GravM}. 
In terms of the 3+1 quantities typically computed in NR codes, 
provided that the Einstein constraint equations are 
satisfied,
the gravitoelectromagnetic tensors in vacuum can be expressed as   
\begin{align}
\mathcal{E}_{ij} &= {^3 R_{ij}}+K K_{ij}-K_{ik}K^k_j  \notag\\
\mathcal{B}_{ij} &= -\epsilon_i^{\ kl} D_kK_{lj} \label{eq:SpatialEB}
\end{align} 
where $K$ is the trace of the extrinsic curvature $K_{ij}$, while ${^3 R_{ij}}$ and $D_k$ are 
the Ricci curvature and connection, respectively, associated with the spatial 
metric $h_{ij}$.

Given the gravitoelectric and gravitomagnetic tensors, a powerful
tool~\cite{Nichols:2011pu,OwenEtAl:2011} for visualizing the curvature of 
spacetime is 
a plot of the ``vortex'' and ``tendex'' lines, which are 
the flow lines of the
eigenvectors of the gravitoelectromagnetic tensors $\mathcal{E}_{ij}$ and 
$\mathcal{B}_{ij}$. The QKF is also related to an
eigenvalue problem involving $\mathcal{E}_{ij}$ and $\mathcal{B}_{ij}$, 
albeit a complex
one involving the complex tensor $\bm{\mathcal{Q}} \equiv
\bm{\mathcal{E}} + i \bm{\mathcal{B}}$. Specifically, 
it was shown in Ref.~\cite{Beetle2005} that the QKF
can be constructed from the eigenvector $\tilde{\sigma}^i$ that
satisfies the eigenvector equation
\begin{align}
\mathcal{Q}^i_j \tilde{\sigma}^j = -2 \hat{\Psi}_2 \tilde{\sigma}^i \label{eq:EigenQ}
\end{align}
where the eigenvalue $-2\hat{\Psi}_2$ has
the value of $-2\Psi_2$ computed on the QKF. 
Here and
throughout the rest of this paper, we adopt the convention of denoting
quantities associated with a QKF (such as the NP tetrad vector 
$\bm{\tilde{l}}$) with an overscript
tilde and quantities associated with the final tetrad, whose spin-boost
degrees of freedom have been uniquely fixed (yielding a preferred QKT),
with an overscript hat (e.g. $\bm{\hat{l}}$). As we will show in
greater detail later in the section, the QKF's Coulomb potential
$\hat{\Psi}_2$ can be constructed out of the curvature invariants $I$
and $J$ of the spacetime and is invariant under spin-boost
transformations; therefore, we denote $\hat{\Psi}_2$ with a hat 
to indicate it has been fixed to its final value.  

\subsubsection{Selecting the correct eigenvalue \label{sec:SelectEVal}}
For any symmetric matrix $\mathcal{M}$, the eigenvalues associated with the eigenvector problem $\mathcal{M}^i_j \xi^j = \lambda \xi^i$ obey the characteristic equation $p(\lambda) = 0$ where $p(\lambda)= \det(\mathcal{M}-\lambda \mathcal{I} ) $ and $\mathcal{I}$ is the identity matrix. For a  $3\times 3$  matrix, the characteristic equation becomes 
\begin{align}
p(\lambda) = &-\lambda ^3+ \lambda^2 \text{tr}(\mathcal{M}) \nonumber \\
	     &+ \frac{1}{2}\lambda \left( \text{tr}(\mathcal{M}^2)- \text{tr}^2(\mathcal{M})\right) + \det(\mathcal{M}),
\label{eq:characteristicP}
\end{align}
If $\mathcal{M}^i_j = \mathcal{Q}^i_j$, 
direct calculation using Eqs.~\eqref{eq:PsiMatrix} and \eqref{eq:IJ} can 
verify that $ \text{tr}( \bm{\mathcal{Q}})=0$,
$\det(\bm{\mathcal{Q}}) = 2J$ and $ \text{tr}(
 \bm{\mathcal{Q}}^2)=2I$, which reduces the characteristic polynomial to
\begin{equation}
p_{\bm{\mathcal{Q}}}(\lambda)= - \lambda^3+ \lambda I + 2 J. 
\end{equation}
The solution to this cubic
equation can be expressed using the speciality index~\cite{Baker:2000zm}
$\mathcal{S}=27J^2/I^3$ as 
\begin{align}
\lambda = \frac{3J}{I}\frac{ W(\mathcal{S})^{1/3}+ W(\mathcal{S})^{-1/3}  }{\sqrt{\mathcal{S}}} \label{eq:lleq} 
\end{align}
where $W(\mathcal{S}) \equiv \sqrt{\mathcal{S}}-\sqrt{\mathcal{S}-1}$.
There are three 
solutions\footnote{The fraction on the right of Eq.~\eqref{eq:lleq} has a
three-sheeted Riemann surface with branch points of order two at
$\mathcal{S}=0$ and $\mathcal{S}=1$, as well as a branch point of order three
at $\mathcal{S}=\infty$. The three different eigenvalues arise from
the values on the three sheets respectively \cite{Beetle2005}.}
corresponding to the three transverse frames, but only one (namely the QKF)
contains the Kinnersley tetrad in the Kerr limit \cite{Nerozzi2005} (and thus 
satisfies criterion \ref{crit:Kinnersley}).

We must now select the correct eigenvalue to define the QKF. 
Only one of the three
eigenvalues has an analytic expansion around $\mathcal{S} = 1$ 
(which holds for all Type-D spacetimes, including Kerr \cite{Baker:2000zm}).
We select this eigenvalue (which we denote $\lambda^0$) 
to define the QKF, and so 
$-2\hat{\Psi}_2=\lambda^0$.
For reference, the series expansion of $\lambda^0$ and also the other two 
eigenvalues $\lambda^{\pm}$
around $\mathcal{S} = 1$ is
\begin{align}
\lambda^0 &= -2 \hat{\Psi}_2 \sim -\frac{2J}{I} \left[ -3 + \frac{4}{3}(\mathcal{S}-1)+\cdots \right] \label{eq:P2Expand}, \\
\lambda^{\pm} &\sim -\frac{2J}{I} \left[ \frac{3}{2} \pm i\frac{\sqrt{3}}{2} \sqrt{\mathcal{S}-1}  - \frac{2}{3}(\mathcal{S}-1)+\cdots \right]. \notag
\end{align}
In practice, this selection criterion is equivalent to choosing the 
eigenvalue with the largest magnitude~\cite{Nerozzi2005}. 

\subsubsection{Constructing the QKF tetrad vectors}

We now summarize the necessary results that allow the reconstruction of the QKF from the eigenvector of the matrix $\bm{\mathcal{Q}}$; for a complete derivation, see Ref.~\cite{Beetle2005}. The eigenvector corresponding to the eigenvalue  $-2\hat{\Psi}_2$ can be expressed as
\begin{align}
\tilde{\sigma}^j = \tilde{x}^j+i \tilde{y}^j \label{eq:SpatialEigenVec}
\end{align}
where the real vectors  $\tilde{x}^j$ and $\tilde{y}^j$ are orthogonal with respect to the spatial metric $h_{ij}$
and their normalization obeys the condition 
\bea \label{eq:NormalizeEigen}
\| \bm{\tilde{x}}\|^2-\|\bm{\tilde{y}}\|^2=1.
\eea
Here and throughout this section, we will use $\| \bm{v} \|$ and $\bm{v} \cdot \bm{w}$ to represent norm and inner product of spatial vectors under $h_{ij}$.
The vectors  $\tilde{x}^j$ and $\tilde{y}^j$ can in turn be used to  define the vectors
\begin{align}
\tilde{\lambda}^i &= \frac{\tilde{x}^i+\epsilon^{ijk}\tilde{x}_j \tilde{y}_k  }{\|\bm{x}\|^2   },&  \tilde{ \nu}^i &= \frac{-\tilde{x}^i+\epsilon^{ijk}\tilde{x}_j \tilde{y}_k  }{\|\bm{x}\|^2   }, \notag\\
\tilde{\mu}^i& =   \frac{\tilde{\lambda}^i+\tilde{\nu}^i+i\epsilon^{ijk}\tilde{\lambda}_j \tilde{\nu}_k  }{1 + \bm{\tilde{\lambda}}\cdot\bm{\tilde{\nu}}   }, 
 \label{eq:CompEigenProj}
\end{align}
where the normalization condition on $\bm{\tilde{\sigma}}$ [Eq.~\ref{eq:NormalizeEigen}]
ensures 
\bea
\|\bm{\tilde{\lambda}}\| = \|\bm{\tilde{\nu}}\| = \|\Re(\bm{\tilde{\mu}})\|^2-\|\Im(\bm{\tilde{\mu}})\|^2=1.
\eea
The resulting  vectors $\bm{\tilde{\lambda}}$, $\bm{\tilde{\nu}}$ and $\bm{\tilde{\mu}}$ turn out to be proportional to the spatial projections of QKF
basis vectors $\bm{\tilde{l}}$, $\bm{\tilde{n}}$ and $\bm{\tilde{m}}$
respectively. To see this, let the spatial vectors be expressed in terms of a spatial triad $E^a_{i}$ which is part of an orthonormal tetrad $E^a_{\alpha}$ with $E^a_0 = T^a$; then, the full QKF tetrad can be constructed as follows:
\begin{equation}\begin{split}
\tilde{l}^a &= \frac{|A|^{-1}}{\sqrt{1-\bm{\tilde{\lambda}}\cdot \bm{\tilde{\nu}}}} (T^a +
\tilde{\lambda}^{i}E^a_i), \\
\tilde{n}^a &= \frac{|A|}{\sqrt{1-\bm{\tilde{\lambda}}\cdot\bm{\tilde{\nu}}}} (T^a + \tilde{\nu}^iE^a_i), \\
\tilde{m}^a &= \frac{e^{i\Theta}}{\sqrt{2}}\frac{\sqrt{1+\bm{\tilde{\lambda}}\cdot \bm{\tilde{\nu}}}}{\sqrt{1-\bm{\tilde{\lambda}}\cdot \bm{\tilde{\nu}}}} (T^a + \tilde{\mu}^{i}E_i^a).\label{eq:SpatialProjTetrad}
\end{split}
\end{equation}
Note that the residual spin-boost freedom [cf. Eq.~\eqref{eq:Type3}] has been made explicit in  Eq.~(\ref{eq:SpatialProjTetrad})
by means of the parameters $A$ and $\Theta$ (which have yet to be determined).

Also note that the equation for $\bm{\tilde{m}}$ above must be modified 
if the normal to the spatial slice $\bm{T}$ lies in the plane spanned
by $\bm{\tilde{l}}$ and $\bm{\tilde{n}}$, since in this special case
the vectors $\bm{\tilde{\lambda}}$ and $\bm{\tilde{\nu}}$ turn out not to be 
independent of each other (as is true generally) but are instead related by 
$\tilde{\lambda}^i = -\tilde{\nu}^i$.
It is unclear whether such a slicing can be found for any
spacetime, but once found, it is closely associated
with a TF.  In this case the vector $\bm{\tilde{\mu}}$ is undefined and
$\bm{\tilde{m}}$ should be constructed from any real unit vector
$\bm{\tilde{r}}$ in the spatial 2-plane orthogonal to $\bm{\tilde{\lambda}}$
and $\bm{\tilde{T}}$ according to
\begin{align}
\tilde{m}^a = \frac{e^{i\Theta}}{\sqrt{2}}(\tilde{r}^i+i \epsilon^{ijk} \tilde{\lambda}_j \tilde{r}_k)E^a_i.
\end{align}

Because the spatial eigenvector problem \eqref{eq:EigenQ} 
 can be solved point-wise, 
the construction of the QKF is a local  procedure and satisfies criterion~\ref{crit:local}. 
Furthermore, the procedure can be applied in the strong field region 
[cf. criterion~\ref{crit:strong}], although the physical interpretation is only clear if the
tetrad can be smoothly extended from there to infinity. 
By choosing our tetrad to be a QKF, we have used up four of the six
possible degrees of tetrad freedom and have uniquely fixed the directions associated with
the real null vectors $\bm{\tilde{l}}$ and $\bm{\tilde{n}}$. We will
address the remaining spin-boost freedom in the next three subsections.

\subsection{The spin-boost tetrad freedom}
\label{sec:spinboost}
After electing to work in the QKF, the residual tetrad freedom is
restricted to a Type III spin-boost transformation [Eqs.~\eqref{eq:Type3} and
  \eqref{eq:SpatialProjTetrad}]. As seen in
Eq.~\eqref{eq:TypeIIIOnNPScalar}, the boost transformation affects the
magnitude of $\tilde{\Psi}_4$, while the spin transformation modifies the
phase of $\tilde{\Psi}_4$. 

To gain some insight into what the spin-boost transformations do
physically, consider a congruence of observers whose world lines are
the integral curves of the $\bm{T}$ field in 
Eq.~(\ref{eq:OrthonormalVsNullTetrad}).
For these observers, a spin
transformation of the tetrad mixes up the two polarizations of gravitational wave 
by the induced phase 
rotation\footnote{Recall that for plane waves on Minkowski background,
we have $\Psi_4 = -\ddot{h}_{+}+i \ddot{h}_{\times}$, where $h$ is the metric
perturbation.}; in practice,
this rotation occurs
because the observers are rotating the orientation of their
coordinates and thus redefining what they consider to be the latitudinal
and longitudinal directions. 
Similarly, the boost transformation in
Eq.~(\ref{eq:Type3}) alters the velocity with which these
observers move along the direction of wave-propagation, causing the
gravitational wave they observe to be redshifted or blueshifted.

In order to identify the
gravitational wave and curvature content contained in $\Psi_4$ in an
unambiguous manner, we need to provide a prescription for fixing 
$A$ and $\Theta$ throughout the spacetime.
Note that $\tilde{\lambda}$ and
  $\tilde{\nu}$  constructed in Eq.~\eqref{eq:CompEigenProj} 
are dependent on the choice of slicing; thus simply setting $A$
  and $\Theta$ in Eq.~\eqref{eq:SpatialProjTetrad} 
to particular values does not select a tetrad in a
  slicing independent manner. Fixing these parameters but altering the 
slicing will
lead to different tetrads in the same frame (the QKF), thus when 
we leave $A$ and $\Theta$ undetermined, the frame as a whole that we 
obtain from Eq.~\eqref{eq:SpatialProjTetrad} is slicing independent.

One example of fixing the
spin-boost freedom in a gauge independent way often used in
mathematical analysis is selecting the so-called canonical transverse
tetrad (CTT)~\cite{Penrose1986V2}, which is defined by the condition that
\begin{equation} \label{eq:CanonicalTrans}
\Psi_1=0=\Psi_3 \quad \mbox{and} \quad \Psi_0 = \Psi_4.
\end{equation}  
The CTT has the property that the super-Poynting
vector given in Eq \eqref{eq:EnFlux} has vanishing magnitude; in this
tetrad, the observers are co-moving with local wavefront in the 
asymptotic region and
consequently measure $\|\bm{\mathcal{P}}\|=0$. 
Since no physical observer can
travel at the speed of light and co-move with the wavefront, we require
a more physically motivated prescription for fixing the spin-boost
freedom. 

Several approaches for providing such a physically motivated prescription 
have been suggested. 
A common approach is to impose conditions on spin
coefficients  (such as $\epsilon=0$~\cite{Kinnersley:1969zza}). 
The Kinnersley tetrad for the Kerr metric has spin 
coefficients\footnote{For how the spin coefficients 
[which are complex scalars] are defined in terms of the 
null tetrad, see e.g. Eq.~(1.286) of Ref.~\cite{ChandrasekharBook}.} 
that obey $\kappa=\sigma=\lambda=\nu=\epsilon=0$. The meaning of some of these coefficients can be gleaned from the equations governing how the tetrad evolves along the $\bm{l}$ direction, namely \cite{ChandrasekharBook}
\begin{eqnarray}
l^{b} l^{a}_{;b} &=& 2 \Re(\epsilon) l^a -\kappa \overline{m}^a - \overline{\kappa} m^a ,\label{eq:SpinCoef1} \\
l^{b} m^{a}_{;b} &=& 2 i \Im(\epsilon) m^a + \overline{\pi} l^a - \kappa n^a.\label{eq:SpinCoef2}
\end{eqnarray}
If $\kappa=0$ for example, the null vector $\bm{l}$ is 
tangent to a geodesic and  
further if $\Re(\epsilon) =0$ this geodesic 
is affinely parameterized.

Note that choosing $\bm{l}$ to be geodesic or $\kappa = 0$ is not necessarily consistent with choosing to work in a TF, although these conditions are 
consistent in the Kerr limit. In a TF, the only freedom available to set the spin coefficients to zero is the spin-boost transformation. 
Since $\kappa$ transforms as $\kappa \rightarrow A^{-2}e^{i\Theta}\kappa$  under Eq.~\eqref{eq:Type3}, the spin coefficient $\kappa$ 
cannot be set to zero. 
The spin coefficient $\epsilon$, on the other hand, transforms as 
\bea \label{eq:AThetaAlongIntegralCurve}
\epsilon \rightarrow A^{-1} \epsilon - \frac{1}{2} A^{-2} l^a \nabla_{a} A + \frac{i}{2}A^{-1} l^a \nabla_a \Theta,
\eea
and can be made to vanish by suitably chosen $A$ and $\Theta$. Equations~\eqref{eq:SpinCoef1} and \eqref{eq:SpinCoef2}, indicate that the condition $\epsilon=0$ can be used to fix the scaling of $\bm{l}$ as well as the phase of $\bm{m}$. 
 Setting $\epsilon = 0$ 
can therefore be used as a means of 
fixing the spin-boost freedom, but this choice has the 
disadvantage that Eq.~\eqref{eq:AThetaAlongIntegralCurve} must be solved in order to obtain $A$ and $\Theta$, which can be expensive numerically.

In the following subsections, we present an alternative method of fixing the spin-boost freedom  by constructing a coordinate system based on the curvature invariants. Differentials of these new coordinates are then used to set the scale or fix the spin degree of freedom of the final QKT. This method avoids the need to solve differential equations by directly imposing local conditions of the  the tetrad basis vectors.

\subsection{A geometrically motivated coordinate system \label{sec:COORDS} }
\begin{figure*}[tbp]
\includegraphics[width=0.85\textwidth]{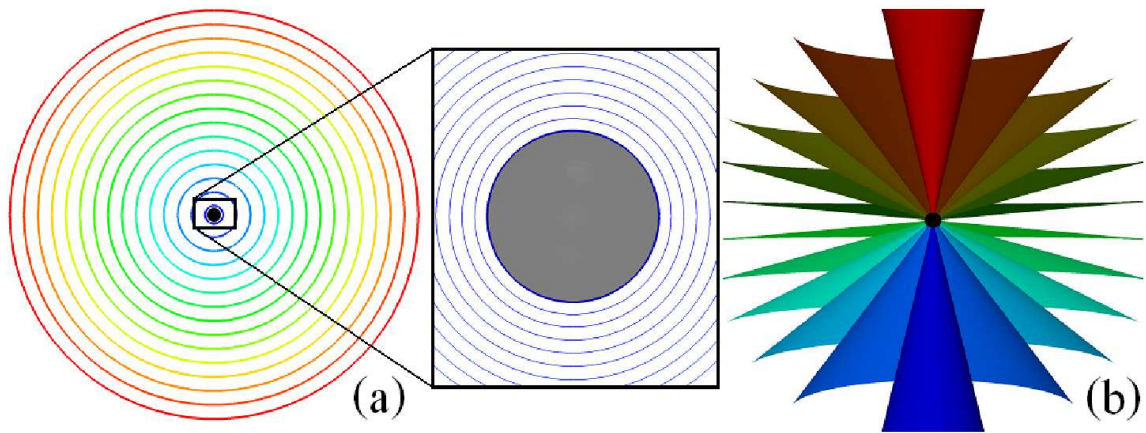}
\caption{Properties of the $(\hat{r}, \hat{\theta})$ coordinates constructed from the Coulomb potential in the QKF. (a) The equatorial plane of a Kerr spacetime in a Kerr-Schild slicing with contours of constant Boyer-Lindquist radius $\hat{r}$ at equal increments. The inset zooms in around the event horizon
(indicated by a transparent black disk). The $\hat{r}$ contour increments in the inset, while still uniform, are smaller than in the main figure, and the thick contour line coinciding with the event horizon matches the Boyer-Lindquist radius $\hat{r}_+$ in Eq.~\eqref{eq:BLHorizon}.
(b) Surfaces of constant latitudinal coordinate $\hat{\theta}$ for the Kerr-Schild slicing.
}
\label{fig:KerrImplied1}
\end{figure*}

In this paper, we fix the spin-boost freedom 
by exploiting the curvature invariant $\hat{\Psi}_2$ 
[identified
  in Eqs.~\eqref{eq:EigenQ} and \eqref{eq:P2Expand} and computed using
  Eq.~\eqref{eq:lleq}] to define geometrically motivated and
unambiguous radial and latitudinal coordinates. The quantity $\hat{\Psi}_2$ can be
interpreted as the Coulomb potential experienced by an
observer~\cite{Szekeres1965}, 
and all observers in a QKF agree on its value.
Our prescription for fixing the spin-boost freedom 
is to effectively tether our observers 
to a fixed position with respect to
the coordinates associated with the
instantaneous background Coulomb potential they experience. 
By doing this, we choose ``stationary'' 
observers that watch gravitational waves pass, 
in contrast to the CTT observers (Sec.~\ref{sec:spinboost}) that co-move 
with the waves.
In Kerr limit, our choice amounts to
selecting a set of stationary observers associated with the Boyer-Lindquist
coordinate system. 

To illustrate this idea more fully, 
note that when we work within the QKF, 
the complex gravitoelectromagnetic tensor from Eq.~\eqref{eq:PsiMatrix} reduces to 
\begin{equation} \label{eq:PsiMatrixT}
\bm{\tilde{\mathcal{Q}}}=\left[ {\begin{array}{ccc}
\hat{\Psi}_2 - (\tilde{\Psi}_0+\tilde{\Psi}_4)/2 & i (\tilde{\Psi}_0-\tilde{\Psi}_4)/2 &
0 \\
i (\tilde{\Psi}_0-\tilde{\Psi}_4)/2 & \hat{\Psi}_2 + (\tilde{\Psi}_0+\tilde{\Psi}_4)/2 &
0 \\
0 & 0 & -2\hat{\Psi}_2
\end{array}} \right],
\end{equation}
making $\bm{\tilde{N}}$ an eigenvector.
Of particular interest is the component 
\bea
\tilde{\mathcal{Q}}_{\tilde{N}\tilde{N}}= -2\hat{\Psi}_2= \mathcal{E}_{\tilde{N}\tilde{N}} + i \mathcal{B}_{\tilde{N}\tilde{N}}.
\eea
As illustrated in detail in \cite{OwenEtAl:2011} and 
particularly in Sec.~IV A of
Ref.~\cite{Nichols:2011pu}, 
within the context of vortexes and tendexes, $\mathcal{E}_{\tilde{N}\tilde{N}}$ measures tidal acceleration and $\mathcal{B}_{\tilde{N}\tilde{N}}$ the differential frame-dragging experienced by a person whose body is aligned 
along the radial $\bm{\tilde{N}}$ eigenvector. The frame dragging induced by 
the angular momentum of the source 
implies a latitudinal coordinate, and the radial tidal acceleration implies 
a radial coordinate.
The Coulomb potential $\hat{\Psi}_2$ thus
contains information about a pair of geometrically motivated 
coordinates $\hat{r}$ and $\hat{\theta}$.

\begin{figure*}[tbp]
\includegraphics[width=0.85\textwidth]{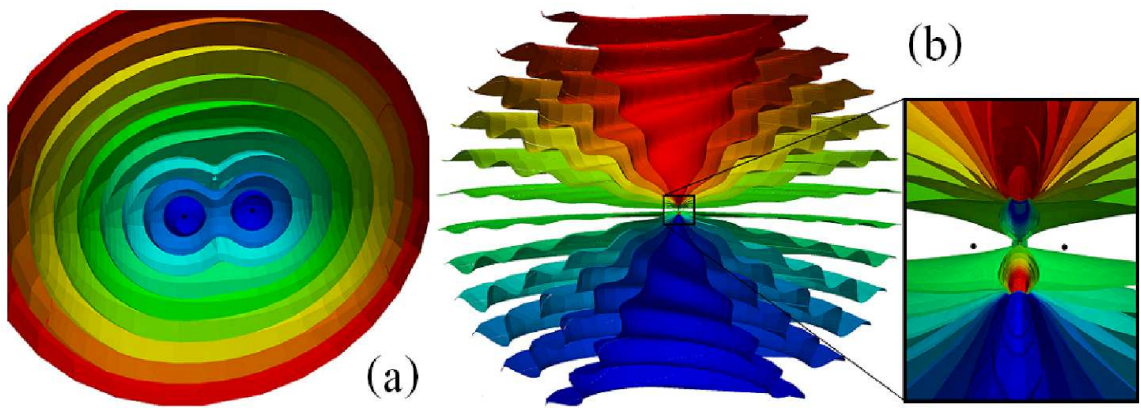}
\caption{
(a) Snapshot surfaces of constant $\hat{r}$ for an equal-mass nonspinning binary merger simulation taken during the inspiraling phase.
Far away from the black holes, the contours represent those expected from a monopole moment. 
When moving closer to the black holes, higher order multipoles 
become important. 
(b) Constant $\hat{\theta}$ surfaces for the same simulation as in (a), 
shows a ``spiral-staircase'' pattern generated by 
rotating deformed cones as discussed in greater detail in Sec.~\ref{sec:Binary}.}
\label{fig:KerrImplied2}
\end{figure*}

To relate the Coulomb potential $\hat{\Psi}_2$ to the
geometric coordinates $\hat{r}$ and $\hat{\theta}$ in a meaningful way
that reduces to the Boyer-Lindquist coordinates in the Kerr limit 
(thus satisfying criterion~\ref{crit:Kinnersley}), we
make use of expressions for the Kerr spacetime
[Eqs.~\eqref{eq:p2k} and \eqref{eq:KerrParaDefs}] to define the coordinates. In other words,
we define 
$\hat{r}$ and $\hat{\theta}$ 
using the complex equation
\begin{equation} \label{eq:GeoCoordsDef}
\begin{split}
\hat{\rho}= \hat{r} - i \hat{a}\cos(\hat{\theta})= \left(\frac{\hat{M}}{\hat{\Psi}_2}\right)^{1/3}
\end{split}  
\end{equation}
where $\hat{M}$ and $\hat{a}$ are real constants that 
become just the mass and spin of the central black hole in the Kerr limit. A discussion regarding these parameters in dynamical simulations follows in Sec.~\ref{sec:Params}. 
Recall that the Coulomb potential $\hat{\Psi}_2$ can be constructed directly from the curvature invariants $I$ and $J$ of the spacetime; 
the construction of the coordinates out of curvature invariants makes them slicing or gauge independent, thus satisfy criterion~\ref{crit:gauge}.  

Figures~\ref{fig:KerrImplied1} and \ref{fig:KerrImplied2} explore some properties of $\hat{r}$ and $\hat{\theta}$. 
The first property is the ability to recover the Boyer-Lindquist radial and latitudinal coordinates from a Kerr spacetime expressed in any slicing. A particular example using Kerr-Schild slicing is shown in Fig.~\ref{fig:KerrImplied1}, where we plot the contours of $\hat{r}$ and $\hat{\theta}$ under Kerr-Schild spatial coordinates $(r,\theta,\phi)$. The resulting figures 
show that the coordinate transformations between $(\hat{r},\hat{\theta})$ and $(r,\theta)$ 
(unlike those for Boyer-Lindquist $t$ and $\phi$) 
do not become singular at the event horizon [cf. criterion~\ref{crit:strong}], which coincide with 
the contour of  
\bea \label{eq:BLHorizon}
\hat{r} = \hat{r}_+ \equiv \hat{M}+\sqrt{\hat{M}^2-\hat{a}^2}
\eea
as expected. 
The $(\hat{r},\hat{\theta})$ coordinate system for a dynamical simulation of two equal-mass, nonspinning black holes during their inspiral phase is shown in  Fig.~\ref{fig:KerrImplied2}. The peanut shaped features in panel (a) makes apparent the fact that the coordinate system is adjusting to the intrinsic geometry of the simulation. The cones of constant angular coordinate $\hat{\theta}$  display a wavy feature when compared to the simulation coordinate $\theta$. This feature and its origin will be discussed in greater detail in Sec.~\ref{sec:Binary}, where we explore the binary simulation 
in more detail.

\subsection{Fixing the spin-boost degrees of freedom \label{sec:SpinBoostFixingByCoords} }
The previous subsection provides us with an unambiguous and geometrically
motivated set of radial and latitudinal coordinates that are valid throughout the spacetime and 
that are
independent of the choice of slicing. Our strategy for
fixing the last two degrees of tetrad freedom is to require that the
tetrad frames can be associated with observers that are in some sense
``stationary'' with respect to our geometrically motivated coordinates while also requiring that the selected tetrad reduces to the
Kinnersley tetrad in the Type-D limit.

To achieve this construction (and thus to provide a global prescription for fixing the spin-boost freedom), note that $d\hat{r}$ provides a measuring rod in the radial direction, relative to the wavefront, against which the scale of the radial component of $\bm{\hat{l}}$  can be fixed. Similarly $d\hat{\theta}$ provides a 
transverse direction which can be used to fix the phase of $\bm{\hat{m}}$.
Let us now begin with any tetrad in the QKF  $\{\bm{\tilde{l}},\bm{\tilde{n}},\bm{\tilde{m}},\bm{\overline{\tilde{m}}}\}$, constructed according to Eq.~\eqref{eq:SpatialProjTetrad}. The prescription we use to fix the parameters $A$ and $\Theta$ associated with the spin-boost degrees of freedom to obtain the final QKT $\{\bm{\hat{l}},\bm{\hat{n}},\bm{\hat{m}},\bm{\overline{\hat{m}}}\}$ is to require that the final tetrad obeys
\begin{align}
\left(d\hat{r}\right)_a \hat{l}^a &= 1; \label{eq:FixingLa} \\
\arg\left[\left(d\hat{\theta}\right)_a \hat{m}^a \right] &= \arg \left[ \hat{\rho} \right] .
 \label{eq:FixingMa}
\end{align}
Note that these conditions are exactly the conditions satisfied by the Kinnersley tetrad in Eq.~\eqref{eq:KinnL} and \eqref{eq:KinnM} except that the Boyer-Lindquist coordinate has been replaced by its corresponding geometrically constructed counterpart introduced in Sec.~\ref{sec:COORDS}. The reduction to the Kinnersley tetrad in the Type-D limit is thus trivial [cf. criterion \ref{crit:Kinnersley}]. Furthermore, conditions~\eqref{eq:FixingLa} and \eqref{eq:FixingMa} contain only local differentiation and algebraic calculations and thus obey criterion~\ref{crit:local}. They also inherit gauge independence from the QKF and the geometric coordinates, thus satisfy criterion~\ref{crit:gauge}.          

It turns out that the final QKT  can  be constructed by starting with a tetrad in the QKF with $A=1$ and $\Theta = 0$  in Eq \eqref{eq:SpatialProjTetrad}, computing the quantities
\begin{eqnarray}
 A &=& \left(d\hat{r}\right)_a \tilde{l}^a, \label{eq:FixingL} \\
   \Theta &=& - \arg\left[ \left(d\hat{\theta}\right)_a \tilde{m}^a \right]  + \arg\left[\hat{\rho}\right], \label{eq:FixingM}
\end{eqnarray}
and then substituting these values back into  Eq.~\eqref{eq:SpatialProjTetrad} to obtain the final tetrad. Our fictitious observers have now oriented and scaled their tetrads according to the Coulomb potential they experience by observing the local changes in tidal acceleration and differential frame dragging. 

\subsection{The effect of $\hat{a}$ and $\hat{M}$ on the tetrad choice  \label{sec:Params}}
In the definition of the geometric coordinates ($\hat{r}$, $\hat{\theta}$) in Eq.~\eqref{eq:GeoCoordsDef}, two constants $\hat{M}$ and $\hat{a}$ corresponding to the mass  and spin of a Kerr black hole in the Type-D limit entered our prescription. We now clarify their influence on the final computed quantities of $\hat{\Psi}_4$ and the constructed tetrad. 

First, we observe that the spin $\hat{a}$ does not affect the spin parameter $\Theta$ in expression (\ref{eq:FixingM}) and can be left undetermined, since only the direction of $d\hat{\theta}$ is required to determine the argument of its inner product with $\bm{\tilde{m}}$.

The final computed quantities are however dependent on the value of $\hat{M}$, 
which enters as a constant factor scaling the boost parameter $A$. The computed $\hat{\Psi}_4$ is simply rescaled by a constant scaling factor if the value of $\hat{M}$ is changed. This allows one to compute all quantities real time during the simulation with (say) $\hat{M}=1$ and to a posteriori rescale the results once the final mass of the remnant black hole is known.

\subsection{The remaining gauge freedom \label{sec:RemainingCoords}}
Using the appropriate combination of the curvature invariants [Sec.~\ref{sec:COORDS}] to prescribe radial and latitudinal coordinates $(\hat{r},\hat{\theta})$ fixes two of the four degrees of gauge freedom, while the choice of a TF [Sec.~\ref{sec:QKF}] and the subsequent fixing of the spin-boost freedom [Sec.~\ref{sec:SpinBoostFixingByCoords}] removes all six degrees of tetrad freedom. What remains is to fix the final two degrees of gauge freedom: the slicing (or time coordinate $\hat{t}$) and the azimuthal coordinate $\hat{\phi}$. 

For a given slicing, ``far enough'' from the strong field region, surfaces of constant $\hat{r}$ and $\hat{\theta}$ intersect in a circle. This can be seen graphically in  Fig.~\ref{fig:KerrImplied2} by superimposing plot (a) and (b) and taking ``far enough'' to mean the region where the  mass monopole and current dipole are the dominant terms in the Coulomb background. The prescription of the azimuthal coordinate $\hat{\phi}$ is then as simple as requiring that given a specific (as yet undetermined) starting point, the proper distance increments $d\hat{\phi}$ along the circle remains constant. 

Fixing the time slicing requires more finesse.
One method of specifying the time slicing indirectly is by means of a congruence of outward propagating affinely parameterized null geodesics [see Sec.~\ref{sec:PeelingInPND} below for a suggested congruence] starting from a fixed radius  $\hat{r}$; the affine parameter $\tau$ is then used as a coordinate. 
This approach is particularly suited to the task of wave extraction where the quantities computed should exhibit the scaling laws predicted by the peeling property \cite{Sachs1961,Sachs1962}. 

The prescriptions given above contain residual freedom. Fixing them 
is beyond the scope of our current work. In this paper, wherever needed, 
we simply use the coordinate time in the simulation and the simulation's 
azimuthal coordinate.

\subsection{The peeling theorem \label{sec:Peeling}}

\subsubsection{Peeling in Newman-Penrose scalars} 
In this section, we consider the peeling property,
  which describes the way in which, for an isolated gravitating system
  that is asymptotically flat, the components of the curvature tensor
  fall off as one moves farther away from the source of the emitted
  gravitational radiation.  At sufficiently large distances, only Type
  N radiation is noticeable; the limiting Type N radiation can be
  identified as the gravitational-wave (GW) content of the spacetime
  (typically denoted as $\Psi_4$ on an affinely parameterized 
  out-going geodesic null tetrad). 
  [Note that gravitational radiation is only rigorously
  defined at future null infinity (denoted $\mathscr{I}^+$).] 
  A caricature of this behavior is given in Fig.~\ref{fig:peeling}. 

Here we review the usual derivation of the peeling property~\cite{Sachs1961,Sachs1962,Penrose1963,Penrose1965,Penrose1986V2},
 commenting on some of the properties of the QKT within this context; 
an alternate derivation of the the peeling property using spinor notation can be found in~\cite{Penrose1965}. 
The basic idea of the usual derivation is to introduce a new `unphysical' metric $d\acute{s}$ that is conformally related to the physical metric $ds$ by $ d\acute{s} = \Omega \ ds$. 
The  metric $d\acute{s}$ is finite and well defined where the physical metric blows up (points on $\mathscr{I}^+$ are infinitely distant from their neighbors \cite{Penrose1986V2}) and allows us to explore the properties of the spacetime at $\mathscr{I}^+$ or at conformal null infinity, where $\Omega \rightarrow 0$. 
All quantities associated with the conformal metric $d \acute{s}$ will be denoted with an acute (e.g. $\acute{ds}$).

The relationship between metric tensors can be expressed as
\begin{align}
\acute{g}_{ab} &= \Omega^2 g_{ab},& \acute{g}^{ab} &= \Omega^{-2} g^{ab}, \label{ConfMet}
\end{align}  
and the topology at $\mathscr{I}^+$ is $S^2 \times \mathbb{R}$. Now let $l^a$ be tangent to an affinely parameterized out-going null geodesic on the real spacetime, with an affine parameter $\tau$ such that $l^a \nabla_a \tau = 1$. Then let $\acute{l}^a$ be tangent to an affinely parameterized geodesic in the conformally related spacetime with affine parameter $\acute{\tau}$.
Note that if we take $l^a = \Omega^{2} \acute{l}^a$, then the geodesic equation in physical spacetime implies its counterpart in the conformal spacetime 
\cite{Penrose1986V2}; furthermore, if we choose $n^a = \acute{n}^a$ at $\mathscr{I}^+$, then we have that the direction of  $(n^a = \acute{n}^a)|_{\mathscr{I}^+}$  does not depend on the geodesic and is tangent to $\mathscr{I}^+$ \cite{Penrose1986V2}.

Substituting these choices into the expressions for the metric [Eq.~\eqref{metexp}] and subsequently into Eq.~\eqref{ConfMet} we have that at $\mathscr{I}^+$ the conformal tetrad relates to the physical tetrad by means of the expressions
\begin{align}
 \label{eq:TetradConform}
\quad l^a = \Omega^{2} \acute{l}^a, \quad m^a = \Omega \acute{m}^a, \quad n^a =
\acute{n}^a.
\end{align}
Departing from $\mathscr{I}^+$ by moving into the manifold, differences in parallel transport in the physical and conformal manifolds lead to 
higher order terms in the $\bm{m}$ and $\bm{n}$ equations (see Eqs.~(9.7.30) and (9.7.31) in Ref.~\cite{Penrose1986V2}). 
By comparing the affine parameter on the
two manifolds along a geodesic 
and imposing Einstein's vacuum field equations, we can show that in general 
$ d \tau = \Omega^{-2} d \acute{ \tau}$
and that for large affine parameter $\tau$ or small conformal affine parameter $\acute{\tau}$ we have \cite{Penrose1986V2}
\begin{align}
\acute{\tau} &= -A^{-2} \tau^{-1} + \sum_{n=2}D_n \tau^{-n}, \\ 
\tau &= -A^{-2} \acute{\tau}^{-1} + \sum_{n=0}C_n \acute{\tau}^{n} , \\
\Omega &= A^{-1} \tau^{-1}+ \sum_{n=2} E_n \tau^{-n} , \label{eq:RelationTauOmega}   \\
\Omega &= -A \acute{\tau}-\sum_{n=3}A_n \acute{\tau}^n, \label{OmegaSeries}
\end{align}
where $A_n$, $C_n$, $D_n$, $E_n$ are constants and $A = -\frac{d \Omega}{d \acute{\tau}}|_{\Omega \rightarrow 0}$  is a non-zero constant.
Any quantity $\acute{\theta}_{\cdots}$ that is $C^h$ continuous at $\mathscr{I}^+$ can be expressed in terms of a series expansion about $\mathscr{I}^+$ as follows  
\begin{align}
\acute{\theta}_{\cdots} &= \sum_{n=0}^h   \acute{\tau}^n \acute{\theta}_{\cdots}^{(n)} + o(\acute{\tau}^h) \notag\\
&= \sum_{n=0}^h   \tau^{-n}
\theta_{\cdots}^{(n)} + o(\tau^{-h}).
 \label{expansioninf}
\end{align}

Since the Weyl tensor is conformally invariant, $C^a_{\ bcd} = \acute{C}^a_{\ bcd}$, or 
\begin{align}
C_{abcd} = \Omega^{-2} \acute{C}_{abcd}, \label{eq:Confweight}
\end{align}
all the relevant quantities  can be computed on the conformal manifold where the metric is finite and well behaved, and then interpreted on the physical manifold where the metric quantities may have diverged.
At $\mathscr{I}^+$ in an asymptotically flat spacetime, the Weyl tensor $\acute{C}_{a bcd}$  vanishes and the dynamics of the gravitational field as one approaches $\mathscr{I}^+$ can be described using a tensor $ \acute{K}_{abcd}$, where 
\begin{align}
\acute{C}_{abcd}=\Omega \acute{K}_{abcd} \label{eqCK}
\end{align}
and the components of $\acute{K}$ expressed on the tetrad basis $\{\bm{\acute{l}},\ \bm{\acute{n}},\ \bm{\acute{m}},\ \bm{\acute{\overline{m}}}\}$ admit expansions in the form of Eq.~$\eqref{expansioninf}$.

The peeling-off property of the Weyl scalars naturally arises when one expresses the quantities related to $\bm{\acute{K}}$  in terms of the physical metric and the tetrad basis $\{\bm{l},\ \bm{n},\ \bm{m},\ \bm{\overline{m}}\}$. 
 Let us take a detailed look at $\Psi_4$: analogous to the definition of $\Psi_4$ in Eq \eqref{eq:Psi4}, let 
\begin{align}
\acute{\Psi}_4 &= - \acute{K}_{abcd} \acute{n}^a \acute{\overline{m}}^b \acute{n}^c \acute{\overline{m}}^d 
\end{align}
The fact that $\bm{\acute{K}}$ is regular as we approach $\mathscr{I}^+$ implies that $\acute{\Psi}_4$ admits a series expansion of the form
\begin{align} \label{eq:ExpConformalPsi}
\acute{\Psi}_4= \sum_{n=0} \tau^{-n} \Psi_4^{(n)},
\end{align}
where in particular $\Psi_4^{(0)} = \acute{\Psi}_{4}|_{\mathscr{I}^+}$. 
Similar expansions can be found for  $\acute{\Psi}_i, i=0,1,2,3$. 
At $\mathscr{I}^+$, the physical $\Psi_4$ [defined by Eq.~\eqref{eq:Psi4}] is related to $\acute{\Psi}_4$ by
\begin{align}
\Psi_4 &= -(\Omega ^{-2} \acute{C}_{abcd}) (\acute{n}^a) ( \Omega \acute{\overline{m}}^b) (\acute{n}^c )(\Omega\acute{\overline{m}}^d) 
= \Omega \acute{\Psi}_4,
\end{align}
where we have used Eqs.~\eqref{eq:TetradConform}, \eqref{eq:Confweight} and \eqref{eqCK}. By a similar argument as used for $\Psi_4$, the differing powers of $\Omega$ appearing in Eq.~\eqref{eq:TetradConform} result in a hierarchy being set up where 
\begin{align} \label{eq:PeelingPowerLaws}
\Psi_i &= \Omega^{5-i} \acute{\Psi_i}.  
\end{align}
This expression is merely a product of the series in Eq.~\eqref{eq:RelationTauOmega} and \eqref{expansioninf}. 
Resumming the product of series implies that the physical Weyl scalars along an affinely parameterized out-going null geodesic can  be expressed as 
\begin{align}
\Psi_i = \tau^{i-5}\sum_{n=0} \tau^{-n}\psi_i^{(n)} \label{eq:powerSeries}
\end{align}
where $\psi_i^{(n)}$ are constant along the geodesic. 

\subsubsection{Peeling in principal null directions \label{sec:PeelingInPND}}    
\begin{figure*}[tbp]
  \includegraphics[width=.36\textwidth]{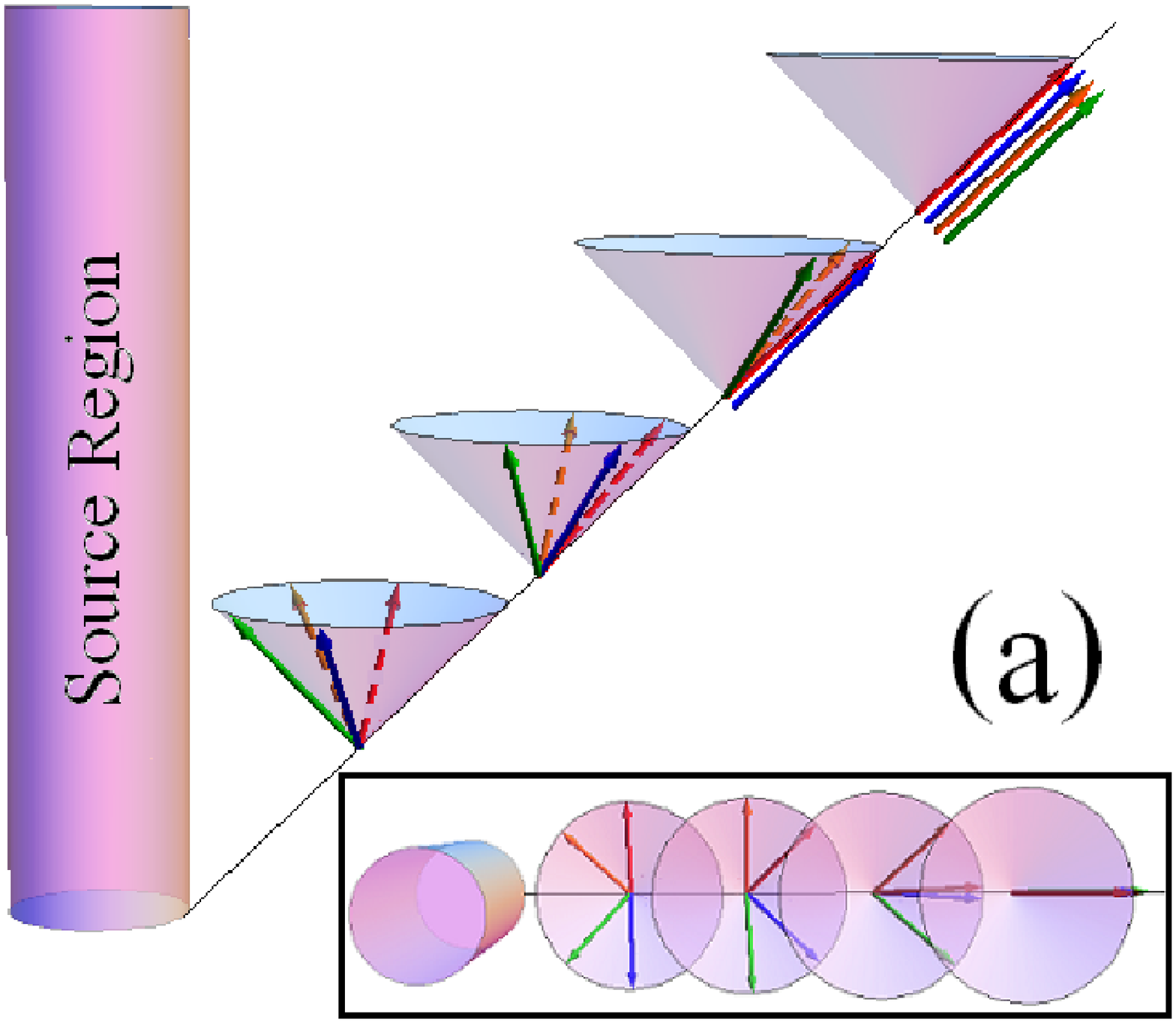}
  \includegraphics[width=0.3\textwidth]{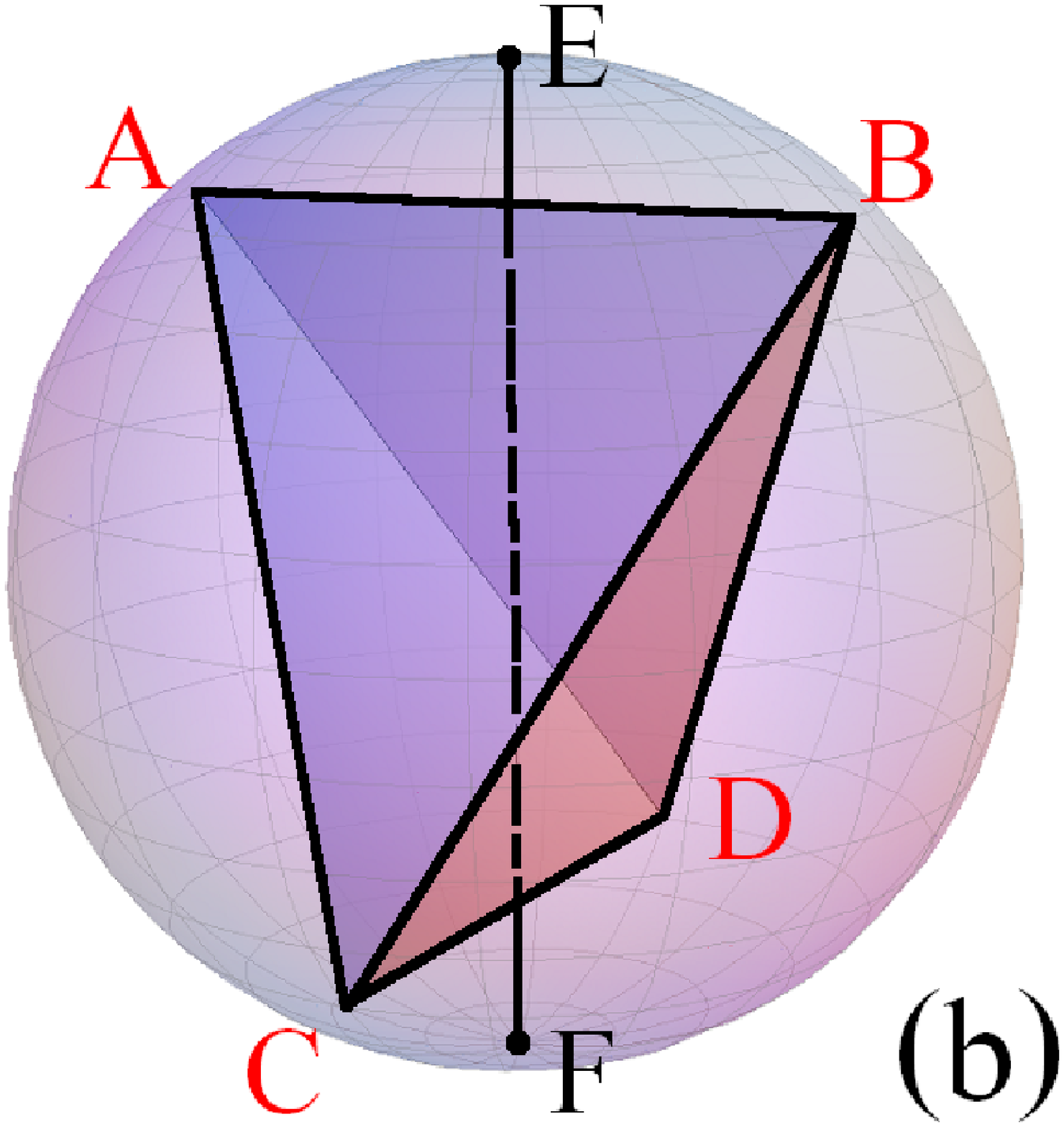}
  \includegraphics[width=.32\textwidth]{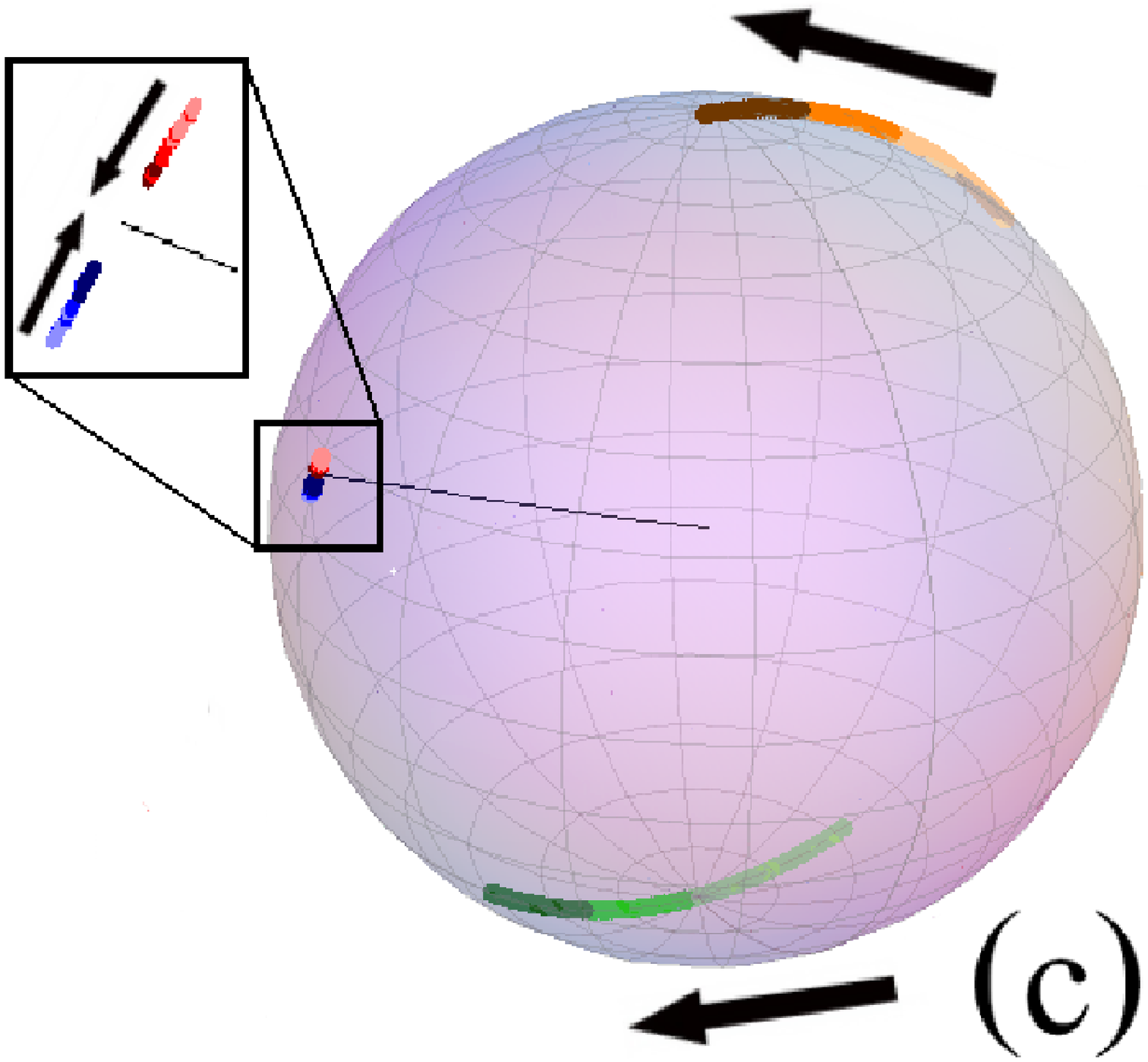}
  \caption{(a) A pictorial representation of the peeling property as bunching
of principal null directions (PNDs)
\cite{Penrose1986V2}, with the inset showing a top-down view.  
(b) The relationship between the PNDs and the quasi-Kinnersley tetrad (QKT).
Points $A,B,C,D$
correspond to PNDs on the 
anti-celestial sphere, and are arranged as the vertexes
of a tetrahedron.  The anti-celestial sphere  can be thought of  as a spatial slice of the future null cone, where each point on the sphere represents a null direction. The line EF linking the mid-points of a pair of opposite edges strike
the anti-celestial sphere at opposite poles which correspond to the null direction $\bm{\hat{l}}$ associated with the QKT
at point (F) and to the null direction $\bm{\hat{n}}$ 
associated with the QKT at point (E) [cf. Fig.~8-5 in Ref.~\cite{Penrose1986V2}].
(c) The four principal null directions recorded during a head-on numerical simulation [described in Sec.~\ref{sec:HeadOn}] are represented
as points (in four different colors) on the anti-celestial sphere.
We begin integrating a null geodesic in the
 $\bm{\hat{l}}$ direction
 and then compute the PNDs at discrete intervals along that geodesic. Darker
colored points correspond to values farther along the geodesic (farther removed from source region). For cleaner visualization, the angular coordinates on the anti-celestial sphere in this figure are simply those of the simulation coordinates and not the abstract ones in Eq.~\eqref{BPOWER}. 
We nevertheless see that the PNDs are distributed in a pairwise symmetric manner
relative to tangent $\bm{\ell}$ of the geodesic (denoted by the black radial line).  
Two of the PNDs stay close to $\bm{\ell}$, whose close-ups are shown in the framed inset.
The other two demonstrate a clear motion toward $\bm{\ell}$, where arrows indicate progress along the null geodesic. The numerical findings are thus consistent with
the bunching behavior depicted in panel (a).
} 
  \label{fig:peeling}
\end{figure*}

Note that the peeling property is not a function of which geodesic is chosen (provided that the geodesic strikes $\mathscr{I}^+$ and is affinely parameterized); on the contrary, it is a feature of the spacetime curvature and the distribution of principal
null directions (PNDs) as one approaches $\mathscr{I}^+$. This feature is illustrated graphically in  Fig.~\ref{fig:peeling} (a): as one moves in toward the source from   $\mathscr{I}^+$ along a null geodesic, the PNDs ``peel off'' away from the geodesic direction \cite{Penrose1965}.

Let us now quantify this behavior more precisely. Starting from the $\bm{l}$ vector associated 
with the out-going null geodesic, perform a Type II Lorentz transformation, so  
from Eqs.~\eqref{eq:Type2} and \eqref{eq:TypeIIOnNPScalar} we have that the four principle null direction (PNDs) can be expressed as:
\begin{align} 
\bm{k}= \bm{l} + \overline{b} \bm{m} + b \bm{\overline{m}} + b\overline{b} \bm{n}, \quad
\label{Knull}\end{align}
 where $b$ takes on the values of the four roots of the complex equation
\begin{align}
 \Psi_0 + 4b\Psi_1 + 6 b^2\Psi_2 + 4 b^3\Psi_3 + b^4 \Psi_4=0. \label{rooteq}
\end{align}
From Eq \eqref{Knull} it becomes apparent that the magnitude of $b$ determines the extent to which the PNDs depart from the null vector $\bm{l}$ since $k^a l_a =-b\overline{b}$.
 By making the   identification  proposed in~\cite{Nerozzi2005} between a pair of spherical coordinates $(\theta, \ \phi)$ and the boost $b$,
\begin{align}
b_{(i)} = \cot\left(\frac{\theta_i}{2} \right)e^{i\phi_i}, \ \quad i \in \{1,2,3,4\}, \label{BPOWER}
\end{align}
we can graphically demonstrate the motion of the PNDs by plotting the four roots on the anti-celestial sphere as shown in Fig. \ref{fig:peeling} (b). (The anti-celestial sphere can be thought of as the space of all possible directions associated with out-going null rays.)  If $\theta_{i} = \pi$, then the magnitude of the boost $b_{(i)}$ vanishes and $\bm{k}=\bm{l}$ is a PND; on the other hand, if $\theta_{i} = 0$ then $\bm{k}\propto \bm{n}$. 

Asymptotically, where the Weyl scalars admit power series expansions such as Eq.~\eqref{eq:powerSeries}, we can obtain the dominant behavior of $b$ by setting 
\begin{equation} \label{eq:ExpansionCeles}
b=\sum_{n=0} \tau^{-n}b^{(n)}
\end{equation}
and substituting this expression into Eq.~(\ref{rooteq}).
We then have that
$b^{(0)} = 0$ and $b^{(1)}$ can be found by finding the four roots of the equation 
\begin{align}
& \psi_0^{(0)} + 4b^{(1)}\psi_1^{(0)}  + 6 \left(b^{(1)}\right)^2\psi_2^{(0)}  \notag\\
& \quad \quad \quad \quad + 4 \left(b^{(1)}\right)^3\psi_3^{(0)} + \left(b^{(1)}\right) ^4 \psi_4^{(0)} =0. \label{rooteq3}
\end{align}
Further higher order terms become more complicated and involve mixtures of higher order terms in the 
expansions of the Weyl tensor components. 

The leading order coefficients $\psi_i^{(0)}$ in Eq.~\eqref{eq:powerSeries} are independent of the choice of geodesic path, while  
higher order terms $\psi_i^{(n)}$ with $n>0$ are path or geodesic-dependent, which implies in turn that the $b^{(n+1)}$ are geodesic-dependent.
This path dependence suggests the possible existence of an optimal null trajectory  along which the series converges most rapidly and from which the GW content can be most effectively extracted. 
One approach to finding the optimal trajectory is to minimize the higher order terms, $\psi_i^{(n)}$ ($n>0$), achieving a rapidly converging series. Possibly the most rigorous method of ensuring rapid convergence would be to identify the Kinnersley tetrad and thus the wave propagation direction at $\mathscr{I}^{+}$ and then to integrate backward in time, but
such a strategy cannot be executed real time during a numerical simulation. 
Instead, the method advocated here  is to align the initial geodesic direction with the wave propagation direction in the computational domain and then to integrate forward in time. This direction  can be identified in a slicing independent way by $\bm{\hat{l}}$ in the QKT as was shown in Sec.~\ref{sec:Direction}. In Sec.~\ref{sec:BinaryBoth}, we will demonstrate numerically the rapid convergence rate that results from this approach.

Choosing the QKT $\bm{\hat{l}}$ as the initial direction is further justified by considering the manner with which PNDs converge onto the outgoing geodesic's tangent direction.
In the QKT $\hat{\Psi}_1= 0 =\hat{\Psi}_3$, which greatly reduces the complexity of Eq.~\eqref{rooteq}. The transformation from $\bm{\hat{l}}$ to PND takes the simplified form 
\begin{align} \label{eq:PairSymmetry}
\hat{b}^2 = \frac{1}{\hat{\Psi}_4} \left( -3 \hat{\Psi}_2 \pm \sqrt{ 9 \hat{\Psi}_2^2- \hat{\Psi}_4 \hat{\Psi}_0    }\right).
\end{align}
The four roots now occur in pairs and can be parameterized using only two angles. 
\begin{equation} \label{eq:PairWiseDist}
\begin{split}
\hat{b}_{(1)} =& \cot\left(\frac{\hat{\theta}_1}{2}\right) e^{i\hat{\phi}_1}, \quad \quad \hat{b}_{(2)}
= \cot\left(\frac{\hat{\theta}_1}{2}\right) e^{i\hat{\phi}_1+i\pi} \\
\hat{b}_{(3)} =& \cot\left(\frac{\hat{\theta}_2}{2}\right) e^{i\hat{\phi}_2}, \quad \quad \hat{b}_{(4)}
= \cot\left(\frac{\hat{\theta}_2}{2}\right) e^{i\hat{\phi}_2+i\pi} 
\end{split}
\end{equation}
The out-going null direction $\bm{\hat{l}}$ of the QKT thus finds itself in the center of the four PNDs due to the added symmetry imposed by the QKT. This situation is depicted graphically in Fig. \ref{fig:peeling} (b). 
By initially selecting a QKT direction in the interior of the computational
domain from which to shoot the geodesics to infinity, we impose an additional symmetry on the manner in which the PNDs approach the geodesic's tangent initially, hoping that this additional symmetry is maintained as the geodesic
approaches $\mathscr{I}^+$ to ensure the clean pairwise convergence of the PNDs to the geodesic's tangent.

Once the geodesic is shot off in the $\bm{\hat{l}}$ direction, there is nothing  to ensure that it remains in the QK out-going null direction. In practice, however, the QK property appears to be maintained to a high degree of accuracy, as is indicated by the symmetric pairwise convergence of the PNDs onto the null geodesic shown in  Fig.~\ref{fig:peeling} (c). For this plot the angle between the QKT direction of $\bm{\hat{l}}$ and the tangent $\bm{\ell}$ to the geodesic remains less than   $4.2 \times 10^{-4} \pi$.

\subsubsection{Peeling of QKT quantities \label{sec:PeelingQKT}}
We close this section on the peeling property by revisiting the geometrically motivated coordinate system (introduced in Sec.~\ref{sec:COORDS}) in the asymptotic region. The curvature invariants $I$ and $J$ (and thus $\hat{\Psi}_2$) can be constructed using the series expressions Eq.~\eqref{eq:powerSeries}. The dominant behavior of the curvature invariants  are 
\bea \label{eq:PeelingInvariants}
I\sim \tau ^{-6}I^{(0)}, \quad J\sim\tau^{-9}J^{(0)}, \quad \hat{\Psi}_2 \sim \tau^{-3} \hat{\psi}_2^{(0)}
\eea
[see Eq.~\eqref{eq:IJ}] where the quantities with a superscript~$\cdot^{(0)}$ are constant along the geodesic. 
Assigning the radial coordinate using Eq \eqref{eq:GeoCoordsDef} sets 
\bea \label{eq:AffineParamVsR}
\hat{r} \sim \tau \Re\left[\left(\hat{M}/\hat{\psi}_2^{(0)}\right)^{1/3} \right]. 
\eea
The peeling property states that the PNDs converge onto the out-going geodesic direction $\bm{\ell}$. Since each pair of PNDs are equidistant from the QKT $\bm{\hat{l}}$, this implies that $\bm{\hat{l}}$ approaches the $\bm{\ell}$ direction. The asymptotic relationship between $\hat{r}$ and $\tau$ given in Eq.~\eqref{eq:AffineParamVsR}, together with the condition Eq.~\eqref{eq:FixingLa} that we use to fix the boost freedom of the QKF, implies that $\bm{\hat{l}}$ not only asymptotes to the \emph{direction} of $\bm{\ell}$, it is also  affinely parameterized in this limit.
The geometrically constructed $\hat{r}$ asymptotically denotes the spherical wavefronts of light-rays approaching $\mathscr{I}^+$.

Lastly, we underscore the fact that using the QKT has the advantage of  identifying a unique affine parameterization of the geodesic as it approaches $\mathscr{I}^+$.
The prescription given in Eq.~\eqref{eq:FixingLa} for 
fixing the boost freedom of the QKT has used the geometry 
of the spacetime implicit in the Coulomb potential to fix 
the parameterization of $\bm{\hat{l}}$ 
in a global manner, removing the freedom to choose a different affine parameter through the transformation $\tau\rightarrow A\tau$.  
These ideas will be revisited in greater detail when we look at extrapolation 
in the context of the numerical simulations  in Sec.~\ref{sec:Extrapolation}.

\section{Numerical implementation \label{sec:Cons}}
In this section, we detail the numerical implementation of the analytic ideas mentioned in the previous sections using the Spectral Einstein Code (SpEC). 
A description of SpEC and the methods it uses are 
given in Ref.~\cite{SpECwebsite} and the references therein.

\subsection{Constructing the QKT \label{sec:QTKcons}}
We construct the QKT in a numerical simulation by first constructing an orthonormal tetrad adapted to
 the simulation's coordinate choice and then the orthonormal tetrad's null counterpart $\{\bm{l},\ \bm{n},\ \bm{m},\ \bm{\overline{m}} \}$ and the associated NP scalars $\Psi_i$. In order to find a QKF $\{\bm{\tilde{l}},\ \bm{\tilde{n}},\ \bm{\tilde{m}},\ \bm{\overline{\tilde{m}}} \}$, the construction described in Sec.~\ref{sec:QKF} can be used; alternatively, the appropriate Type I and Type II transformations [Eqs.~\eqref{eq:Type1} and \eqref{eq:Type2}] to the QKF can be found. Finally, we construct the geometrically motivated coordinate system $(\hat{r},\ \hat{\theta})$ described in Sec.~\ref{sec:COORDS}, and we use these coordinates to fix the remaining Type III tetrad freedom to obtain the QKT.
 
\subsubsection{Implementing a coordinate tetrad \label{sec:coordTetrad}} 
Specifically, we begin our construction by noting that the SpEC code stores the spacetime metric $g_{ab}$ on a Cartesian coordinate basis $\{x^a\}=\{t,\ x,\ y,\ z\}$. (Note that henceforth the index $0$ refers to the time coordinate.) 
We can also define a set of related spherical coordinates $\{t,\ r,\ \theta,\ \phi\}$ by using the standard definitions
\begin{equation}
x = r\sin \theta \cos\phi, \quad y = r \sin \theta \sin \phi, \quad z = r \cos\theta. \label{eq:sphericalcoords}
\end{equation}
We further define the time-like unit normal to the spatial slicing and radially outward-pointing vector as
\begin{equation} \label{eq:CoordTetrad1}
T^{a} = \frac{\delta^{a}_0 - \beta^a}{\alpha}, \quad
N^{a} = \frac{r^a}{\sqrt{r^b r_b}}, 
\end{equation} 
respectively, where $\alpha$ is the lapse and $\beta^a$ is the shift, and $\bm{r}$ is the spatial location vector. 
Inserting these orthonormal vectors into 
Eq.~\eqref{eq:OrthonormalVsNullTetrad} yields 
$\mathbf{l}$ and $\mathbf{n}$, two legs of the null tetrad tied to the simulation's coordinates.

We next construct the remaining two tetrad legs $\{\mathbf{m},\ \mathbf{\overline{m}}\}$, ensuring that the normalization conditions of Eq.~\eqref{eq:NullNormalization} are satisfied. In other words, we seek to construct the null vector $\bm{m} = 1/\sqrt{2}(\bm{E^2}+ i\bm{E^3}) $ where $\bm{E^2}$ and $\bm{E^3}$ are orthogonal to $\bm{T}$ and $\bm{N}$ and to each other and obey the normalization condition
\bea
\|\bm{E^2}\|^2=\|\bm{E^3}\|^2&=&1. \quad
\eea
Our construction begins by computing the vectors
\bea
\bm{K} = \frac{1}{r\sin\theta}\frac{\partial}{\partial \phi} ,\quad
\bm{F} = \frac{1}{r} \frac{\partial}{\partial \theta}, \quad 
\eea
where $\theta$, $\phi$ are spherical coordinates defined in Eq.~\eqref{eq:sphericalcoords}.  Then, we ensure orthogonality by means of the Grams-Schmidt-like construction 
\begin{eqnarray}
(\hat{F})^{a} &=& F^{a} + F^{b}l_{b} n^{a} + F^{b} n_{b} l^{a} \label{eq:mImTrans}, 
\end{eqnarray}
rescaling appropriately to obtain the correct normalization as follows:
\begin{align} \label{eq:CoordTetrad2}
(E^2)^a =\frac{ \hat{F}^{a}  }{\sqrt{ \hat{F}^{a} \hat{F}_{a} } }. 
\end{align}
Similarly, for the final tetrad leg, we construct the orthogonal vector 
\begin{align}
(\hat{K})^{a} &=& K^{a} + K^{b}l_{b} n^{a} + K^{b} n_{b} l^{a} \label{eq:mReTrans}
            - K^{b} E^2_b(E^2)^{a},
\end{align}
normalizing it as follows:
\begin{align} \label{eq:CoordTetrad3}
(E^3)^a =\frac{ \hat{K}^{a}  }{\sqrt{  \hat{K}^{a} \hat{K}_{a} } } 
\end{align} 

\subsubsection{Obtaining a tetrad in the QKF}

Given the orthonormal coordinate tetrad $\{T^a,\ N^a,\ (E^2)^a,\ (E^3)^a\}$, we next construct  
a tetrad $\{\bm{\tilde{l}},\ \bm{\tilde{n}},\ \bm{\tilde{m}},\ \bm{\overline{\tilde{m}}} \}$ 
in the QKF by using the results of Sec.~\ref{sec:QKF}, in particular Eqs.~\eqref{eq:EigenQ}, \eqref{eq:SpatialEigenVec},  
\eqref{eq:CompEigenProj} and \eqref{eq:SpatialProjTetrad}. We can alternatively construct a QKF tetrad 
by explicitly rotating our initial coordinate tetrad into a transverse one via Type I and II transformations [Eqs.~\eqref{eq:Type1} and \eqref{eq:Type2}].
We have implemented both constructions numerically and verified that they agree; in the remainder of this subsubsection, we discuss details of each  implementation in turn. 

The hyper-surface approach of~Sec.~\ref{sec:QKF} requires us to solve the complex eigenvector problem in Eq.~\eqref{eq:EigenQ}, with $\bm{Q}$ calculated either from Eq.~\eqref{eq:SpatialEB} or from Eq.~\eqref{eq:PsiMatrix}.
Using Eq.~\eqref{eq:PsiMatrix}, the eigenvector problem can be solved analytically. After computing the desired eigenvalue $\lambda = -2\hat{\Psi}_2$, which is the root of Eq \eqref{eq:lleq}
that  admits the  expansion \eqref{eq:P2Expand} (in practice, it suffices to select the eigenvalue with the largest norm as suggested by Beetle et al.~\cite{Beetle2005}), the corresponding un-normalized eigenvector $\bm{\tilde{\Sigma}}$ 
of matrix (\ref{eq:PsiMatrix}) is 
\begin{equation}
\begin{split}
\tilde{\Sigma}_{E^2} =& 2\lambda^2-\lambda \left(\Psi_0+\Psi_4-2\Psi_2\right) \\
& +2\left[ \left( \Psi_1 + \Psi_3\right)^2 - \Psi_2\left( \Psi_0 + \Psi_4 + 2\Psi_2 \right) \right], \\
\tilde{\Sigma}_{E^3} =& i \left[ \lambda\left( \Psi_0 - \Psi_4 \right) 
+ 2 \left(\Psi_3^2 - \Psi_1^2 + \Psi_2\left( \Psi_0-\Psi_4 \right) \right) \right], \\
\tilde{\Sigma}_{N} =& 2\left[ \left( \Psi_0 + \Psi_2 \right) \Psi_3 + \lambda\left( \Psi_1- \Psi_3 \right) 
- \Psi_1\left( \Psi_2 + \Psi_4 \right)\right]. 
\end{split}
\end{equation}
where the $\Psi_i$ values are those extracted on the coordinate tetrad. 
(Note that this formula fails when $\Psi_1 = 0 = \Psi_3$, but in this case the coordinate tetrad is already in the QKF.)
To normalize $\bm{\tilde{\Sigma}}$ into $\bm{\tilde{\sigma}}$ that satisfies Eq.~\eqref{eq:NormalizeEigen},
we multiply it with a suitable complex number, namely 
\bea
\bm{\tilde{\sigma}} = \left[ -\frac{1}{\sqrt{2}}\left(\frac{|\alpha|}{\alpha} \right)\frac{\sqrt{\beta+\gamma}}{\gamma} + \frac{\sqrt{2} |\alpha|}{\gamma \sqrt{\beta+\gamma}} i \right] \bm{\tilde{\Sigma}}
\eea
where
\bea
\tilde{\Sigma}^a &=& X^a + i Y^a, \quad \alpha = X^a  Y_a, \\
\beta &=& ||\bm{X}||^2 - ||\bm{Y}||^2, \quad \gamma = \sqrt{\beta^2 + 4 \alpha^2}.
\eea

Alternatively, we can construct the QKF using the Type I and II transformations applied to the coordinate frame as follows.
Starting from a general Petrov Type I spacetime with five non-vanishing Weyl scalars, we perform a Type I rotation, introducing a parameter $\overline{a}$, followed by a Type II rotation that introduces a parameter $b$. These parameters can then be chosen to set $\Psi_1=\Psi_3=0$ by  solving the resulting system of two equations for the two parameters $\overline{a}$ and $b$.
Reference~\cite{Nerozzi2005} shows that the appropriate choice of parameters can be found by defining the intermediate quantities 
\begin{align}
H = \Psi_0 \Psi_2 - \Psi^{2}_1, &\quad G& = \Psi^{2}_0\Psi_3 -
3\Psi_0\Psi_1\Psi_2 + 2\Psi^{3}_1 
\end{align}
and then setting  
\begin{align} 
\Psi_1 + \Psi_0 \overline{a} = \frac{ G \pm \sqrt{ G^2 + (\Psi_0 \lambda - 2H)^2(H
+\Psi_0 \lambda) } }{ \Psi_0 \lambda - 2H } \label{eq:SolveForA} \\
 b = - \frac{ \Psi_3 + 3\overline{a} \Psi_2 + 3\overline{a}^{2}\Psi_1 + \overline{a}^{3}\Psi_0
}{\Psi_4 + 4\overline{a} \Psi_3 + 6\overline{a}^{2}\Psi_2 + 4\overline{a}^{3}\Psi_1 + \overline{a}^{4}\Psi_0 } \label{eq:SolveForB}  
 \end{align}
Note that this prescription becomes ill defined when $\Psi_0$ on the the initial tetrad approaches zero or when $ \Psi_0 \lambda - 2H  = 0$, making it difficult to find $\overline{a}$ by solving Eq.~\eqref{eq:SolveForA};
this problem is easily resolved by first applying a Type II transformation that takes the initial tetrad into one in which these pathologies do not arise.
Furthermore, we have two possible solutions for $\overline{a}$ resulting from the freedom to interchange the $\bm{\tilde{l}}$ and $\bm{\tilde{n}}$ legs associated with the transverse frame; the convention we use is to choose the root that gives $(\tilde{l}-\tilde{n})_a l^a > 0$, i.e. we choose $\bm{\tilde{l}}$ to be outgoing in the simulation
coordinates.  

\subsubsection{Obtaining the quasi-Kinnersley tetrad from the geometric coordinates\label{sec:Quasi}}
 
With a QKF in hand, we next seek to specialize to the particular QKT described in Sec.~\ref{sec:SpinBoostFixingByCoords}, where we use 
geometrically motivated coordinates $(\hat{r},\hat{\theta})$ given by Eq.~\eqref{eq:GeoCoordsDef} to fix the final Type III degrees of freedom.
In order to  fix these freedom using Eqs.~\eqref{eq:FixingLa} and \eqref{eq:FixingMa}, we must calculate the one-forms $d\hat{r}$ and $d\hat{\theta}$. We compute the spatial derivatives spectrally, and we compute the time derivatives using the  Bianchi identities in the $3+1$ form~\cite{Friedrich96}
\begin{equation} \label{eq:Bianchi31}
\begin{split}   
\partial_t \mathcal{E}_{ij} =& \mathcal{L}_{\beta} \mathcal{E}_{ij} 
                 + \alpha \left[D_k \mathcal{B}_{l(i}\epsilon_{j)}^{\phantom{i}kl} - 3
\mathcal{E}^k_{(i}K_{j)k}  \right. \\ 
                 & \left. + K^k_k \mathcal{E}_{ij} - \epsilon_i^{\phantom{i}kl}
\mathcal{E}_{km} K_{ln} \epsilon_j^{\phantom{i}mn}
                 + 2 a_k \mathcal{B}_{l(i}\epsilon_{j)}^{\phantom{i}kl}\right] \,, \\
\partial_t \mathcal{B}_{ij} =& \mathcal{L}_{\beta} \mathcal{B}_{ij} 
                +\alpha  \left[ -D_k \mathcal{E}_{l(i}\epsilon_{j)}^{\phantom{i}kl} - 3
\mathcal{B}^k_{(i}K_{j)k} \right. \\  
                & \left. + K^k_k \mathcal{B}_{ij} 
                - \epsilon_i^{\phantom{i}kl} \mathcal{B}_{km} K_{ln}
\epsilon_j^{\phantom{i}mn}
                - 2 a_k \mathcal{E}_{l(i}\epsilon_{j)}^{\phantom{i}kl} \right] \,,
\end{split}
\end{equation}
where $\mathcal{L}$ denotes Lie derivative, $D$ is induced 3-D covariant derivative operator, $\alpha$ denotes the lapse, $\bm{\beta}$ the shift, $\bm{K}$ the extrinsic curvature, and $a_k = \partial_k \text{ln}\alpha$. 
The time derivative of the metric $\partial_t g_{ij}$ is already known from the numerical evolution of the spacetime.
Using the above  equations and applying the chain rule, we compute the time derivatives of $\hat{r}$ and $\hat{\theta}$:
\bea
\left( \partial_t g_{ij}, \partial_t \mathcal{E}_{ij},\partial_t \mathcal{B}_{ij}
\right) \xrightarrow{\text{Eqns. }(\ref{eq:IJ}),
(\ref{eq:lleq}),(\ref{eq:GeoCoordsDef})} \left(\partial_t \hat{r},
\partial_t \hat{\theta} \right). \nonumber 
\eea
Equipped with all the components of $d\hat{r}$ and $d\hat{\theta}$, we can apply Eqs.~\eqref{eq:FixingL} 
and \eqref{eq:FixingM} to fix spin-boost degree of freedom, finally obtaining the QKT on which we
 can then extract Newman-Penrose scalars $\hat{\Psi}_{i}$ via Eqs.~(\ref{eq:Psi0}-\ref{eq:Psi4}).

We note that it may not always be possible 
to define the $\hat{r}$ and $\hat{\theta}$ coordinates using  Eq.~\eqref{eq:GeoCoordsDef} for spacetimes with additional symmetries. 
For example, in axisymmetric spacetimes 
admitting a twist-free azimuthal Killing vector, $\hat{\Psi}_2$ is real,
and as a result the $\hat{\theta}$ coordinate cannot be computed using Eq.~\eqref{eq:GeoCoordsDef}.
In fact, for Minkowski spacetimes, we cannot even define the $\hat{r}$ coordinate, because $\hat{\Psi}_2=0$. 
In such cases, the symmetries of the spacetime typically provide a set of preferred coordinates, which one would naturally adopt in a numerical simulation. In our QKT implementation, we presume that any such preferred coordinates are adopted, and we replace 
$(d\hat{r},d\hat{\theta})$ by their simulation-coordinate counterparts when degeneracies occur.

\subsection{Extrapolation \label{sec:Extrapolation}}
We now turn to extracting the asymptotic gravitational wave content at $\mathscr{I}^+$ by using the peeling property, i.e., to \emph{extrapolation}, which necessarily involves information from several spatial slices in the spacetime. Our procedure is to shoot a null geodesic affinely parametrized by $\tau$ toward $\mathscr{I}^+$, monitoring $\hat{\Psi}_4$ along the geodesic. The best possible polynomial in $1/\tau$ is fitted to the result. The existence of this polynomial follows from the peeling property, which is made explicit in Eq.~\eqref{eq:powerSeries}. We  identify the coefficient of the $1/\tau$ term or $\psi_4^{(0)}$ with the radiation content at $\mathscr{I}^+$.

In contrast to the usual method (extrapolating $\Psi_4$ as computed 
using a tetrad parallel-transported along an outgoing null geodesic), 
note that here we choose to extrapolate $\hat{\Psi}_4$ (defined using the QKT), which we expect to also display the correct peeling behavior [see Sec.~\ref{sec:PeelingQKT}]. In addition, the initial direction of the outgoing null geodesic is along $\bm{\hat{l}}$, so 
at the geodesic's starting point $\Psi_4=\hat{\Psi}_4$, and [Sec.~\ref{sec:Peeling}], 
at $\mathscr{I}^+$ also the outgoing null geodesic is along $\bm{\hat{l}}$ 
so that $\Psi_4=\hat{\Psi}_4$. In practice, 
as we integrate along these outgoing null 
geodesics, we monitor the difference between the null vector $\bm{\ell}$ tangent to 
the outgoing geodesic and $\bm{\hat{l}}$ from the QKT, and we find that this 
difference remains small (cf. Fig.~\ref{fig:peeling} and the surrounding 
discussion).
Therefore $\Psi_4$ and $\hat{\Psi}_4$ are not significantly different for the simulations we examined.  
When we extract the $\hat{\Psi}_4$ waveform, it converges 
rapidly to its asymptotic value with increasing extraction radius [Fig.~\ref{fig:SinglePointWaveFormConvergence}].

Selecting the initial tangent of the geodesics to be $\bm{\hat{l}}$ determines the parameterization
of these geodesics upto an additive constant $B$  corresponding to the freedom to shift the zero point of the
affine parameter, $\tau \rightarrow \tau +B$.  The asymptotic waveform is insensitive to the choice of the field $B$.  
Nevertheless, to provide an exact prescription we fix $B$ by recalling that in the Kerr limit,
the affine parameter is just the Boyer-Lindquist $\hat{r}$~\cite{ChandrasekharBook}. We thus
choose $B$ on the initial world tube (where we start shooting out null geodesics) to be such that
$\tau = \hat{r}$. 

\subsection{Sensitivity of QKT method to numerical error}
The numerical implementation of the QKT described in this section keeps the computation ``as local as possible'' in the following sense: the bulk of the calculation requires only local derivatives and knowledge of the metric and the extrinsic and intrinsic curvature of the spatial slice. 
However, this says nothing about the \emph{accuracy} of our method, which depends on how susceptible our method is to numerical noise.

To begin addressing this issue, we first recall exactly how many numerical derivatives are to be taken.
Equation~\eqref{eq:SpatialEB}, which is used to construct the gravitoelectromagnetic tensors $\bm{\mathcal{E}}$ and $\bm{\mathcal{B}}$, requires i) second spatial derivatives of the spatial metric in order to get  the intrinsic Ricci curvature, in addition to ii) the first spatial derivatives of the extrinsic curvature of the slice. Once the gravitoelectromagnetic tensor is obtained and the resulting curvature invariants $I$ and $J$ are computed, another derivative is required to compute the gradients of the coordinates $(\hat{r}, \hat{\theta})$ that then fix the Type III freedom of the tetrad.
Note that the first step, i.e. the computation of the gravitoelectromagnetic tensors only requires spatial derivatives, which we can compute spectrally (i.e., inexpensively and accurately, since we expect 
to observe exponential convergence in spatial derivatives with increasing spatial resolution). However, taking the gradient of the coordinates constructed out of the curvature invariants requires both spatial derivatives and a time derivative. Fortunately, this time derivative can be computed using the Bianchi identities as described in Sec.~\ref{sec:Quasi}, which again reduces the operation to spatial differentiation (although here the accuracy of the derivatives are also limited by the accuracy at which the constraint equations are satisfied).

What we find in practice is that the higher derivatives needed by our QKT method can at places 
 have a significantly higher amount of 
numerical noise than the numerical derivatives directly used in the actual evolution system.
This is a significant challenge to our method, since SpEC presently evolves the Einstein equations in first-order form, i.e., as a set of coupled partial differential equations containing only first derivatives in space and time.
Therefore, the evolution equations themselves will only guarantee the existence of one derivative
of the evolution variables (e.g., of the metric).  Constraints show convergence which 
means, among other things, that the auxiliary variables (defined during the reduction of 
second order differential equations to first order)
do converge to the appropriate metric derivative
quantities.  However, the evolution system, although quite capable at constraining the size of numerical error, 
does not necessarily force it to be smooth (differentiable to higher orders) 
at subdomain boundaries.

Consider the hypothetical example of adding white noise to a smooth analytical metric, such as the Kerr metric \eqref{eq:KerrMetric}. 
No matter how small the magnitude of the noise, it would prevent us from taking derivatives analytically. 
Numerically, under-resolving 
the high-frequency noise would smooth out the data and 
allow differentiations to proceed without significantly amplifying the added noise; 
therefore, we expect that filtering (the spectral
equivalent of finite-difference dissipation) would improve the smoothness of the numerical data 
and thus reduce difficulty in taking higher numerical derivatives. 
However, such filtering can effectively under resolve not only noise but also physical information. In other words, 
overly dissipative
schemes tend to be \emph{less} accurate; therefore, the current choice in SpEC is to dissipate as little as possible
while still maintaining robust numerical stability. This criterion is different from the use of filtering to damp out
on short time scales any high frequency modes that would be produced during an evolution. 

\begin{figure}[tbp]
  \begin{center}
    \includegraphics[width=0.99\columnwidth]{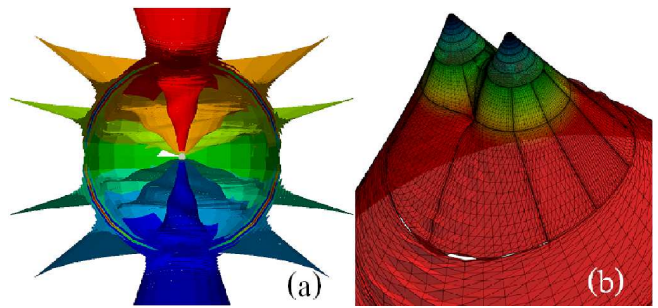}
  \end{center}
  \caption{(a) $\hat{\theta}$ contours displaying a pulse of high frequency junk radiation propagating outward. Ahead of the pulse, the geometric coordinate contours are consistent with
the Kerr-like initial data, but behind the pulse of junk radiation, the spacetime settles down
to an actual binary inspiral with a signature ``spiral-staircase'' pattern also
seen in Fig.~\ref{fig:KerrImplied2}(b). (b) The surface is a 2-D spatial slice containing the symmetry axis in a
head-on simulation, warped and colored according to $\hat{r}$ value. The sub-domain
boundaries are marked out with dense black lines, and appear to be a source of non-smooth noise.
These noisy features are reduced by increasing resolution. 
}
  \label{fig:NoiseDiagram}
\end{figure}

A better approach for reducing non-smooth numerical error is to go directly to their source. 
The lack of smoothness in the constraints observed in a typical SpEC evolution
is partly due to the penalty algorithm, which is known to produce convergent but non-smooth 
numerical errors at subdomain boundaries. [See Fig.~\ref{fig:NoiseDiagram}(b) for an illustration
of the penalty-algorithm induced non-smooth error.] 
Because this non-smoothness converges away with increased resolution, 
our method is observed to be viable given a sufficiently high numerical resolution; 
however, it remains to be seen whether ``sufficiently high'' means ``significantly higher'' 
than typical resolutions currently in use.
Alternatively, improvement to non-smooth numerical error could come through the use of newer
inter-patch boundary algorithms, such as Discontinuous Galerkin
methods \cite{Hesthaven2008}.
There also exists an ongoing effort to bring a (currently experimental) first-order-in-time, second-order-in-space version of SpEC \cite{Taylor:2010ki} 
into a state suitable for accurate gravitational-wave production, with the hope of added efficiency and of achieving numerical
error of higher differential order. Such possibilities as these, however, are future work, well outside the scope of this paper. 

Lastly, we consider the non-smooth noise sensitivity 
of our QKT quantities from another point of view: it can be used as a diagnostic of high-frequency, 
non-smooth numerical error. 
For instance, one source of non-smooth constraint violation in numerical simulations is the high-frequency, spurious ``junk'' radiation present at the beginning of numerical simulations (because of how the initial data are constructed), 
which poses a particularly difficult numerical problem.  The frequency of these modes
is of $O(M)$, orders of magnitude higher than that of the orbital motion (and the associated
gravitational waves).  This makes resolving the junk radiation a difficult task.   
In the effort to reduce junk radiation, the geometric coordinates can be 
used as a visualization tool.  Fig.~\ref{fig:NoiseDiagram}(a) is an illustration of how the $\hat{\theta}$ contours,
plotted as a function of code coordinates,  
react to the junk travelling through the grid, while adjusting themselves to reflect a more realistic spacetime.
By comparing the difference ahead and behind the easily identifiable 
junk pulse in Fig.~\ref{fig:NoiseDiagram}(a), one gets a 
glimpse of the missing pieces in the initial data.  

\section{Numerical Tests of the QKT scheme \label{sec:Numtests}}
We now consider several numerical tests used to gauge the
effectiveness of our proposed QKT scheme for waveform extraction. 
Most of these tests are
motivated by analytic solutions and are used to verify that our
choices of geometric coordinates and the QKT are yielding the
expected results. These tests broadly fall into two classes: i) 
non-radiative spacetime tests and ii) radiative spacetime tests. Each 
will be considered in turn in the following subsections.
 
\subsection{Non-radiative spacetimes \label{sec:Num}}
\subsubsection{Kerr black hole in translated coordinates \label{sec:ShiftedKerr}}
\begin{figure}[tbp]
  \includegraphics[width=0.95\columnwidth]{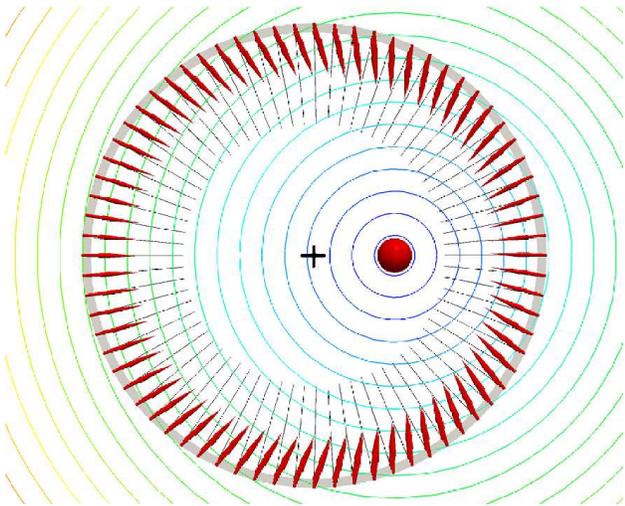}
  \caption{A Kerr-Schild black hole with $J/M^2=0.5$, with 
  the
    coordinates translated a distance $9M$ along $x$
    axis. The red sphere indicates location of the black hole's horizon,
    and the black cross indicates the
    coordinate origin. The colored circles are constant
    geometric radius $\hat{r}$ contours; these demonstrate the
    ability of our geometric coordinates to select an origin based on
    the Coulomb potential of the QKF, i.e., an origin which reflects
    the gravitational curvature of the spacetime. Also shown are the
    spatial projections of the $\bm{n}$ direction of the coordinate and
    quasi-Kinnersley tetrads, at points on a narrow strip marked by
    a grey ring. The black lines indicate the $\bm{n}$ direction
    associated with the coordinate tetrad, which point toward
    the coordinate origin, and the red lines/arrows are QKT $\bm{\hat{n}}$
    directions that identify the black hole as geometric origin, away from
    which the gravitational waves travel. 
}
 \label{fig:ShiftOrigin}
\end{figure}

\begin{figure}[tbp]
  \begin{center}
    \includegraphics[width=0.95\columnwidth]{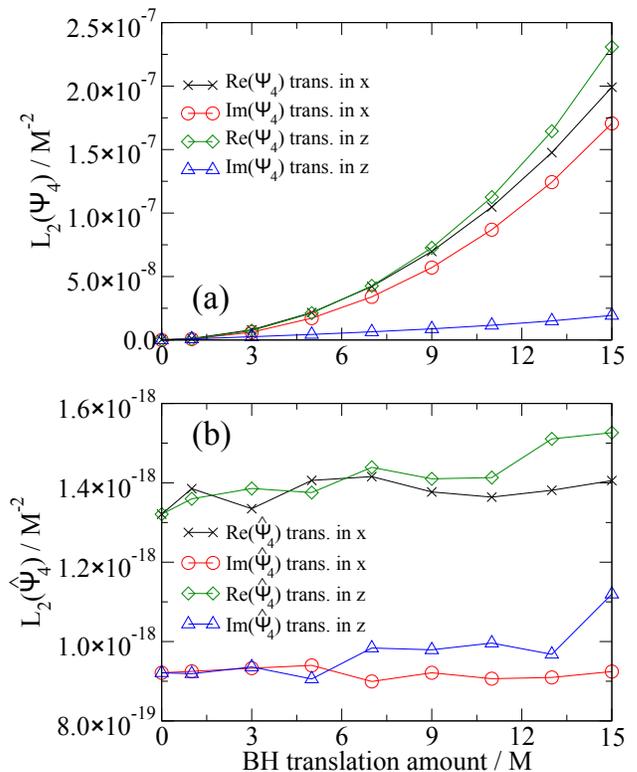}
  \end{center}
  \caption{(a): The $L_2$ norm [Eq.~(\ref{eq:L2NormDef})] of Newman-Penrose scalar $\Psi_4$
    computed (between radii $50$M and $140$M) for the 
    Kerr-Schild black hole in translated coordinates on the coordinate tetrad. 
    (b): The Newman-Penrose scalar $\hat{\Psi}_4$ computed in 
    the quasi-Kinnersley tetrad. Note the $L_2$ norm 
    is eleven orders of magnitudes smaller for
    the QKT when compared to that of the coordinate
    tetrad, showing that the QKT correctly adapts to the underlying curvature of the spacetime. 
}
  \label{fig:Psi4KerrShift}
\end{figure}

The spacetime in this test is a Kerr black hole in Kerr-Schild
coordinates, but the coordinate origin is translated away from the black
hole along $x$ or $z$ axis. Here we work in units of the black hole mass, 
and the dimensionless spin is $J/M^2 = 0.5$ 
pointing in the $z$ direction.  Tetrads determined only by our
simulation coordinates [see Eqs.~\eqref{eq:CoordTetrad1}-\eqref{eq:CoordTetrad3}] 
would not be aware of the translation, and the
spatial projection of $\bm{n}$ 
would point toward the coordinate center
instead of the black hole itself. In contrast, the QKT should
adjust to the displaced origin, picking up the true geometrical origin
of the gravitating system determined by the Coulomb potential of the
QKF. Figure \ref{fig:ShiftOrigin} shows the direction of spatial projection of $\bm{n}$ and
$\bm{\hat{n}}$ associated with the two tetrads. The QKT identifies the black hole at the
center of the circular shape, as do the geometrically motivated coordinate $\hat{r}$.

Figure \ref{fig:Psi4KerrShift} compares $\Psi_4$ extracted using the coordinate and
quasi-Kinnersley tetrads, respectively, using the so-called ``$L_2$ norm'' as a measure. 
The $L_2$ norm of a quantity $X$ is defined here as 
\begin{align}\label{eq:L2NormDef}
L_2(X)= \sqrt{\sum_{i=1}^{N_{tot}}\frac{X(x_i)^2}{N_{tot}}},
\end{align}
where $x_i$ are the spectral collocation points of a pseudo-spectral grid and $N_{tot}$ is the total
number of points. The present study uses four spherical shells between radii $50$M and $140$M with $N_{tot} \approx 4 \times 45^3$ collocation points. 
The QKT correctly produces vanishing $\hat{\Psi}_4$
(up to numerical round-off error), while the coordinate tetrad   
fails to identify the correct out-going direction and as a result misinterprets $\hat{\Psi}_2$ as gravitational radiation content in $\Psi_4$. 
(We observe similar behavior for $\Psi_0$.)
Using such a coordinate tetrad in a simulation with a displaced center
will result in spurious effects being picked up in the extracted radiation, of a magnitude not necessarily smaller than the physical gravitational wave content of the spacetime.

In a simulation of a dynamical spacetime, a similar effect should be expected when
the ``center of mass'' (e.g. in a Newtonian approximation) of the system does not coincide
with coordinate center. For example, consider  
a binary merger of unequal mass holes with the coordinate origin placed at the midpoint between the black holes;  
$\Psi_4$ extracted at finite radii would pick up a slowly varying offset at an integer multiple of the orbital frequency, 
and this contribution would complicate the extrapolated waveform.

\subsubsection{A Schwarzschild black hole with translated coordinates and a gauge wave \label{sec:Evil}}
We further explore the effects of coordinate choice or gauge by
introducing a time dependent gauge wave into a Schwarzschild solution
whose origin has been translated by a constant amount.
The resulting metric components now have
an explicit time dependence, and we expect the coordinate 
tetrad to produce a false gravitational wave signal, even though the Schwarzschild
spacetime is static and emits no physical radiation.

The exact analytic solution we use for this test is constructed from the Schwarzschild solution 
in ingoing Eddington-Finkelstein coordinates. We then 
apply a time-dependent coordinate transformation that yields a metric of the form 
\beq
\begin{split}
ds^2 =&-(1+C)^2\left(1-\frac{2M}{r}\right)dt^2 \\
       &+ 2(1+C)\left[\frac{2M}{r}-\left(1-\frac{2M}{r}\right)C\right]dtdr \\
       &+(1+C)\left[1+\frac{2M}{r}-\left(1-\frac{2M}{r}\right)C\right]dr^2 +
r^2d\Omega^2
\end{split}
\eeq
where $C(r,t)$ is the radial waveform of the introduced gauge wave. For our test we select generically chosen parameters  
\bea
C = 0.7\sin(0.03(t+r) + 3.1), \quad M = 1
\eea
Note that again we translate the black hole off the coordinate origin by a constant amount (here $r = 20$M) as described in the previous subsubsection. 

\begin{figure}[tbp]
  \begin{center}
    \includegraphics[width=0.99\columnwidth]{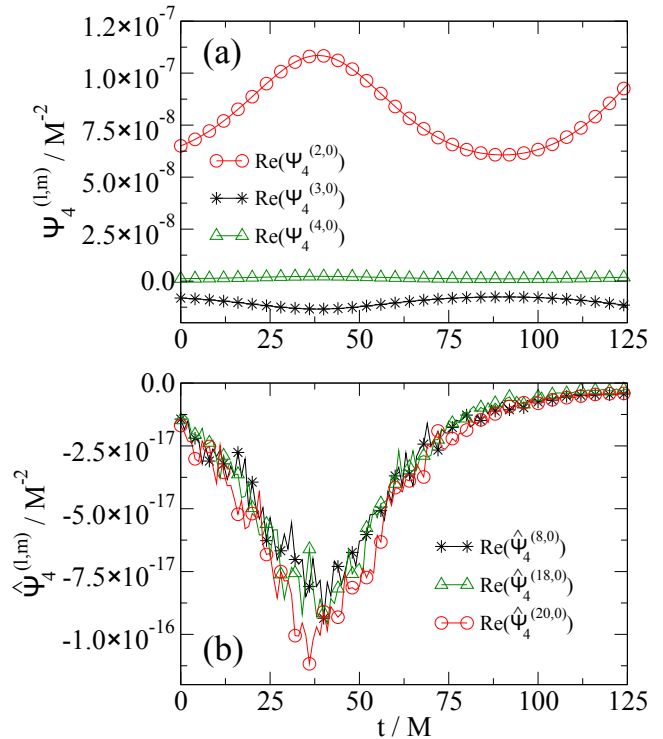}
  \end{center}
  \caption{
Spherical harmonic coefficients of Newman-Penrose $\Psi_4$ computed for the gauge wave solution, on a sphere of radius $r=120$M centered on the black hole. (a): $\Psi_4$ extracted on the coordinate tetrad.
(b): $\hat{\Psi}_4$ extracted on the QKT.
Only the three largest $(l,m)$ spin-weighted spherical harmonic modes (up through $l=35$) are shown. Note the scaling on the two figures differ by nine orders of magnitude.
}
  \label{fig:Psi4_3}
\end{figure}

Figure \ref{fig:Psi4_3} shows spin-weighted spherical harmonic expansion coefficients $\Psi_4^{(l,m)}$ of $\Psi_4$, computed using the 
 coordinate and quasi-Kinnersley tetrads. While only the three largest amplitudes are shown, we have computed all amplitudes up through $l=35$.
These scalars are computed on a sphere at a radius of $120$M from the
black hole, with the poles of harmonics aligned with the direction in which the black hole is shifted. 
As expected, the waveform extracted using the coordinate tetrad picks up a time dependence
associated with the gauge wave, while the QKT returns vanishing values, correctly identifying the static spacetime solution. 

In the generalized harmonic form of the Einstein field equations,
the gauge may be set by the covariant wave equations 
\bea
\Box x^a = H^a
\eea
where $H$ is either a specified or evolved source function~\cite{Lindblom2007, Pretorius2005a, Pretorius2006,
Lindblom2009c}. 
It is thus probable that gauge modes similar to the one considered in this example
may be present in fully dynamical
simulations. 
Consider a gauge wave that generates a deviation between the coordinate tetrad basis vectors
$\{\bm{l},\ \bm{n},\ \bm{m},\ \overline{\bm{m}}\}$ and their counterparts in the QKT. Such differences  
can be represented by a sequence of type II, I and then III transformations parameterized by the time 
dependent transformation parameters $b(t)$, $a(t)$ and $\mathcal{A}(t)$ that appear in Eqs.~\eqref{eq:Type2},
 \eqref{eq:Type1} and \eqref{eq:Type3} respectively. If we restrict ourselves to asymptotic regions 
where $\hat{\Psi}_4$ dominates over other NP scalars, then according to Eqs.~\eqref{eq:TypeIIOnNPScalar}, 
\eqref{eq:TypeIOnNPScalar} and \eqref{eq:TypeIIIOnNPScalar}, we have that to leading order in $a$ and $b$ 
the coordinate $\Psi_4$ is given by
\bea
\Psi_4 = \left(1 + 4\overline{a}(t)b(t)\right) \mathcal{A}^{-2}(t)\hat{\Psi}_4.
\eea
If the gauge wave falls off when we move away from the source region, 
then we may have 
\bea
\left(1 + 4\overline{a}(t)b(t)\right) \mathcal{A}^{-2}(t) \rightarrow 1
\eea 
and its effect can in principle be extrapolated away. However, for some cases, such as a plane gauge wave, the time dependent 
perturbation introduced into $\Psi_4$  could persist in the extrapolated waveform. 
Therefore, minimizing any such gauge-dependent content in $\Psi_4$ extracted at finite radii
is preferable to relying on extrapolation to remove them;
some pathological gauge modes might not fall off sufficiently quickly with radius.

\subsection{Radiative spacetimes \label{sec:Num2}}
Having observed that the QKT correctly reflects the curvature content of non-radiative spacetimes, including in the presence of a gauge wave, we next apply the QKT to spacetimes emitting gravitational radiation.
In this subsection, we verify that the scheme is consistent with analytic perturbation theory results. 

The QKT by construction reduces to the Kinnersley tetrad
in the Kerr limit. Therefore, if we perturb a Kerr black hole by a small amount, the
$\hat{\Psi}_4$ computed on the QKT should reproduce the analytic
perturbation theory results computed on the
Kinnersley tetrad associated with the unperturbed Kerr background.
Verifying this correspondence provides us
with the means to quantitatively test whether the QKT extracts the
correct waveform and that we have all normalization conventions implemented correctly.
The idea of ensuring the correspondence between the computed waveform
and the perturbation theory results is what motivated the authors of Ref.~\cite{Beetle2005}
to adopt transverse tetrads in the first place;   Chandrasekhar~\cite{ChandrasekharBook} also used
the transverse tetrad in his metric reconstruction program, where he explicitly computed the perturbed tetrad
and curvature perturbations on the tetrad, obtaining the expected correspondence.
For simplicity, here we perturb 
a Schwarzschild black hole with an odd-parity Regge-Wheeler-Zerilli (RWZ) perturbation, as
described in Ref.~\cite{Sarbach2001}.

\begin{figure}[tbp]
  \begin{center}
    \includegraphics[width=0.97\columnwidth]{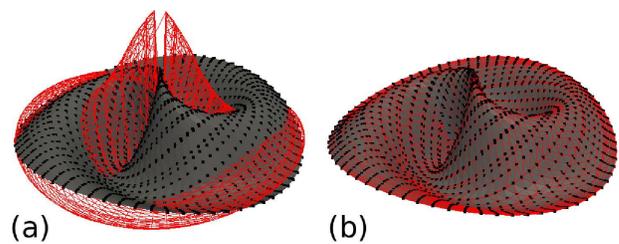}
  \end{center}
  \caption{$\Re(\Psi_4)$ resulting from a traveling-wave perturbation on the equatorial plane of
the computational domain. Panels (a) and (b) correspond to results obtained with and without the coordinate transformation \eqref{eq:cubic}.  
We use $\underline{\Psi_4}$, $\Psi_4$ and $\hat{\Psi}_4$ to respectively denote the analytical result and the values computed on the coordinate and quasi-Kinnersley tetrads. The height of the surface in the vertical direction indicates the value of $\Re(\Psi_4)$,  the solid grey surface denotes $\Re(\underline{\Psi}_4)$, the red wire-frame $\Re(\Psi_4)$ and the black
dots $\Re(\hat{\Psi}_4)$. The amplitude of the red wireframe has been suppressed  by a factor of $10^3$ in (a) so that it fits into the figure. 
The suppression factor has not been applied to Panel (b). 
}
  \label{fig:warp}
\end{figure}
\begin{figure*}[tbp]
  \begin{center}
    \includegraphics[width=0.85\textwidth]{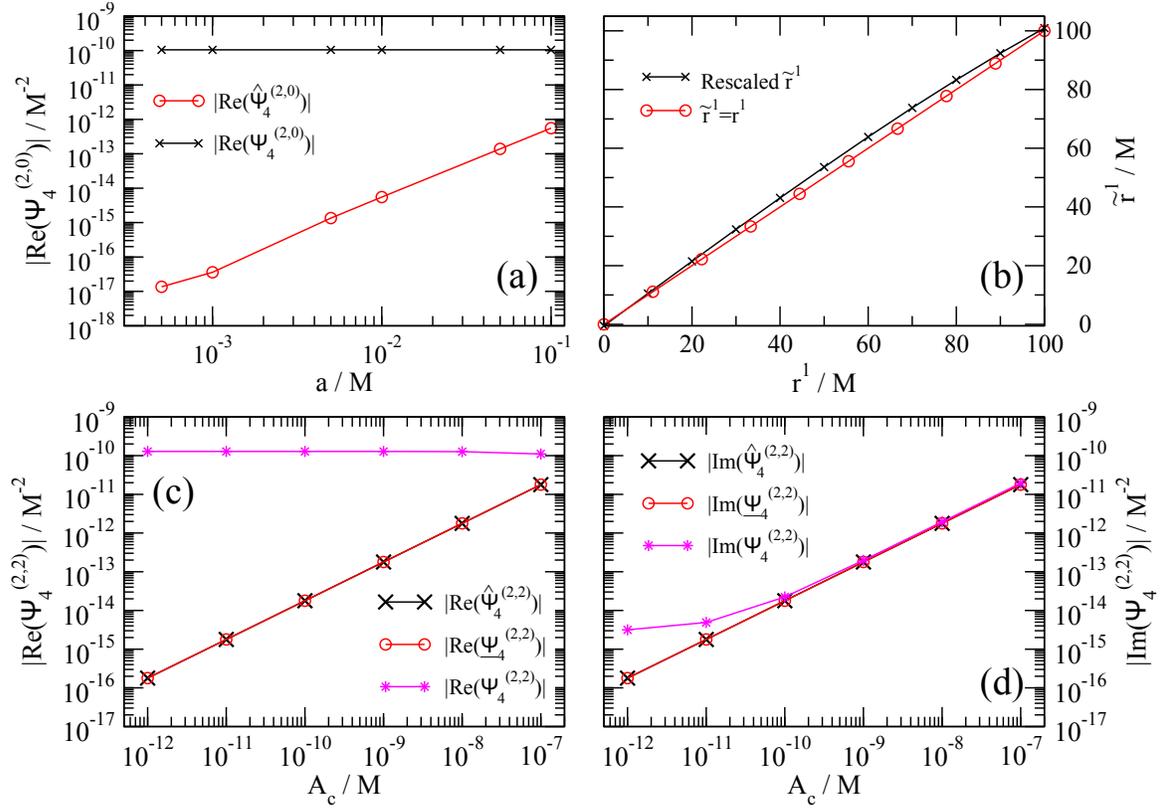}
  \end{center}
  \caption{Testing the QKT's ability to recover perturbation theory results.  An $l=2,m=\pm 2$ perturbation, with magnitude $A_c$, is added as a function
of retarded time in addition to  an $l=1$, $m=0$  perturbation, and finally a cubic coordinate distortion is applied as described in Eq.~\eqref{eq:cubic}. The spherical harmonic coefficients of the coordinate-tetrad $\Psi_4$ and QKT $\hat{\Psi}_4$ are then extracted at Boyer-Lindquist radius $95$M and compared with the analytic results 
$\underline{\Psi}_4$.
(a): The magnitude of the $\Psi_4$, $(l,m)=(2,0)$ mode as a function of spin parameter $a$, with the
radiative $(2,\pm 2)$ perturbation held fixed at $A_c=10^{-9}$M. 
(b): Exploring the effect of cubic rescaling. The rescaled first component 
of $\bm{\tilde{r}}$ vector, $\tilde{r}^1$, 
is plotted against the original component $r^1$. 
The identity map ($\tilde{r}^1 = r^1$) is given for comparison.
(c) and (d): The real and imaginary parts of the ${}_{-2}Y_{22}$ coefficient vs the amplitude $A_c$ of the radiative perturbation. The $l=1,m=0$ mode amplitude is chosen so that the resulting angular momentum perturbation is held constant at $a=0.001$M. 
}
  \label{fig:Perturb}
\end{figure*}

We start with a background Schwarzschild metric in Schwarzschild coordinates expressed
in the standard form of~\cite{Sarbach2001}   
\bea
ds^2 &=& (-\alpha^2 + \gamma^2 \beta^2)dt^2 + 2 \gamma^2 \beta dtdx + \gamma^2 dx^2 \nonumber \\
     &&    + r^2(d\theta^2 + \sin^2 \theta d\phi^2)
\eea
where
\begin{equation} \label{eq:BLCoordPertTest}
\begin{split}
x &= r, \quad \alpha(r) = \sqrt{1-\frac{2M}{r}}, \\
\beta(r) &= 0, \quad \gamma(r)=\frac{1}{\alpha(r)}. 
\end{split}
\end{equation}
We then introduce $l=2,m=\pm 2$ radiative perturbations.
The full RWZ formalism giving the explicit calculation of the perturbed metric is expounded concisely in Appendix A of~\cite{Sarbach2001}; it turns out that
the construction of the perturbed metric and the associated perturbed curvature quantities (such as $\Psi_4$) hinges on one function, the RWZ function $Z$, which obeys the RWZ equation 
\begin{align}
\frac{\partial^2 Z}{\partial t^2} = c_1 \frac{\partial^2 Z}{\partial t \partial r}
+ c_2 \frac{\partial^2 Z}{\partial r^2} + c_3\frac{\partial Z}{\partial t} +
c_4\frac{\partial Z}{\partial r} - \alpha^2 V Z  \label{eq:RWZEQ}.
\end{align}
In the RWZ equation, the coefficients $c_i$ are functions of $\alpha$, $\beta$ and $\gamma$, which for our chosen values [Eq.~\eqref{eq:BLCoordPertTest}] become 
\bea
c_1(r) &=& 0 = c_3(r), \\
c_2(r) &=& \left(1-\frac{2M}{r}\right)^2, \\ 
c_4(r) &=& \frac{2M}{r^2}\left(1-\frac{2M}{r}\right), \\
\alpha^2 V &=& \left(1-\frac{2M}{r}\right)\left(l(l+1) -
\frac{6M}{r}\right)\frac{1}{r^2}. 
\eea
Given  $Z$, the  analytic solution $\underline{\Psi_4}$ for the gravitational wave content in the spacetime can be computed to first order using 
\begin{align}
\underline{\Psi}_4^{(1)} = -\sum_{lm}\left[\frac{i}{r} (\tilde{\Delta} + 2\gamma +
2\mu)\tilde{\Delta} Z_{lm}\right]C_l \left[{}_{-2}Y^{lm}\right] \label{eq:P4Analytic3},
\end{align}
where $C_l = \sqrt{(l-1)l(l+1)(l+2)/4}$, the operator $\tilde{\Delta}$ is $\underline{n}^a \nabla_a$ 
with $\bm{\underline{n}}$ being a null direction associated with the background Kinnersley tetrad, and 
$\gamma$ and $\mu$ are spin coefficients associated with the same background tetrad, which for our case are 
\cite{ChandrasekharBook}
\bea
\mu &=& -\frac{\Delta}{2\Sigma \rho}, \quad \gamma = \mu + \frac{r-M}{2\Sigma} 
\eea
where $\Delta$, $\Sigma$ and $\rho$ are defined in Eq.~(\ref{eq:KerrParaDefs}). 

Note that in the discussion that follows, $\underline{\Psi_4}$ denotes the analytic result while $\Psi_4$ and $\hat{\Psi}_4$ are, respectively, the computed values on the coordinate and quasi-Kinnersley tetrads in the numerical implementation.

In order to solve the linear second order partial differential equation \eqref{eq:RWZEQ} for $Z$, an initial value and time derivative for $Z$ must be specified. For our investigation, we make use of a traveling-wave perturbation of the form
\begin{equation} \label{eq:Perturb}
\begin{split}
Z(t_0,r) = A_c e^{i \omega (t_0-r_*)}, \quad 
\frac{\partial Z}{\partial t}(t_0,r) = i \omega A_c e^{i \omega (t_0-r_*)}
\end{split} 
\end{equation}
where $r_*$
is the usual tortoise coordinate defined by $dr_*/dr=r/(r-2M)$, while $t_0=0$, $\omega = 0.1$ and $A_c$ is a constant initial amplitude. For our test, we also set $M=1$. 
This perturbation is graphically depicted in Fig. \ref{fig:warp}.
The waveform constructed from the perturbation has the classical profile for $\Psi_4$, 
often observed during numerical binary black hole mergers; this is to be expected, since $l=2$ is the dominant mode contributing to the gravitational radiation emitted by a binary.

Next, we numerically compare the coordinate-tetrad $\Psi_4$ and the 
QKT $\hat{\Psi}_4$ with the analytic perturbation-theory result 
$\underline{\Psi}_4$.  
In this test, we adopt the Boyer-Lindquist coordinates; therefore,  
the corresponding coordinate orthonormal tetrad [see Eq.~\eqref{eq:CoordTetrad1}] 
happens to coincide with a  Kinnersley frame of the background spacetime 
[although it is boosted with respect to the Kinnersley tetrad in Eqs.~\eqref{eq:KinnL}-\eqref{eq:KinnM}]. 

To illustrate this more clearly, observe that the standard coordinate tetrad we constructed in Sec.~\ref{sec:coordTetrad} 
results in orthonormal vectors $T^a$ and $N^a$ that are respectively
\bea
T^{a} = \left[\frac{1}{\alpha},\, 0,\, 0,\,0\right],\quad N^{a} = \left[0,\,\alpha,\,0,\,0\right]
\eea  when expressed on the coordinate basis, where 
$\alpha=\sqrt{1-2M/r}$ is the lapse of the background metric. 
 The resulting null vector  $l^a$  constructed according to Eq.~\eqref{eq:OrthonormalVsNullTetrad} is
\bea l^a = \frac{1}{\sqrt{2}}(T^a+N^a) = \frac{1}{\sqrt{2}}\left[\frac{1}{\alpha},\alpha,0,0\right].
\eea
In the static limit, the Kinnersley tetrad 
 [Eqs.~\eqref{eq:KinnL} and \eqref{eq:KinnN}] reduces to 
\begin{equation} \label{eq:BLCoordTetrad}
\hat{l}^a = \left[ \frac{1}{\alpha^2} ,\, 1,\, 0,\, 0\right], \quad \hat{n}^a =  \frac{1}{2}\left[1 ,\, -\alpha^2,\, 0,\, 0\right]
\end{equation}
so $\hat{l}^a = \sqrt{2}l^a/\alpha$ and there exists a relative boost factor of $A = \alpha/\sqrt{2}$ between the coordinate and Kinnersley tetrads.

Therefore, to account for the difference,
we will multiply the coordinate-tetrad $\Psi_4$  
by $(1-2M/r)/2$ throughout this subsection to facilitate comparison with analytical and QKT values. 
With this adjustment, the extracted quantities $\Psi_4$ and $\hat{\Psi}_4$ 
both match the analytically calculated $\underline{\Psi}_4$ from perturbation theory. 
These results are graphically depicted in Fig.~\ref{fig:warp}(b).

Next, we explore the gauge dependence of the QKT result. 
To this end, we introduce a coordinate transformation into some other gauge. As a result, the 
coordinate tetrad associated with the new ``non-privileged gauge'' differs from the QKT, resulting
in a mismatch
between the coordinate $\Psi_4$ and the analytic perturbation theory result $\underline{\Psi}_4$ [see Fig.~\ref{fig:warp}(a)]. 
 The QKT $\hat{\Psi}_4$ implemented in the code should 
 then be able to recover 
the analytic result. 
As an illustrative example, for a 
gauge transformation we choose a cubic rescaling of the spatial coordinates, which takes the radial vector $r^a$ expressed on a Cartesian coordinate basis
defined in Sec.~\ref{sec:QTKcons} to a vector with components $\tilde{r}^a$ using the equation
\begin{align}
\tilde{r}^a = \left( \nu + \frac{\nu^0-\nu}{R^2}|r-r_0|^2\right)(r^a-r_0^a) + r_0^a \label{eq:cubic}. 
\end{align}
where we choose $r_0^a = (5M,0,0)$, $R=100M$, $\nu_0=1$ and $\nu=1.1$. Panel (b) of Fig.~\ref{fig:Perturb}  compares
the coordinates before and after rescaling.

We now calculate the perturbed metric
in the new distorted coordinates and extract the gravitational waves using both the coordinate
tetrad and the QKT, comparing the results with the analytical
$\underline{\Psi}_4$ calculated in the Boyer-Lindquist coordinates and visually portrayed in Fig.~\ref{fig:warp}.
When no coordinate distortions have been introduced [Fig.~\ref{fig:warp} (b)], the coordinate tetrad and
the QKT both generate $\Psi_4$ that matches the analytical
prediction. When we apply the cubic coordinate distortion described above however, the
coordinate tetrad result deviates from $\underline{\Psi}_4$, but the QKT still
recovers the analytical value [Fig.~\ref{fig:warp} (a)].  

In addition to the coordinate transformation we also add a $l=1$, $m=0$ mode [explicit expressions for metric perturbation due to this mode can be found in Appendix A1a of Ref.~\cite{Sarbach2001}]. This mode should make no contribution to the detected radiation in $\hat{\Psi}_4$, but should introduce a small angular momentum perturbation affecting the spin of the spacetime, thus avoid degeneracy in $\hat{\theta}$.  The amplitude of the perturbation is usually set so that the spin of the resulting spacetime is  $a=J/M=0.001$.
After imposing the coordinate distortion and adding the   $l=1$ $m=0$ mode, we extract ${}_{-2}Y_{lm}$ coefficients of $\Psi_4$ and $\hat{\Psi}_4$ on a sphere of Boyer-Lindquist radius $95$M from the black hole. We begin by exploring the effect of the $l=1$, $m=0$ mode.
Figure \ref{fig:Perturb} (a) shows the effect of increasing the
strength of the  $l=1$, $m=0$ perturbation on the $l=2$, $m=0$ mode of extracted waveform; 
recall that we did not introduce an $l=2$, $m=0$ mode into the metric perturbation. 
The coordinate quantity $\Psi_{4}^{(2,0)}$ shows a constant value possibly originating from the cubic coordinate distortion.  The quantity $\hat{\Psi}_{4}^{(2,0)}$, on the other hand, shows a strong dependence on the $l=1$ perturbation
amplitude; this could be due to the fact that the projection onto spherical harmonics, as opposed to spheroidal harmonics, is no longer correct when the perturbing  spin is introduced.  The effect is however small when compared to the magnitude of the $l=2$, $m=2$ modes.

The lower panels of Fig. \ref{fig:Perturb} explore the effect of increasing  the $l=2$, $m=\pm 2$ perturbation amplitude $A_c$ on the extracted $l=2$, $m=\pm 2$ modes.
In general the QKT shows very good
agreement with analytical result over a range of perturbation magnitudes as desired, while the quantities computed on the coordinate tetrad disagree significantly with the analytic perturbative result. 

\section{Application of the QKT to numerical simulations of binary black holes \label{sec:BinaryBoth}}
We now turn to exploring the properties and effectiveness of the QKT scheme when applied to more generic 
numerical simulations involving the collision of two black holes.
 We consider two examples: 
a circular inspiral of two equal-mass, nonspinning black holes 
[Sec.~\ref{sec:Binary}] and 
the head-on collision of two nonspinning, equal-mass black holes 
[Sec.~\ref{sec:HeadOn}]. 
\subsection{Equal-mass, nonspinning binary-black-hole inspiral 
\label{sec:Binary}}
In this subsection, we apply our QKT method to a fully dynamical
simulation of 
two equal-mass, nonspinning black holes that 
inspiral through 16 orbits, merge, and ring down.  
We summarize some
of the physical parameters of this simulation in Table~\ref{tab:scheel} 
(which is a reproduction of Table II of Ref.~\cite{Scheel2009}); 
further details of this simulation and the numerical 
method used are given in Ref.~\cite{Scheel2009} and the references 
therein.
\begin{table}[h]
\begin{tabular}{lr}
\hline
\thickhline
Initial orbital eccentricity: & $e \sim 5 \times 10^{-5}$ \\
Initial spin of each hole: & $S_i/M^2 \leq 10^{-7}$ \\
Duration of evolution: & $\Delta T/M = 4330$ \\
Final black hole mass: & $M_f/M = 0.95162\pm 0.00002$ \\
Final spin: & $S_f/M_f^2 = 0.68646 \pm 0.00004$ \\ 
\thickhline
\hline
\end{tabular} 
\caption{Physical properties of the equal-mass, nonspinning binary-black-hole 
inspiral reported in Ref.~\cite{Scheel2009}. Here $M$ is the sum of the 
Christodoulou masses of the initial holes, and $M_f$ is the Christodoulou 
mass of the final hole.\label{tab:scheel}}
\end{table}

We examine two aspects of the QKT that we have 
considered in previous sections: i)
that the direction $\bm{\hat{l}}$ identified by the QKT corresponds to the 
wave-propagation direction (as discussed in Sec.~\ref{sec:prop}) and the
implications this has for the geometric coordinates $\hat{r}$ and
$\hat{\theta}$,  and ii) that 
the falloff rates of the Newman-Penrose scalars
are consistent with the peeling property (as described in 
Sec.~\ref{sec:Peeling}). 

\subsubsection{Wave-propagation direction}
\begin{figure}[tbp]
  \begin{center}
    \includegraphics[width=0.99\columnwidth]{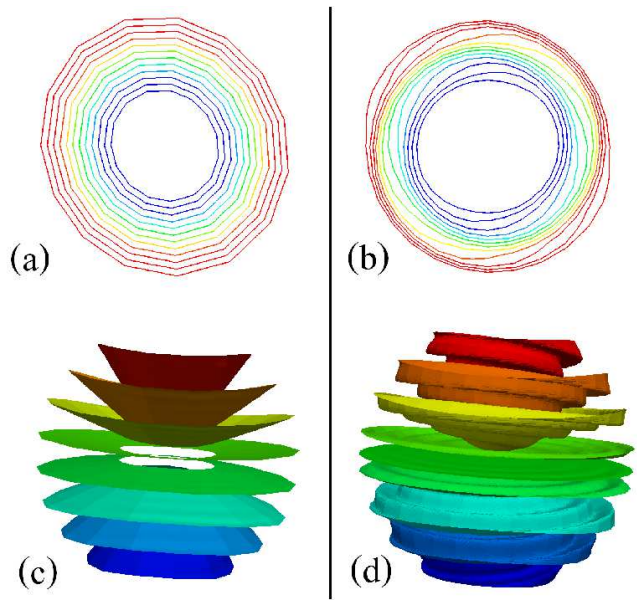}
  \end{center}
  \caption{Geometrical coordinates obtained from the QKF
    $\hat{\Psi}_2$. (a) Contours of $\hat{r}$ on a slice of the null
    preferred characteristic surface (PCS) generated
    by the geodesic developments of $\bm{\hat{l}}$.  (b) Contours of
    $\hat{r}$ on a constant-simulation-time Cauchy slice. (c) Contours
    of $\hat{\theta}$ on the same null surface as (a). (d)
    $\hat{\theta}$ contours in the same Cauchy slice as (b).}
  \label{fig:GeomCoordsProp}
\end{figure}

If (as claimed in Sec.~\ref{sec:prop}) 
the vector $\bm{\hat{l}}$ associated with the QKT
correctly identifies the out-going wave-propagation direction, one
would expect that as one follows the wavefront out to infinity, 
the spacetime curvature
along this trajectory and the associated derived quantities
should become quite simple. To illustrate this, we consider an $S^2$
coordinate sphere
 in the original
 Cauchy slice and identify the correct null direction $\bm{\hat{l}}$ associated
 with the QKT at each point.  We then 
 integrate the null geodesic equations
 outward to produce a null hyper-surface and consider this null
 hyper-surface to be a new slicing and a preferred characteristic
 surface (PCS) of the spacetime. Identifying the geometrical
 coordinates $\hat{r}$ and $\hat{\theta}$ within this slicing, we plot
 their contours in the left-hand panels [plots (a) and (c)
   respectively] of Fig.~\ref{fig:GeomCoordsProp}. For comparison, 
 we also show $\hat{r}$ and $\hat{\theta}$
 computed within the original Cauchy slice.  Note the
 simplicity of the computed geometric quantities within the PCS
 associated with the out-going wavefront as opposed to the
 corresponding quantities computed within the Cauchy surface.

The structure observed within the Cauchy surface can be understood as
follows.
The holes generate 
a rotating mass quadrupole approximately 
given by the Newtonian relation
\begin{align}
\mathcal{I}_{ij}&(t,r) \approx M q^{i}(t-r) q^{j}(t-r) - \frac{M R^2}{12} \delta_{ij} \notag\\
&= \frac{MR^2}{24} \bma 3\cos\left[ 2\Omega t' \right] +1 & 3 \sin\left[ 2\Omega t' \right] & 0 \\
3 \sin\left[ 2\Omega t' \right] & -3\cos\left[ 2\Omega t' \right] +1 & 0 \\
0 & 0 & -2
\ema,
\end{align}
where $M$ is total mass of the binary, $R$ is the separation between
the two black holes, $\bm{q}$ is the location of one of the black holes, and
the choice of coordinates is such that the other black hole is located at
$-\bm{q}$. In the matrix, $\Omega = \sqrt{M/R^3}$ is the orbital angular
velocity and $t' = t-r$.  This quadrupole moment deforms the
$\hat{r}$ contour into an ellipsoid (or peanut shape when
closer to the two holes [see Fig.~\ref{fig:KerrImplied2} (a)]), while
its time dependence causes the orientation of the ellipsoid to rotate
at a frequency of $2\Omega$.

On the PCS, the structure is much simpler.  The inner contour sets the
basic shape for constant $\hat{r}$ surfaces, which are roughly
ellipsoidal. These surfaces then expand, retaining their orientation
as the distortion is propagated outward at the speed of light along
the wavefront. Figure \ref{fig:GeomCoordsProp} (a) shows a concentric
pattern of $\hat{r}$ contours on the null hyper-surfaces, in contrast
to the rotating contours on a spatial Cauchy hyper-surface that slices
through many PCSs, as depicted in Figure \ref{fig:GeomCoordsProp} (b).
The angular $\hat{\theta}$ coordinates similarly display a relative
simplicity on the PCS, taking on the shape of a slightly deformed
(squashed sideways)
cone. Figures \ref{fig:GeomCoordsProp} (c) and (d) show the
$\hat{\theta}$ surfaces on a PCS and in a spatial slicing
respectively; the orientation of the deformed constant $\hat{\theta}$
cones is independent of the distance to black hole in the PCS, but
rotates around when moving outwards on the spatial slice, forming a
``spiral-staircase'' pattern.
\subsubsection{Peeling property}
Next, we explore the falloff rate of the
Newman-Penrose scalars computed on the QKT as one moves outward along
the PCS generators (i.e., along the null geodesics 
tangent to the QKT $\bm{\hat{l}}$ where they originate). 
This rate allows us to quantify to what extent the QKT
obeys the peeling property derived in Sec.~\ref{sec:Peeling} and to
what extent the computed quantities are suitable for use in the
extrapolation procedure prescribed in Sec.~\ref{sec:Extrapolation}.
\begin{figure}[tbp]
  \begin{center}
    \includegraphics[width=0.99\columnwidth]{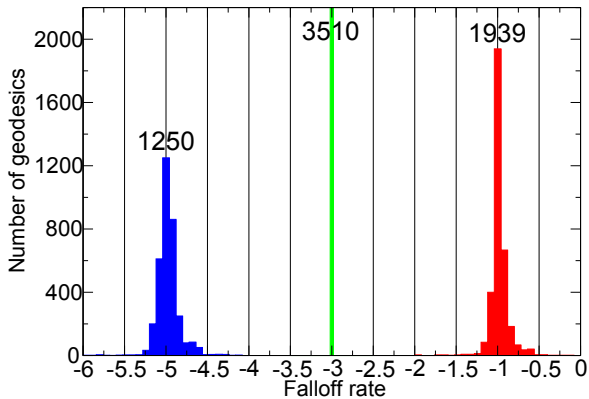}
  \end{center}
  \caption{The distribution of power-law falloff
    rates of Newman-Penrose scalars $\hat{\Psi}_i$
    against affine parameter $\tau$. The three concentrations (colored
    blue, green and red) from left to right indicate 
    falloff rates of $\hat{\Psi}_0$,
    $\hat{\Psi}_2$ and $\hat{\Psi}_4$, respectively. The vertical axis
    indicates the number of geodesics (totaling 3510) with falloff
    rate falling inside bins of width $0.08$, and the number of geodesics
    for the center-most bins are shown.}
  \label{fig:FallOffDist}
\end{figure}

To this end, we start with $3510$ null geodesics from the grid points
of a mesh (of $351$ points) covering a sphere of radius $\hat{r}
\approx 150M$ surrounding the source region.  Over a small time
interval of $10M$, a new set of geodesics are shot off every $1M$. The
affine parameter $\tau$ is initially set to $\hat{r}$ and the
geodesics are evolved for around $150M$. The
Newman-Penrose $\hat{\Psi}_i$'s are recorded at intervals of $\Delta \tau = 1M$
along the geodesics. A histogram of the best fits 
for the power-law falloff (i.e., of the
 slopes of
the $\ln(|\hat{\Psi}_i|)$ vs $\ln(\tau)$ graphs) 
for the $3510$
geodesics are plotted in Fig. \ref{fig:FallOffDist}. 

Recall that in fixing the spin-boost or Type III freedom of the QKF to
obtain the QKT, the $\hat{\bm{l}}$ vector was scaled so that
$\hat{\Psi}_2\propto (\hat{r})^{-3}$. The very sharply defined peak at
$-3$ in Fig.~\ref{fig:FallOffDist} 
provides direct
numerical evidence that 
the relation $\tau \propto \hat{r}$ [cf.~Secs.~\ref{sec:Peeling} and 
\ref{sec:Extrapolation}]
remains
valid at leading order for the considered range of the computational domain. 
 
Figure~\ref{fig:FallOffDist} also indicates that $|\hat{\Psi}_0|$ and
$|\hat{\Psi}_4|$ scale as $\tau^{-5}$ and $\tau^{-1}$, respectively, as
expected from Eq.~(\ref{eq:powerSeries}). 
Here, the peaks are not as sharply defined, since 
we do not by construction enforce the power-law scalings of $\hat{\Psi}_0$ and 
$\hat{\Psi}_4$ (as we do for $\hat{\Psi}_2$).

\subsection{Head-on nonspinning binary merger \label{sec:HeadOn}}
\label{sec:Headon}
To further examine the properties of the QKT and the geometrical
coordinates we now take a detailed look at the numerical simulation of
a head-on merger. The physical parameters of the
simulation are given in Table~\ref{tab:Headon}.

\begin{table}[h]
\begin{tabular}{lr}
\thickhline
Initial separation: & $d/M = 20$ \\
Initial spin of each hole: & $S/M^2 \leq 2 \times 10^{-12}$ \\
Duration of evolution: & $\Delta T / M = 600 $ \\
Final black hole mass: & $M_f/M = 0.987 \pm 2 \times 10^{-3}$ \\
Final black hole spin: & $S_f / M_f^2 = 3. \times 10^{-7} \pm 2 \times 10^{-7}$ \\
\thickhline
\end{tabular}
\caption{Physical parameters of the head-on binary-black-hole merger 
considered in Sec.~\ref{sec:Headon}. Here $M$ is the sum of the initial 
black hole Christodoulou masses, 
and all initial quantities are measured at the initial 
time $t=0$ of the simulation.
\label{tab:Headon}}
\end{table}

\begin{figure}[tbp]
  \begin{center}
    \includegraphics[width=0.95\columnwidth]{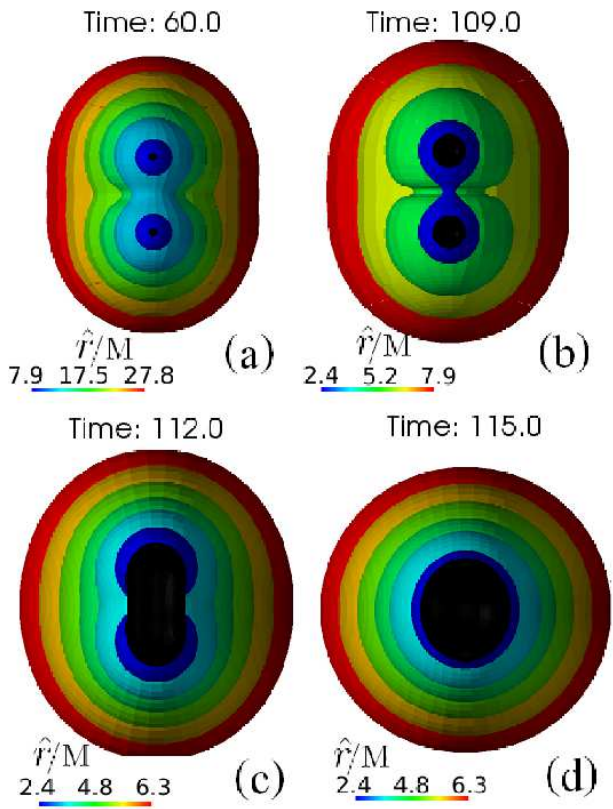}
  \end{center}
  \caption{
    Here we show the evolution of $\hat{r}$
    contours during the merger in the near zone; the black
    surfaces in these plots are the apparent horizons, and at $t=112M$,
    the common apparent horizon forms. 
      }
  \label{fig:HeadonBLr}
\end{figure}

The axisymmetric head-on collision of two nonspinning black holes has
been studied extensively~\cite{Fiske2005,Alcubierre:2004bm,
  Sperhake2005,Baker:2000zh,Baker:2002qf}; in many respects,
these collisions serve as a simple, strongly
nonlinear test of numerical relativity codes.  The
existence of a twist-free azimuthal Killing vector on this spacetime
implies that the metric does not explicitly depend on the azimuthal
coordinate $\phi$ defined about the symmetry axis
and that the angular momentum of the spacetime is zero.
We note that because of the symmetry of this
  configuration, the Coulomb potential associated with the transverse
frame is real and thus that only one geometric
coordinate, the radial coordinate $\hat{r}$, can be determined from
it. Therefore, we fix the latitudinal coordinate to
the simulation coordinate $\theta$.  

\subsubsection{Geometric radial coordinate}

We now explore some of the properties of radial coordinate $\hat{r}$,
the emitted radiation profile, and the waveform.  We show contour
plots of the $\hat{r}$ coordinate at various times near merger in
Fig.~\ref{fig:HeadonBLr}. The characteristic peanut shape expected
from the merger event is clearly visible, and surfaces of constant
$\hat{r}$ coordinate trace both the individual apparent horizon
surfaces at early times and the final apparent
horizon surface at late times. 
Far from the source, constant $\hat{r}$ surfaces become roughly spherical,
indicating that there the geometrical concept of
radius and the gauge choice for the radial coordinate in the
simulation coincide well. For the head-on collision, as
well as in the more dynamical spacetimes depicted in 
Fig.~\ref{fig:KerrImplied2}, plotting surfaces of constant $\hat{r}$ turns
out to be a useful 
tool for visualizing the spacetime
geometry in a way that corresponds to an intuitive feel of the Coulomb
potential's behavior.
\subsubsection{Gravitational waveform \label{sec:HeadonWaveform}}
\begin{figure}[tbp]
  \begin{center}
    \includegraphics[width=0.75\columnwidth]{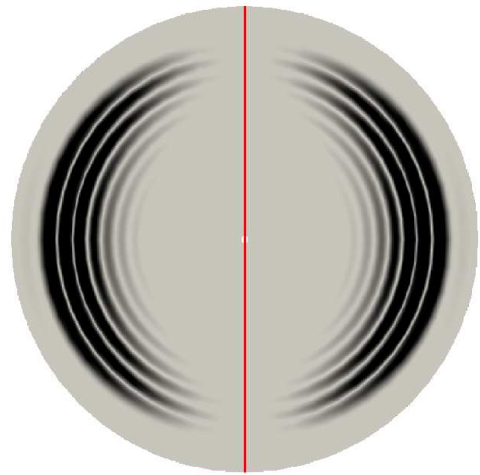}
  \end{center}
   \caption{ 
A snapshot (at $t=242.25M$) of latitudinal distribution of radiation emitted by the head-on
collision. The coloring is according to $|\Re(\hat{\Psi}_4)|$, 
with large values corresponding to darker color. The disk is a vertical 
slice of the computational domain with the thick red line denoting the symmetry
axis.  
\label{fig:HeadonWaveSurface}
}
\end{figure}

\begin{figure}[tbp]
  \begin{center}
    \includegraphics[width=0.95\columnwidth]{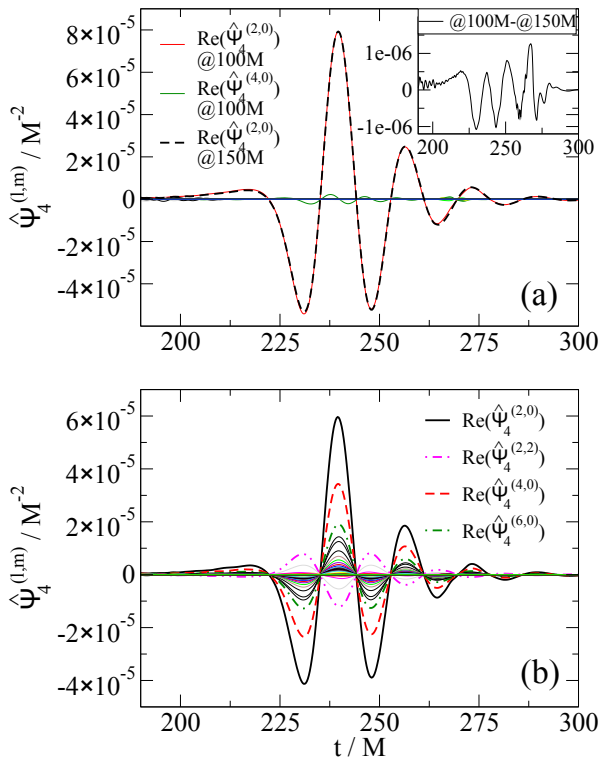}
  \end{center}
  \caption{(a) The $\hat{\Psi}_4$ waveform for the head-on simulation
    extracted at coordinate radii of $r = 100M$ and $r=150M$. For the
    waveform at $r=100M$, all ${}_{-2}Y_{lm}$ modes up to $l=35$ are
    shown. Only $\Re(\hat{\Psi}^{(2,0)}_4)$ and
    $\Re(\hat{\Psi}^{(4,0)}_4)$ are discernibly non-vanishing, with
    the former clearly dominating.  For the $r=150M$ waveform, only
    $\Re(\hat{\Psi}^{(2,0)}_4)$ is shown, it is shifted temporally and
    rescaled by $\approx 1.5$ so that its maximum peak location and
    magnitude match those of the $r=100M$ waveform. The difference
    between the two waveforms is shown in the top-right inset.  (b)
    The spherical harmonic decomposition for $\hat{\Psi}_4$ at
    $r=100M$ obtained by choosing the poles of harmonics along a
    direction orthogonal to the symmetry axis. Multiple modes are
    visible, and a few with the largest magnitudes are labelled.
    }
  \label{fig:HeadonWaveForm}
\end{figure}

For twist-free axisymmetric spacetimes, one can show in
general~\cite{Fiske2005} that if the imaginary part of the tetrad null
vector $\bm{m}$ or $\bm{E^3}$ (as defined in Sec.~\ref{sec:tetradsss})
has the same direction as the azimuthal Killing vector, then $\Psi_4$
expressed on this tetrad is real.  Figure~\ref{fig:HeadonWaveSurface}
depicts $\Re(\hat{\Psi}_4)$ for the  
head-on collision presently under consideration.
Note the absence of radiation along the
symmetric axis in Fig \ref{fig:HeadonWaveSurface}; this is a feature
we will examine further later in this section in the context of PNDs.

We show two possible
spherical harmonic decompositions of $\hat{\Psi}_4$ in
Fig.~\ref{fig:HeadonWaveForm}. Panel (a) corresponds to
the case where the azimuthal Killing vector determines the $\theta =
0$ direction; because 
the symmetry-axis corresponds to the $\theta = 0$ direction,
axisymmetry implies that there are no $l=2$, $m=\pm
2$ modes in the spherical harmonic decomposition, only $m=0$ modes
exist, and of those the $l=2$, $m=0$ mode makes the dominant
contribution.

\begin{figure*}[tbp]
  \begin{center}
    \includegraphics[width=0.95\textwidth]{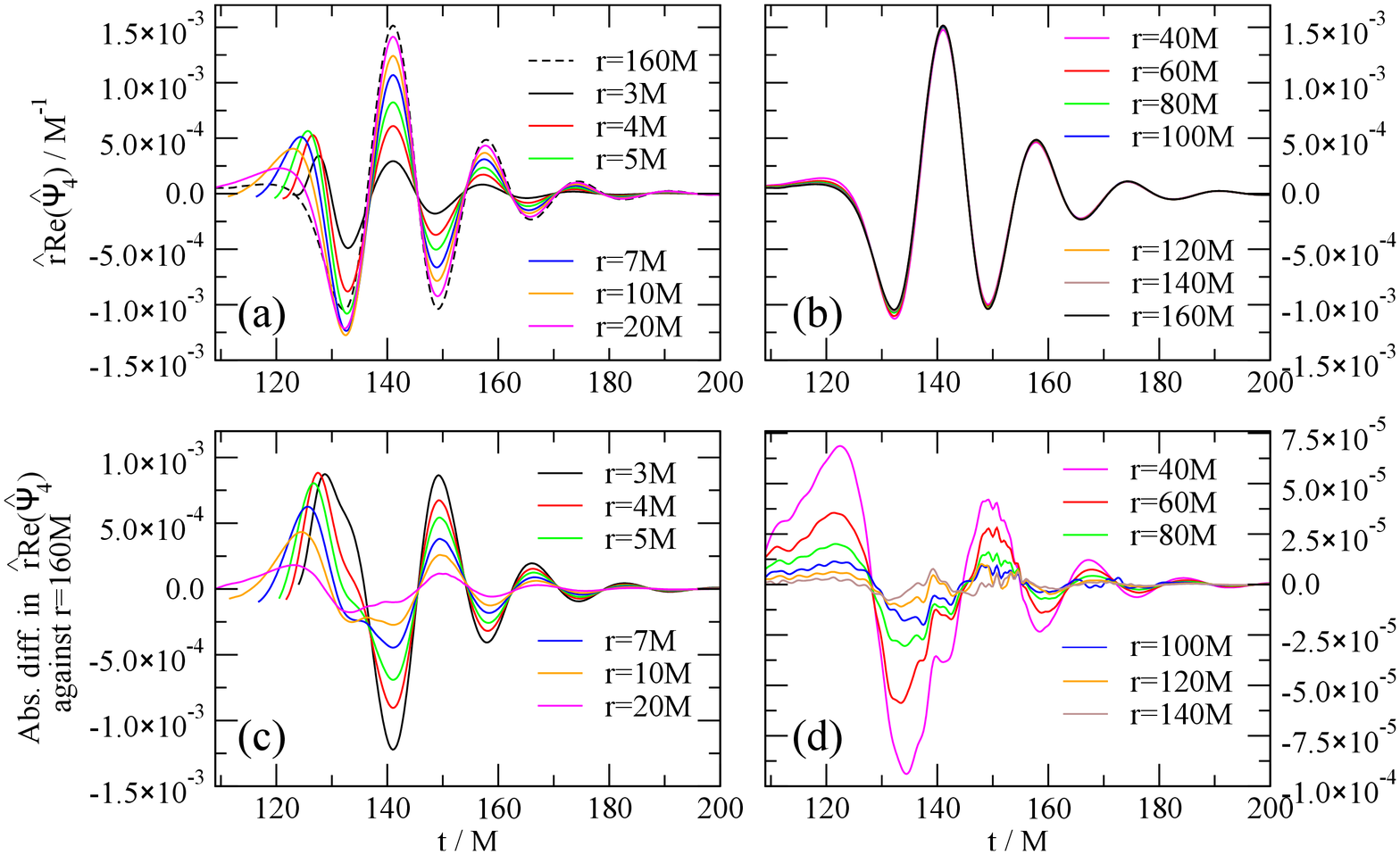}
  \end{center}
  \caption{Panels (a) and (b): The gravitational-wave
      signal
    $\hat{r}\Re(\hat{\Psi}_4)$ curves extracted at several fixed
    spatial (simulation) coordinates. Panels (c) and (d): The absolute
    difference between these waveforms with a reference waveform
    computed at $r=160M$. 
  }
  \label{fig:SinglePointWaveForm}
\end{figure*}

On the other hand, if one relabels the $\theta$ and
  $\phi$ coordinates on the extraction sphere, the spherical harmonic
  decomposition of the same waveform is very different. 
  Panel (b) of 
  Fig.~ \ref{fig:HeadonWaveForm} instead chooses the
  $\theta=0$ line to be orthogonal to 
  the axisymmetry axis rather than along it [as in 
  panel (a)]; a significant 
  $l=2$, $m=2$ mode appears. 
  This is a simple example illustrating the well-known fact that
  unless a clear prescription for the preferred axis of a simulation
  is given, the $l=2$, $m=2$ mode is an ambiguous description of the
  radiation. Solutions to this problem for
  generic black hole binary simulations
  (which include precession) have been proposed in literature. For example, 
  one may first choose a ``radiation axis''~\cite{Schmidt2010,OShaughnessy2011}
  to maximize the component of
  the angular momentum along itself and secondly choose a preferred rotation
  about that axis~\cite{Boyle:2011gg}. Although not yet fully explored, 
  our geometric coordinates suggest 
  an alternative resolution. Namely we can use the extrema of the
  computed geometric $\hat{\theta}$ coordinates to identify the polar
  regions of a simulated spacetime.  

Another question especially relevant to wave extraction is 
how rapidly the waveform computed from
$\hat{\Psi}_4$ in the computational domain converges to ``the''
correct asymptotic waveform.  In Sec.~\ref{sec:Peeling}, we 
argue that asymptotically the QKT quantities on the correct out-going
geodesics (as described in Sec.~\ref{sec:Extrapolation}) should converge
very rapidly to the desired result. We now explore this statement 
quantitatively for the emitted radiation on the equatorial
plane in the head-on binary-black-hole merger we are considering. The
 goal is to determine at which radius a reliable
approximation of the asymptotic waveform is attained.

\begin{figure}[tbp]
  \begin{center}
    \includegraphics[width=0.95\columnwidth]{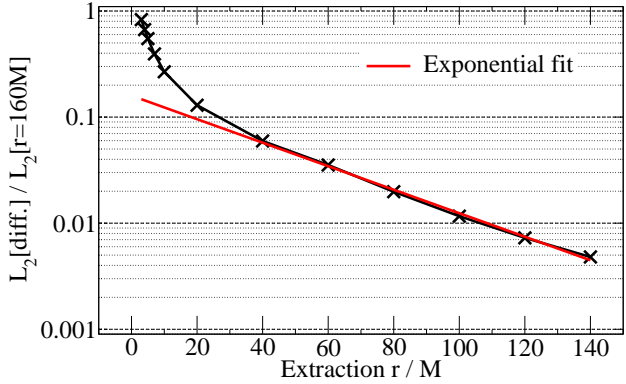}
  \end{center}
  \caption{Exponential convergence of the QKT waveform extracted in the interior to 
the reference waveform measured at $r=160M$. This plot shows the
    $L_2$ norm [defined in Eq.~\ref{eq:L2NormHeadon}] of the absolute
    difference between the waveform at a particular extraction radius and the 
reference waveform at $r=160M$ normalized by dividing by the $L_2$ norm of the
reference waveform. 
   Also shown is the red fitted exponential curve for radii $r>40M$, 
which has a slope of $-0.011$.}
  \label{fig:SinglePointWaveFormConvergence}
\end{figure}

To locate a good cut-off radius for wave extraction, consider a set of
non-inertial observers hovering at different fixed spatial
(simulation) radii in the equatorial plane. For each observer, 
in Fig.~\ref{fig:SinglePointWaveForm} we record
the $\hat{r}\hat{\Psi}_4$ value as a function of time and plot the
resulting curves, with the origin shifted so that the
central maxima of all the curves coincide at around $t=140M$.  
For clarity, we
have divided the curves into two sets, those originating at $r < 30M$
and at $r > 30M$, which we display in panels (a)
and (b), respectively. In both panels the curve traced out at
$r=160 M$ is given for comparison, and we will refer to it as the 
reference waveform.  Panels (c) and (d) show
the absolute difference between the curves extracted at the various
interior points and the reference waveform.

In Fig.~\ref{fig:SinglePointWaveFormConvergence},
we plot the fractional difference between reference waveform and the interior
waveform as a function of extraction radius. The quantity plotted is the $L_2$ 
norm of the absolute difference between the two waveforms divided by the $L_2$ norm of the reference waveform. 
The $L_2$ norm is defined here as 
\bea \label{eq:L2NormHeadon}
L_2[f] = \sqrt{\int_{t=125}^{200} f^2(t) dt}.
\eea
Comparing Fig.~\ref{fig:SinglePointWaveFormConvergence} and panel (b) in 
Fig.~\ref{fig:SinglePointWaveForm}, we find that
for radii greater than $r=40M$ the extracted waveform
corresponds  closely,  within $\sim 5\%$, 
to the reference waveform. 
Figure~\ref{fig:SinglePointWaveFormConvergence} quantifies 
this further: for $r<40 M$, the errors in the
extracted $\hat{\Psi}_4$ waveform are large but the convergence to the
reference waveform is super-exponential, while for $r>40 M$, the errors 
converge exponentially.
This provides quantitative justification 
for the rapid-convergence claims we made in
Secs.~\ref{sec:Peeling} with
respect to the Newman-Penrose scalars calculated on the QKT.
We conclude that the radius 
$\hat{r} = 40M$ appears to be a good minimal extraction radius for QKT
quantities for head-on binary-black-hole collisions.  
It would be
interesting to explore, using a similar analysis, 
whether the exponential convergence properties and the
value for a good minimal extraction radius change considerably when 
applied to generic spacetimes (i.e., to spacetimes with less symmetry than the 
head-on collision we consider here).

\subsubsection{Principal null directions}

\begin{figure}[tbp]
  \begin{center}
    \includegraphics[width=0.95\columnwidth]{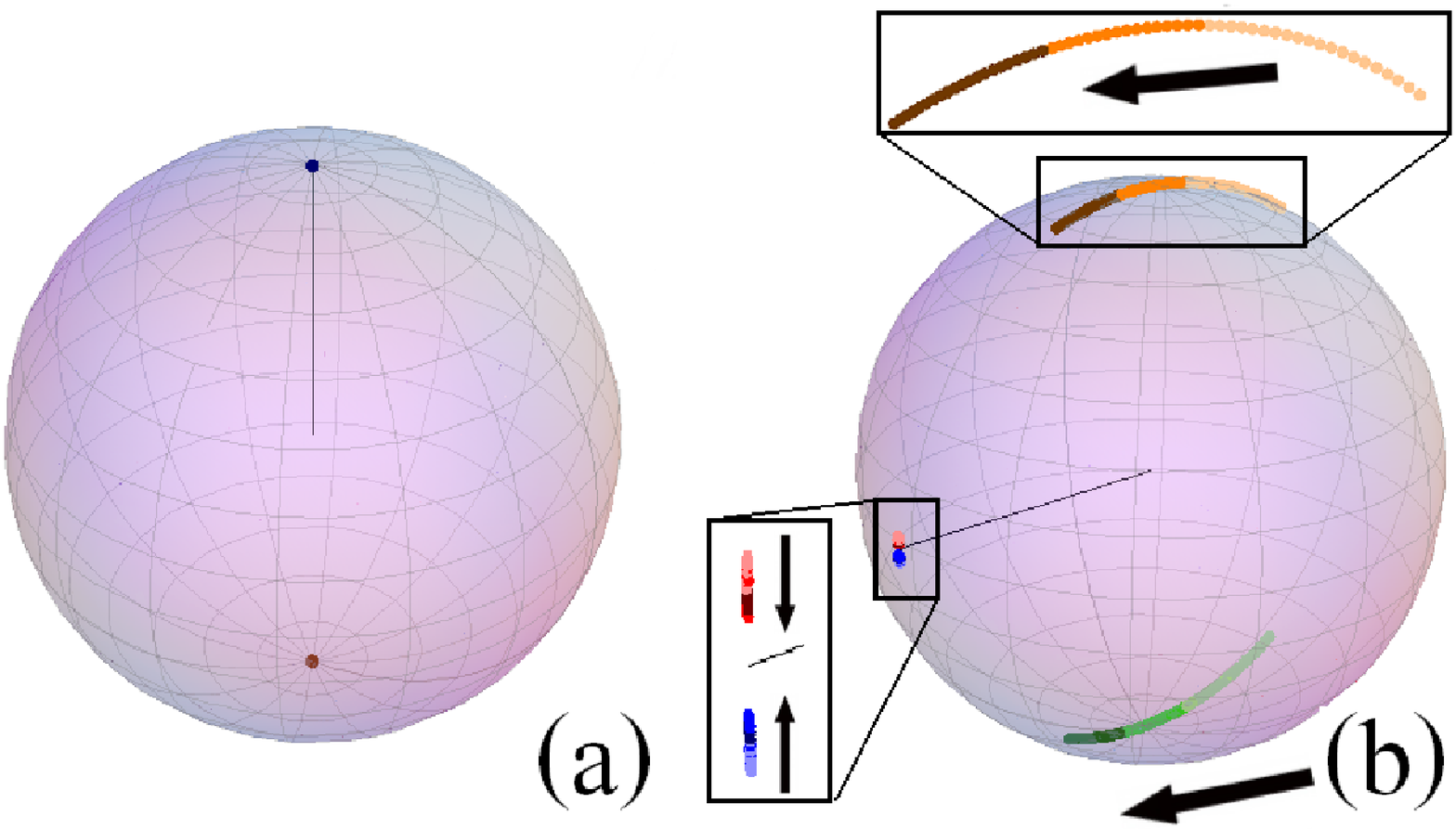}
  \end{center}
  \caption{A graphical representation of the peeling-off behavior of the
   PNDs along out-going null geodesics for the head-on collision. 
	The null geodesics start at $r = 30$M at time $t=152$M. 
	The tangent $\bm{\ell}$ to the geodesics coincide with $\bm{\hat{l}}$ along the geodesic 
	and is denoted in the figures by black radial lines. 
	The geodesic shown in panel (a) travels outward along the symmetry axis 
	where no gravitational radiation is emitted.
	The geodesic shown in panel (b) starts off on the equatorial plane, and  
	all PNDs converge toward the out-going $\bm{\hat{l}}$ direction pointing toward 
	the front of the sphere.
	In this figure, the arrows indicate the movements of the PNDs as one travels further along the geodesic. 
	Darker coloring indicates points calculated at larger affine parameter along  the geodesic. 
}
  \label{fig:HeadonPeeling}
\end{figure}

We conclude our investigation of the spacetime associated with the
head-on collision of two black holes by exploring the behavior of the
PNDs on and off the axis. As we have observed 
and shown graphically
in Fig.~\ref{fig:HeadonWaveSurface}, no radiation is emitted along the
symmetry axis.  This lack of radiation suggests that the PNDs do not
all converge into a Type N pure radiation configuration as seen in
Fig.~\ref{fig:peeling} (c); instead, 
we would expect the PNDs to remain in a
 Type-D configuration, with two
pairs of PNDs trapped at antipodal points of the
anti-celestial sphere. On the axis, the only non-vanishing
Newman-Penrose scalar on the QKT is $\hat{\Psi}_2$. 
The four solutions
in Eq.~\eqref{eq:PairSymmetry} then divide into a pair whose value
diverge and a pair that approach zero. The diverging
solutions give us two PNDs pointing along the $\bm{\hat{n}}$ direction, 
while the
vanishing solutions coincide with $\bm{\hat{l}}$. This situation is
depicted in Fig. \ref{fig:HeadonPeeling} (a):
the two PNDs
at the bottom of the sphere
point away from $\bm{\hat{l}}$ (represented by the radial line) and do not converge onto
the other two PNDs that coincide with
$\bm{\hat{l}}$ at the top of the anti-celestial sphere.
All the PNDs on the top of the sphere form angles smaller than 
$4\times 10^{-7}\pi$
with the spatial projection of $\bm{\hat{l}}$,
while the other two PNDs form angles with the spatial projection of $\bm{\hat{n}}$
that are smaller than $4\times 10^{-6} \pi$.
For comparison, in Fig.~\ref{fig:HeadonPeeling}(b), we 
show the PND behavior on the celestial sphere as one moves along a geodesic
that points away from the symmetry axis. The geodesic shown in this plot starts out at an orientation of
$0.508\pi$ to the symmetry axis. (A similar plot is made in
Fig.~\ref{fig:peeling}(c) where we showed a geodesic starting at
$0.396 \pi$ to the symmetry axis). 
In these two cases, the presence of
radiation causes the two PNDs pointing away 
from $\bm{\hat{l}}$
to converge onto the other two PNDs surrounding $\bm{\hat{l}}$ 
as one moves outward along the geodesic. 
The dominant rate of convergence is
$1/\tau$ [cf. Eq.~\eqref{rooteq3}].

The existence of ``critical'' directions
as demonstrated here in the special case of axisymmetric spacetimes, is a generic feature of all dynamical spacetimes:
it is a topological necessity [see~\cite{Zimmerman2011} and page 173 of~\cite{Penrose1965}]. 
Specifically, in Ref.~\cite{Zimmerman2011}, the authors 
explain this feature as follows: 
In the asymptotic region, gravitational radiation is transverse 
and can be represented by  tendex and vortex lines tangent to spheres of constant $\hat{r}$. 
Then, the Poincar\'{e}-Hopf theorem dictates that there must be locations 
on the sphere where the tendicity associated with the two transverse 
eigenbranches of the gravitoelectric tensor $\bm{\mathcal{E}}$ become degenerate. 
The trace-free property of $\bm{\mathcal{E}}$ then further 
constrains the tendicity to be zero---i.e., requires the gravitational radiation to vanish at the critical points.

Another useful characterization of dynamical numerical simulations is a measure of how rapidly 
the spacetime settles down to Petrov Type D at late times 
[see for example~\cite{Campanelli:2008dv,Owen:2010vw}]. As a graphic depiction of this evolution of spacetime,
one may generate PND diagrams similar to Fig.~\ref{fig:HeadonPeeling} 
along timelike worldlines (instead of null geodesics).  
If one desires a 
quantitative estimate of the rate at which this ``settling down'' occurs, 
a metric on the anti-celestial sphere has to be defined 
in order to calculate distance between the PNDs.   
However a unique prescription of this metric requires a unique prescription of the tetrad, since
 Lorentz transformations on tetrad result in conformal transformations on the anti-celestial sphere.
The QKT construction is useful in this context, because
it uniquely prescribes all the tetrad degrees of freedom 
simultaneously across spacetime, including along timelike worldlines. 

\section{Conclusion}
\label{sec:Conclusion}
As the numerical relativity codes mature, it is becoming increasingly important 
to introduce a protocol that allows us to extract and compare physics from these 
codes in an unambiguous fashion. In particular, the quantities computed should be 
independent of the gauge  or the formulation used. 
Ideally, such a protocol should be valid in the strong-field and wave zones 
and meet the physical criteria outlined in Sec. \ref{sec:Ana}. 

In this paper, we have suggested one such approach. Based on the Newman-Penrose formalism, 
our method fully specifies the tetrad degrees of freedom using purely geometric considerations, 
and two of the gauge degrees of freedom are also uniquely fixed using the curvature invariants $I$ and $J$.
In particular, our tetrad construction makes use of the 
quasi-Kinnersley frame (QKF) \cite{Beetle2005,Nerozzi2005,Burko2006,Nerozzi:2005hz,Burko:2007ps}, 
which is a transverse frame (TF) that 
contains the Kinnersley tetrad in the Kerr limit.

By exploiting the relationship between QKF and eigenvectors 
of the matrix representation $\bm{\mathcal{Q}}$ 
[as in Eq.~\eqref{eq:PsiMatrix}] of the Weyl tensor, one can arrive
at several insights regarding the physical properties of the QKF: i)
its null vector $\bm{\tilde{l}}$ 
has a spatial projection pointing along the
super-Poynting vector [Sec.~\ref{sec:Direction}] and thus along the
direction of wave propagation, and ii) there is [Sec.~\ref{sec:Peeling}] a close
relationship [Fig.~\ref{fig:peeling}(b)] between the QKF null basis and principal null directions
(PNDs) that makes the QKF naturally suited to measuring 
how quickly PNDs bunch together (as they converge onto $\bm{\tilde{l}}$). 
These features help Newman-Penrose
scalars extracted using a QKT to fall off
correctly in accordance with predictions by the peeling theorem. 

In the QKF, the eigenvalue $\hat{\Psi}_2$ of the complex matrix 
$\bm{\mathcal{Q}}$ corresponding to the eigenvector that 
gave us QKF is a curvature invariant thus independent of the slicing in which the calculation was performed.
The physical interpretation of  $\hat{\Psi}_2$ 
is that it represents  the Coulomb background portion of Weyl
tensor; using this quantity, we define a pair of geometric coordinates $\hat{r}$
and $\hat{\theta}$ [Sec.~\ref{sec:COORDS}]. 
These geometric coordinates vividly depict the multipolar 
structure in the Coulomb potential (as can be seen from 
Figs.~\ref{fig:GeomCoordsProp} and \ref{fig:HeadonBLr}). 
For example, they were used to demonstrate that
far enough away from their source
[see Sec.~\ref{sec:HeadOn} for an empirical cutoff],
the Coulomb background $\hat{\Psi}_2$ [Fig.~\ref{fig:GeomCoordsProp}]
 appears to propagate with an almost invariant form along preferred characteristic surfaces
whose generators are geodesics started off in the QKT $\bm{\hat{l}}$ direction.
Besides fixing the gauge freedom,
we have also used the differentials $d\hat{r}$ and $d\hat{\theta}$ [Sec.~\ref{sec:SpinBoostFixingByCoords}] 
to eliminate the spin-boost
freedom remaining in QKF, yielding a final, gauge-invariant 
quasi-Kinnersley tetrad (QKT).

As our QKT is constructed from the gauge-invariant characteristic structure of
Weyl tensor, it can be used to explore the 
physical features of numerical spacetimes in a gauge-invariant 
way.
We have demonstrated this desirable property of our QKT 
with i) a stationary black hole spacetime where the hole is offset from the 
origin [Sec.~\ref{sec:ShiftedKerr}] as well as ii) a gauge wave 
[Sec.~\ref{sec:Evil}] or
iii) physical wave [Sec.~\ref{sec:Num2}] added to the spacetime of a 
Schwarzschild black hole. These examples serve as useful test beds for codes seeking to 
unambiguously extract the physically real effects as opposed to gauge induced false signals. 

We have also used the QKT to analyze two equal-mass, nonspinning 
binary-black-hole merger
simulations. In the first, the two black holes inspiral [Sec.~\ref{sec:Binary}] toward each other, while in the second they plunge
head-on [Sec.~\ref{sec:HeadOn}]. We have confirmed that the
Newman-Penrose scalars under the QKT do indeed fall off at the rates
expected from peeling theorem in these simulations
[Fig.~\ref{fig:FallOffDist} and Fig.~\ref{fig:SinglePointWaveForm}], and 
we have explicitly examined the special
peeling behavior along ``critical'' directions whose existence is ensured by topology 
[Fig.~\ref{fig:HeadonPeeling}].

The gauge invariant 
feature of the proposed framework lends itself to several uses. 
One possible application is that 
they could help eliminate ambiguities such as the pole direction of 
harmonics used to express gravitational waves [see Sec.~\ref{sec:HeadonWaveform}]. 
A further application is to use the QKT to reduce the ambiguity in 
measuring how quickly 
a spacetime settles down to a Type-D spacetime [see the end of Sec.~\ref{sec:HeadOn}]. 
The QKT also is promising  
as a wave extraction method that can
be performed real time and ensures that  waveform approaches
its asymptotic value at infinity as rapidly as possible 
(this is illustrated in Fig.~\ref{fig:SinglePointWaveForm}).
For future work, we plan to make a comparison between QKT-based wave extraction and other wave
  extraction techniques (such as Cauchy
  Characteristic Extraction~\cite{Bishop1996,Bishop:1997ik,Bishop1998,Reisswig:2009rx,Winicour2009,Babiuc:2010ze})
 using various numerical simulations with generic initial conditions.

The validity of geometric coordinates and the QKT throughout spacetime, 
including the strong field regions, suggests that they could be effectively utilized 
as a visualization and diagnostic tool capable of tracking
the evolution of dynamical features of the spacetime. 
For example, in Fig.~\ref{fig:NoiseDiagram} (a), the geometric coordinates point out the missing rotating quadrupolar
moment in the initial data. We expect this type of visualization to prove valuable in ongoing efforts to construct more
realistic initial data and reduce spurious ``junk'' radiation. 
Other utilities for the geometric coordinates include,
e.g., their potential for helping to improve boundary matching algorithms
[see Fig.~\ref{fig:NoiseDiagram} (b)].
We would further like to examine in greater depth, with the help of QKT, the mechanics behind the changes 
in waveform, as one moves closer to the source region [see Fig.~\ref{fig:SinglePointWaveForm}]. 
 
Finally, we note that there should exist  a close relationship between
  the QKT and the tendex and vortex infrastructure introduced in~\cite{OwenEtAl:2011,Nichols:2011pu},  
which is based on the real
  eigenvectors and eigenvalues of $\bm{\mathcal{E}}$ and $\bm{\mathcal{B}}$. Our geometric coordinates and QKT are, in contrast, based on the complex eigenvalues and eigenvectors of $\bm{\mathcal{Q}}$. We expect this connection to yield important insights such as the slicing dependence of the tendexes and vortexes.

\acknowledgments
We would like to thank Luisa Buchmann for alerting the NR group at Caltech about the potential of the QKTs to aid wave extraction, and 
we are indebted to Andrea Nerozzi and Rob Owen for many useful discussions and to Mark Scheel for numerous assistance with the SpEC code. 
This research was supported by
NSF grants PHY-0653653, 
PHY-0601459, PHY-1068881, and PHY-1005655 at Caltech and by NSF 
grants PHY-0969111, PHY-1005426 at Cornell,
by NASA grants NNX09AF97G and NNX09AF96G,
and by the Sherman Fairchild Foundation to Caltech and Cornell and the Brinson Foundation to Caltech. The numerical computations in this paper were completed using the Caltech computer cluster \textsc{zwicky}, which was funded by the Sherman Fairchild Foundation and the NSF MRI-R2 grant PHY-0960291 to Caltech.

\bibliography{PseudoKinnersleyPaper.bbl}
\end{document}